\documentclass[12pt]{article}
\usepackage{graphicx}
\usepackage{bm}
\usepackage{amsmath}
\usepackage{hhline}
\usepackage{amssymb}
\usepackage{times}
\newlength{\dinwidth}
\newlength{\dinmargin}
\begin{document}
\newcommand{\Pomeron}{I\!\!P}
\begin{titlepage}
\begin{center}
\begin{huge}
{\bf Small--$x$ Physics: \\[.3ex] 
From HERA to LHC and beyond}\end{huge} \\
\vspace{1.5cm}
\begin{Large}
Leonid  Frankfurt  \\
\end{Large}
School of Physics and Astronomy, Tel Aviv University, 69978 
Tel Aviv, Israel
\\[.7cm]
\begin{Large}
Mark Strikman  \\
\end{Large}
Department of Physics, Pennsylvania State
University, University Park, PA 16802, USA
\\[.7cm]
\begin{Large}
Christian Weiss \\
\end{Large}
Theory Group, Jefferson Lab, Newport News, VA 23606, USA
\vspace{0.5cm}
\end{center}
\vspace{.5cm}
\begin{abstract}
\noindent
We summarize the lessons learned from studies of hard scattering
processes in high--energy electron--proton collisions at HERA and
antiproton--proton collisions at the Tevatron, with the aim of
predicting new strong interaction phenomena observable in
next--generation experiments at the Large Hadron Collider
(LHC). Processes reviewed include inclusive deep--inelastic scattering
(DIS) at small $x$, exclusive and diffractive processes in DIS and
hadron--hadron scattering, as well as color transparency and nuclear
shadowing effects.  A unified treatment of these processes is
outlined, based on factorization theorems of quantum chromodynamics,
and using the correspondence between the ``parton'' picture in the
infinite--momentum frame and the ``dipole'' picture of high--energy
processes in the target rest frame. The crucial role of the
three--dimensional quark and gluon structure of the nucleon is
emphasized.  A new dynamical effect predicted at high energies is the
unitarity, or black disk, limit (BDL) in the interaction of small
dipoles with hadronic matter, due to the increase of the gluon density
at small $x$. This effect is marginally visible in diffractive DIS at
HERA and will lead to the complete disappearance of Bjorken scaling at
higher energies. In hadron--hadron scattering at LHC energies and
beyond (cosmic ray physics), the BDL will be a standard feature of the
dynamics, with implications for \textit{(a)} hadron production at
forward and central rapidities in central proton--proton and
proton--nucleus collisions, in particular events with heavy particle
production (Higgs), \textit{(b)} proton--proton elastic scattering,
\textit{(c)} heavy--ion collisions. We also outline the possibilities
for studies of diffractive processes and photon--induced reactions
(ultraperipheral collisions) at LHC, as well as possible measurements
with a future electron--ion collider.
\end{abstract}
\end{titlepage}
\tableofcontents
\newpage
\section{Introduction}
\label{sec:intro}
In understanding the nature of strong interactions, progress has
mostly come from the investigation of certain ``extreme'' kinematic
regions, in which the dynamics simplifies. One such region is
high--energy hadron--hadron scattering, in which the center--of--mass
energy is significantly larger than the masses of the hadronic systems
in the initial and final state.  Historically, this was the first area
in which powerful mathematical methods, such as dispersion relations
and Reggeon calculus, could be applied to strong interaction
phenomena. They are based on the general principles of unitarity of
the scattering matrix (conservation of probability) and analyticity of
scattering amplitudes (causality). These methods have given us
important insights into general properties of high--energy processes,
such as the increase of the radius of interaction with energy
predicted by V.~Gribov \cite{Gribov:ex,Gribov:1973jg}, the Froissart
bound for the growth of total hadronic cross sections with energy
\cite{Froissart}, and the Pomeranchuk theorem of asymptotic equality
of particle--particle and particle--antiparticle cross sections.

\begin{figure}[t]
\begin{center}
\includegraphics[width=12cm]{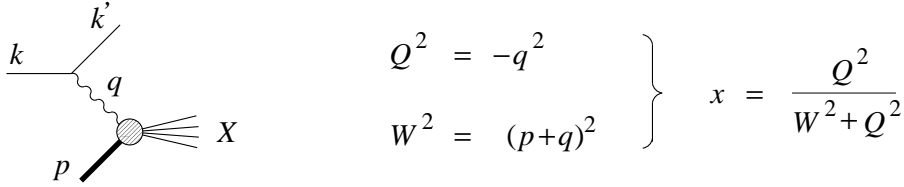}
\end{center}
\caption{The kinematics of deep--inelastic lepton--hadron scattering (DIS).
The interaction proceeds by exchange of a virtual photon, whose 
four--momentum is given by the difference of the lepton momenta, 
$q = k' - k$. The hadronic scattering process is characterized by two 
kinematic invariants, the photon virtuality, $q^2 \equiv - Q^2 < 0$, 
and the photon--proton center--of--mass energy, $W$,
or, alternatively, the Bjorken scaling variable, $x$.
(We neglect the target mass.)
\label{fig:dis}}
\end{figure}

Further progress came with the study of ``hard'' scattering processes,
characterized by a momentum transfer significantly larger than the
typical mass scale associated with hadron structure, $\mu$ (a
reasonable numerical value for this scale is the $\rho$ meson mass).
Such processes can be described in quantum chromodynamics (QCD), the
field theory of interacting quarks and gluons, the fundamental
property of which is the smallness of the effective coupling constant
in small space--time intervals (asymptotic freedom)
\cite{Gross:1973id,Politzer:1973fx}. Hard processes happen so ``rapidly''
that they do not significantly change the environment of the
interacting quarks and gluons inside the hadrons. This allows one to
calculate their amplitudes using a technique called factorization ---
a systematic separation into a hard quark--gluon scattering process
and certain functions describing the distribution of quarks and gluons
in the participating hadrons.  The simplest such process is
deep--inelastic lepton--hadron scattering (DIS) in the so--called
Bjorken limit, $Q^2 \sim W^2 \gg \mu^2$, see
Fig.~\ref{fig:dis}. Historically, the observation of scaling behavior
in the structure functions of inclusive DIS \cite{Bjorken} gave the
first indication of the presence of quasi--free, pointlike
constituents in the proton \cite{Feynman}.  Another class of processes
for which factorization is possible are certain hard processes in
hadron--hadron scattering, such as the production of jets with large
transverse momenta or large--mass dilepton pairs.

A particularly interesting region of strong interactions are hard
scattering processes in the region where the center--of--mass energy
becomes large compared to the momentum transfer, $W^2 \gg Q^2 \gg
\mu^2$. In DIS this limit corresponds to values of the Bjorken
variable $x \ll 1$ (see Fig.~\ref{fig:dis}), whence this field is
known as ``small--$x$ physics.'' On one hand, because of the large
momentum transfer, such processes probe the quark and gluon degrees of
freedom of QCD. On the other hand, they share many characteristics
with high--energy hadron--hadron scattering, such as a large spatial
extension of the interaction region along the collision axis (this
will be explained in detail in Sec.~\ref{subsec:spacetime}). The
treatment of such processes generally requires a combination of the
methods of QCD factorization and ``pre--QCD'' methods of high--energy
hadron--hadron scattering for modeling the dynamics of the hadronic
environment of the quarks and gluons participating in the hard
process. From the point of view of QCD, the high--energy (small--$x$)
region corresponds to a greatly increased phase space for gluon
radiation as compared to $x \sim 1$. QCD predicts a fast rise of the
gluon density in the nucleon with decreasing $x$, and thus a strong
increase of the DIS cross section with energy
\cite{Gross:ju,DGLAPd}.  A challenging question, which is
presently being addressed in different approaches, is the role of
unitarity of the scattering matrix in such processes at high energies.
More generally, one hopes to eventually understand the quark--gluon
dynamics underlying such general high--energy phenomena as the growth
of the radius of interaction with energy, and the Froissart bound.
This dynamics would correspond to a strongly interacting quark--gluon
system at small coupling constant, and represent a fascinating new
form of ``QCD matter'' which could be produced in the laboratory.
%
%

An intuitive understanding of the dynamics of hard scattering
processes and the basis for QCD factorization can be developed by
following the space--time evolution of the reactions in certain
reference frames.  At high energies (small $x$), the space--time
evolution of DIS can be discussed in two complementary ways.  In a
frame where the proton is fast--moving one obtains the well--known
parton picture of hard processes, in which the hard scattering process
involves quarks and gluons carrying a certain fraction (here, $x$) of
the proton's momentum. In the proton rest frame, on the other hand,
the DIS process takes the form of the scattering of a quark--antiquark
dipole from the target, with the dipole formed a long time before
reaching the target, and having a distribution of transverse sizes
extending down to values $\sim 1/Q$.  This representation reveals a
close relation between DIS at small $x$ and the so--called ``color
transparency'' phenomenon --- the transparency of hadronic matter to
the propagation of spatially small color--singlet configurations, as
observed \textit{e.g.}\ in the suppression of the interaction of heavy
quarkonia with hadronic matter.  The correspondence between the
``parton'' and the ``dipole'' picture of small--$x$ processes is a
powerful tool for analyzing the dynamics of strong interactions in
this regime. (A pedagogical introduction to these concepts will be
given in Sec.~\ref{sec:QCD}.)

The experimental investigation of small--$x$ processes became possible
with the advent of high--energy colliders (colliding beam
facilities). Extensive studies of DIS at small $x$ have been performed
at the HERA electron--proton ($ep$) collider at DESY.  Measurements of
inclusive cross sections have spectacularly confirmed the rise of the
gluon density in the proton at small $x$, as predicted by QCD, down to
values $x \sim 10^{-4}$.  Measurements of exclusive processes in DIS,
such as heavy and light vector meson production ($J/\psi, \rho$),
provide information about the spatial distribution of partons in the
transverse plane (``generalized parton distributions'') and allow us
to construct a three--dimensional image of the quark and gluon
structure of the nucleon. Finally, measurements of diffractive
processes in DIS, in which the produced hadronic system is separated
from the target remnants by a large rapidity gap, allow one to probe
the interaction of various small--size color--singlet configurations
with the proton in much more detail than inclusive DIS.

Another --- potentially much more powerful --- laboratory for studying
small--$x$ physics are high--energy proton--proton ($pp$) and
antiproton--proton ($\bar p p$) colliders, such as LHC at CERN (under
construction) and the Tevatron at Fermilab. QCD factorization can be
applied to $pp/\bar p p$ collisions with hard processes, such as the
production of dijets with large transverse momenta or heavy particles
($W^{\pm}$ bosons, Higgs bosons, \textit{etc}.), which originate from
binary collisions of partons in the two colliding hadrons.  At LHC,
such processes can probe parton distributions down to values of 
$x \sim 10^{-7}$. Even higher energies are reached in collisions of
cosmic--ray particles near the Greisen--Zatsepin--Kuzmin cutoff
\cite{gzk} with atmospheric nuclei.  In $pp$ scattering, as compared
to $ep$, one is dealing with collisions of two objects with a complex
internal structure. This results \textit{e.g.}\ in a high probability
of multiple hard scattering processes at high energies, and a much
richer spectrum of soft hadronic interactions.  Thus, while QCD
factorization can still be applied to certain hard processes in
$pp/\bar p p$ collisions, the modeling of the hadronic environment of
the quarks and gluons participating in the hard process (or processes)
is generally much more challenging than in $ep$ scattering.

This review is an attempt to summarize what has been learned about
small--$x$ physics from experiments at HERA and the Tevatron and
related theoretical studies, and use this information to make
predictions for new QCD phenomena observable at LHC. With LHC about to
be commissioned, and the HERA program nearing completion, this is a
timely exercise.  It is not our aim to give a comprehensive overview
of the existing collider experiments and their numerous implications
for our understanding of QCD. Rather, we identify certain specific
``lessons'' which are of particular importance in making the
transition from HERA to LHC:
\begin{itemize}
\item
\textit{Black--disk limit in dipole--hadron interactions.} 
Because of the fast rise of the gluon density at small $x$, the
strength of interaction of small color--singlet configurations
(dipoles) with hadronic matter can approach the maximum value allowed
by $s$--channel unitarity. We quantify this effect by formulating an
optical model of dipole--hadron scattering, in which the unitarity
limit corresponds to the scattering from a ``black disk'', whose
radius increases with energy (black disk limit, or BDL).  We argue
that the onset of the BDL regime can be seen in the diffractive DIS
data at the upper end of the HERA energy range, as well as in elastic
$pp/\bar p p$ scattering at Tevatron energies.  In DIS at higher
energies, the BDL leads to a breakdown of Bjorken scaling.  In
hadron--hadron collisions at LHC energies and beyond (cosmic ray
physics), the dynamics will be deep inside the BDL regime, with
numerous consequences for the hadronic final states.
\item
\textit{Small transverse area of leading partons.} 
Studies of hard exclusive processes at HERA show that partons in the
nucleon with $x > 10^{-2}$ and significant transverse momenta are
concentrated in a small transverse area, $\ll 1 \, \text{fm}^2$,
substantially smaller than the area associated with the nucleon in
soft (hadronic) interactions at high energies. The resulting
``two--scale picture'' of the transverse structure of the nucleon is
essential for modeling the hadronic environment of the colliding
partons in high--energy $pp$ collisions with hard processes.
\end{itemize}
Based on these observations, we make several predictions for new
strong interaction phenomena observable in $pp$, $pA$
(proton--nucleus), and $AA$ (nucleus--nucleus) collisions at LHC:
\begin{itemize}
\item
\textit{Hard processes as a trigger for central $pp$ collisions.} 
In $pp$ scattering at LHC, hard QCD processes involving binary
collisions of partons with momentum fractions $x_{1, 2} > 10^{-2}$
occur practically only in $pp$ events with small impact parameters
(central collisions).  This makes it possible to trigger on central
$pp$ events by requiring the presence of a hard dijet (or double
dijet) at small rapidities.
\item
\textit{Black--disk limit in central $pp/pA/AA$ collisions at LHC.}
The approach to the BDL at high energies will strongly affect the
dynamics of central $pp/pA/AA$ collisions at LHC energies and above.
A crucial point is that in hadron--hadron collisions at such energies
one is dealing mostly with gluon--gluon dipoles, whose cross section
for scattering from hadronic matter is $9/4$ times larger than that of
the quark--antiquark dipoles dominating $ep$ scattering.  We argue
that as a consequence of the approach to the BDL the leading partons
in central $pp$ collisions will acquire large transverse momenta
($p_\perp^2 \sim \text{several 10 GeV}^2$) and fragment independently,
resulting in the disappearance of leading hadrons with small
transverse momenta at forward/backward rapidities, increased energy
loss, and increased soft particle production at central
rapidities. These observable effects allow for experimental studies of
this fascinating new regime of ``strong gluon fields'' at LHC.  We
also outline the role of the BDL in heavy--ion collisions and cosmic
ray physics.
\item
\textit{Diffraction in high--energy $pp$ collisions.}
LHC offers the possibility to study a wide variety of diffractive
processes in high--energy $pp$ scattering, which probe the interaction
of small--size color singlets with hadronic matter and can be used to
map the gluon distribution in the proton. Such processes involve a
delicate interplay between hard (partonic) and soft (hadronic)
interactions. A crucial ingredient in understanding the dynamics is
the information about the transverse spatial distribution of gluons
obtained from exclusive vector meson production in DIS at HERA.
\end{itemize}
We also comment on the potential of LHC for parton distribution
measurements at small $x$, and for studies of small--$x$ dynamics via
photon--induced reactions in ultraperipheral $pp/pA/AA$ collisions.
Finally, we discuss the opportunities for studies of small--$x$
dynamics provided by the planned electron--ion collider.

The primary purpose of LHC is the search for new heavy particles
(Higgs bosons, supersymmetry) in high--energy $pp$ collisions.  The
small--$x$ phenomena we describe here directly impact on this
program. Heavy particles are produced in hard partonic collisions.
For the reason described above, the production of heavy particles in
inelastic $pp$ collisions happens predominantly in central collisions,
which are strongly affected by the approach to the BDL, and the strong
interaction background may be completely different from what one would
expect based on the naive extrapolation of existing data (Tevatron).
Likewise, the search for Higgs bosons in diffractive $pp$ events
depends crucially on the understanding of the strong interaction
dynamics in these processes.
\section{QCD factorization and the space-time evolution of small--$x$ 
scattering}
\label{sec:QCD}
\subsection{QCD factorization of hard processes}
\label{subsec:factorization}
We begin by introducing the basic concept of QCD factorization of hard
processes, and outlining the space--time evolution of small--$x$
scattering processes at high energies. Our main point is the
correspondence between the ``parton'' picture of hard processes in a
frame in which the nucleon is moving fast, and the ``dipole'' picture
of high--energy processes in the target rest frame.  In this Section
we illustrate this correspondence using as an example the simplest
high--energy process, inclusive DIS at small $x$.  Below we shall
apply these results to analyze the HERA DIS data
(Sec.~\ref{sec:inclusive}), and generalize them to processes with
exclusive (Sec.~\ref{sec:exclusive}) and diffractive
(Sec.~\ref{sec:diffraction}) final states.  The correspondence between
the two pictures plays a crucial role in formulating the the approach
the unitariy limit at high energies (Sec.~\ref{sec:BDL}), and for
understanding the dynamics of high--energy hadron--hadron collisions
(Secs.~\ref{sec:hadron} and \ref{sec:harddiff}).

\begin{figure}[t]
\begin{center}
\includegraphics[width=10cm]{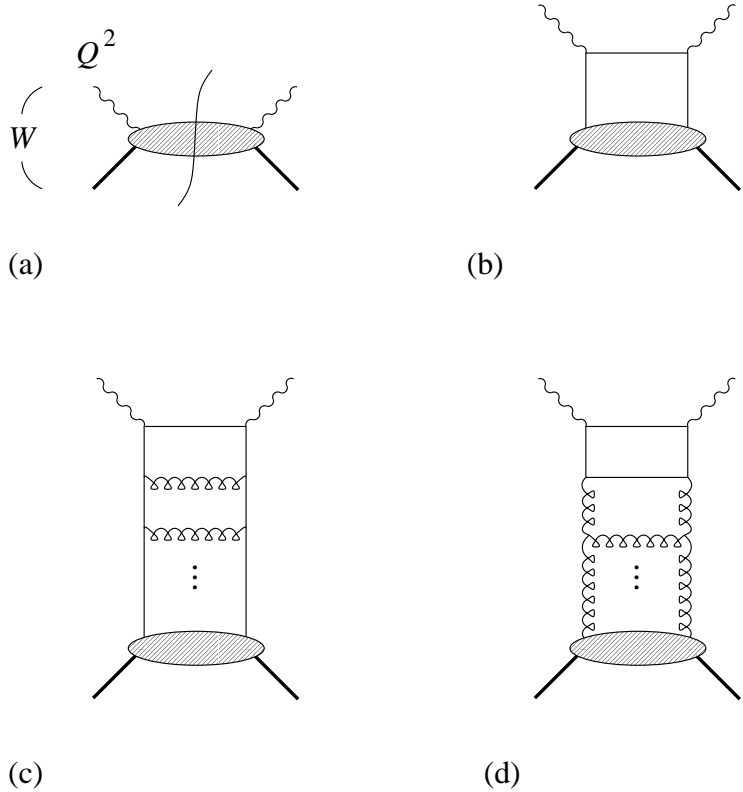}
\end{center}
\caption{(a) The total cross section for $\gamma^\ast p$ scattering 
is given by the imaginary part of the forwad scattering amplitude.
(b) The total cross section in the parton model (Bjorken scaling).
(c,d) QCD radiative corrections, giving rise to 
the leading scaling violations in $\alpha_s \ln (Q^2 / Q_0^2)$.
\label{fig:parton}}
\end{figure}

DIS is essentially the scattering of a virtual photon ($\gamma^\ast$)
from a hadronic target, see Fig.~\ref{fig:dis}. By the optical theorem
of quantum mechanics, the total $\gamma^\ast p$ cross section is given
by the imaginary part of the forward scattering amplitude (virtual
Compton amplitude), see Fig.~\ref{fig:parton}a. We consider the
Bjorken limit, in which both the photon virtuality and the
$\gamma^\ast p$ center--of--mass energy become large compared to the
typical hadronic mass scale, $Q^2 \sim W^2 \gg \mu^2$. As a
consequence of the asymptotic freedom of QCD, DIS in this limit can be
described as the scattering of the virtual photon from quasi--free
quarks (and antiquarks) in the proton. In the simplest approximation,
one neglects the interactions of the quarks altogether, and considers
the scattering from a free quark, see Fig.~\ref{fig:parton}b.  This is
equivalent to the space--time picture of DIS expressed in the parton
model \cite{Feynman}.  Its basic assumption is that, in a reference
frame where the proton moves with a large velocity, the interaction of
the $\gamma^\ast$ with the quarks (``partons'') is instantaneous
compared to the characteristic time of their internal motion in the
proton.  In this picture, the total cross section for $\gamma^\ast p$
scattering in the Bjorken limit is given by
\begin{equation}
\sigma^{\gamma^\ast p \rightarrow X} (Q^2, W) \;\; = \;\; 
\frac{4 \pi^2 \alpha_{\text{em}}}{Q^2 (1 - x)} \; F_2 (x),
\hspace{3em}
F_2(x) \;\; = \;\;
\sum_f e_f^2 \; x \left[ q_f (x) + \bar q_f (x) \right] ,
\label{F_2_parton}
\end{equation}
where $\alpha_{\text{em}}$ is the electromagnetic fine structure
constant.  Here, $e_f$ are the quark charges ($f = u, d, s \ldots$
labels the quark flavor), and $q_f(x)$ and $\bar q_f(x)$ are the
parton densities, describing the number density of quarks and
antiquarks carrying a fraction, $x$, of the fast--moving proton's
momentum.  The transverse momenta of the quarks and antiquarks are of
the order $k_\perp^2 \sim \mu^2$, and are integrated
over. Eq.~(\ref{F_2_parton}) exhibits the famous property of Bjorken
scaling \cite{Bjorken}, \textit{i.e.}, the structure function, $F_2$,
depends on the kinematic invariants characterizing the initial state
only through the dimensionless Bjorken variable (we neglect the
nucleon mass),
\begin{equation}
x \;\; \equiv \;\; \frac{Q^2}{W^2 + Q^2} .
\label{x_def}
\end{equation}
This is a direct consequence of the scattering from pointlike, 
quasi--free particles.

The parton model assumption of widely different timescales for the
$\gamma^\ast$--quark interaction and the internal motion of the quarks
in the proton becomes invalid in quantum field theory, where the
ultraviolet divergences introduce a scale larger than $Q^2$.  At its
most elementary, this is the reason why Bjorken scaling is violated in
quantum chromodynamics --- an argument originally due to V.~Gribov.
The leading scaling violations in $\alpha_s \ln (Q^2 / Q_0^2)$ arise
from gluon bremsstrahlung, as described by the ladder--type Feynman
diagrams shown in Fig.~\ref{fig:parton}c and d, and can be summed up
in closed form.  Here, $\alpha_s$ is the strong coupling constant, and
$Q_0^2$ is an arbitrarily chosen initial scale in the region of
approximate Bjorken scaling. The result can be expressed in the form
of a $Q^2$--dependence of the parton densities, governed by a
differential equation, the
Dokshitzer--Gribov--Lipatov--Altarelli--Parisi (DGLAP) evolution
equation \cite{DGLAPa,DGLAPb,DGLAPd}.  This formulation allows one to
retain the basic space--time picture of the parton model while
incorporating QCD radiative corrections by way of a $Q^2$--dependence
of the parton densities. Note that through evolution the gluon density
in the proton effectively enters into the structure functions of
$\gamma^\ast p$ scattering, see Fig.~\ref{fig:parton}d.
%
%

The basic structure of the $\gamma^\ast p$ total cross section in the
Bjorken limit in QCD is that of a product of a ``hard'' photon--parton
cross section (involving virtualities $\sim Q^2$) and a ``soft''
matrix element (involving virtualities $\sim \mu^2$), describing the
distribution of partons in the proton.  QCD radiative corrections can
be incorporated by a systematic redefinition of the ``hard'' and
``soft'' factors.  This property is referred to as factorization.
Factorization for the $\gamma^\ast p$ total cross section in QCD has
been formally demonstrated in several different approaches, including
operator methods in which the parton densities appear as nucleon
matrix elements of certain non-local light-ray operators, and the
evolution equations coincide with the QCD renormalization group
equations for these operators.  The calculations have also been
extended to sum up next--to--leading (NLO) corrections in 
$\alpha_s \ln (Q^2 / Q_0^2)$.  We shall see below that the basic
technique of factorization can be applied also to exclusive
(Sec.~\ref{sec:exclusive}) and diffractive
(Sec.~\ref{sec:diffraction}) final states in $\gamma^\ast p$
scattering, as well as to certain hard processes in hadron--hadron
scattering (Sec.~\ref{sec:hadron}).

More generally, QCD factorization allows one to perform an asymptotic
expansion of the DIS structure functions in the Bjorken limit. The
contribution from the diagram of Fig.~\ref{fig:parton}b,
Eq.~(\ref{F_2_parton}), determines the leading power behavior at large
$Q^2$, with an additional logarithmic dependence appearing due to
radiative corrections, Fig~\ref{fig:parton}c and d. In the context of
operator methods this is known as the leading--twist approximation.
Power corrections of the order $\mu^2/Q^2$ (higher--twist corrections)
arise from taking into account the effect of the quark transverse
momentum on the hard scattering process and the interaction of the
``struck'' quark with the non-perturbative gluon field in the proton;
the two effects are intimately related because of gauge invariance in
QCD \cite{Ellis:1982cd}.

From a mathematical perspective, Bjorken scaling of the moments of the
DIS structure functions can be seen as a consequence of the conformal
invariance of the QCD Lagrangian. The ultraviolet divergences
associated with radiative corrections give rise to anomalous
representations of the conformal group, with a logarithmic scale
dependence. Later we shall see that at high energies a new dynamical
scale appears in QCD, related to the gluon density in the nucleon and
its transverse area, which breaks the conformal invariance, and thus
leads to the complete disappearance of Bjorken scaling --- the black
disk limit (BDL), see Sec.~\ref{sec:BDL}.
\subsection{Space--time evolution of small--$x$ scattering
in the target rest frame}
\label{subsec:spacetime}
We now turn to DIS at high energies, $W^2 \gg Q^2 \gg \mu^2$, which
corresponds to values of the Bjorken variable $x \ll 1$.  While this
processes can be discussed within the standard QCD factorization
approach described above, one faces the practical question at which
point higher--twist ($1/Q^2$--) corrections enhanced at small $x$, or
radiative corrections beyond the DGLAP approximation giving rise to
factors $\ln (1/x)$, become important.  These and other questions can
be addressed in a transparent way by considering the time evolution of
DIS in the target rest frame, where the process takes the form of the
scattering of a small--size $q\bar q$ dipole from a hadronic target.
More generally, this formulation suggests a new understanding of QCD
factorization, closely related to the so--called ``color
transparency'' phenomenon observed in diffractive processes in
hadron--hadron scattering.

\begin{figure}[t]
\begin{center}
\includegraphics[width=16cm]{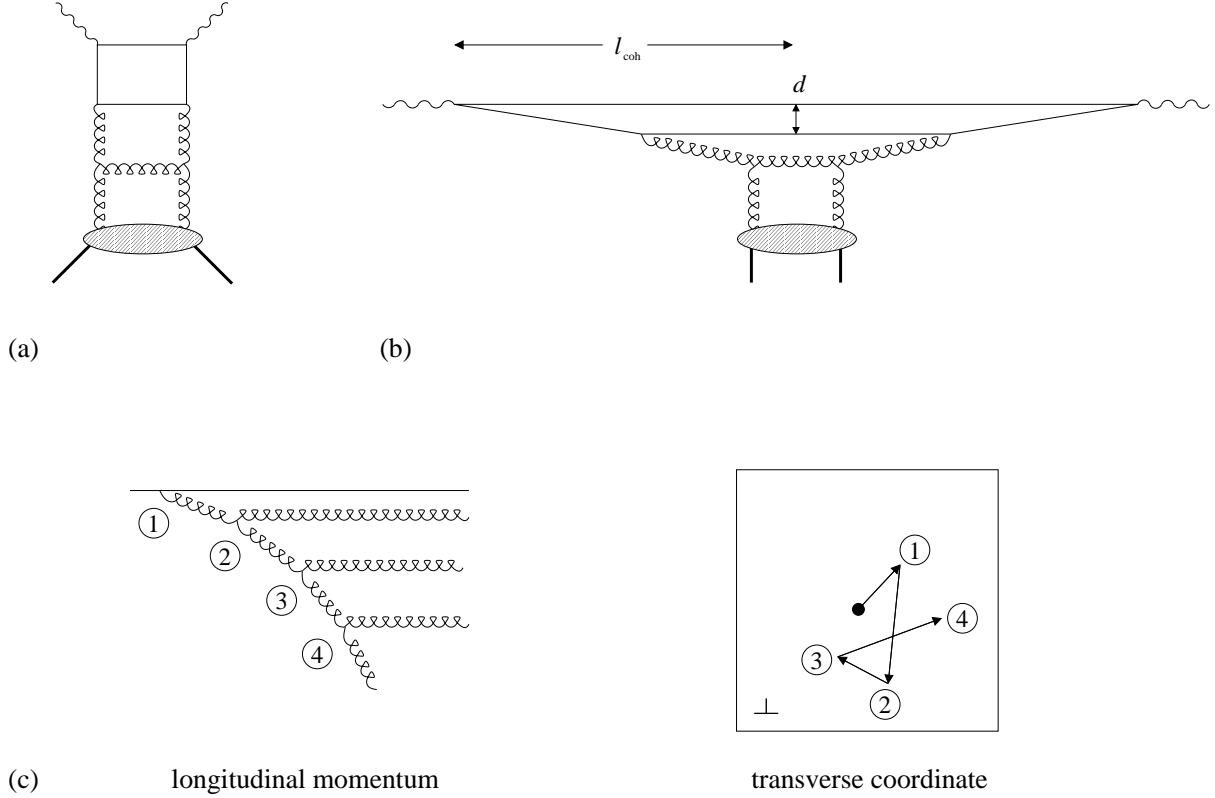}
\end{center}
\caption{(a) A typical Feynman diagram for inclusive $\gamma^\ast p$
scattering at small $x$. (b) Time evolution of small--$x$ scattering
in the target rest frame.  (c) The decay of a fast parton in the
$q\bar q$ dipole.  The degradation of longitudinal momenta is
accompanied by a random walk in the transverse coordinate.
\label{fig:spacetime}}
\end{figure}

QCD factorization in DIS and the DGLAP approximation have been
formulated using the covariant language of Feynman diagrams. A typical
Feynman diagram relevant at small $x$ is shown in
Fig.~\ref{fig:spacetime}a.  In order to arrive at a space--time
interpretation one needs to perform the integration over the
``energy'' variable using the residue theorem. It is this step which
actually introduces the dependence of the amplitudes on the reference
frame. Alternatively, one may directly trace the space--time evolution
using the language of time--ordered perturbation theory. In this
formulation, the time scales of the processes are determined by the
energy denominators associated with the various transitions, via the
energy--time uncertainty relation, $\Delta t = 1/\Delta E$.

%
%
In the target rest frame, the virtual photon in DIS at small $x$ moves
with 3--momentum $P \approx Q^2/(2 m_N x)$. Consider its conversion
into a $q\bar q$ pair with longitudinal momenta $zP$ and $(1-z)P$ and
transverse momenta $\pm \bm{k}_\perp$ (``longitudinal'' and
``transverse'' are defined relative to the direction of motion of the
photon). The energy denominator for this transition is
\begin{equation}
\Delta E \;\; =\;\; \frac{M_{q\bar q}^2 +Q^2}{2 P} ,
\end{equation}
where $M_{q\bar q}^2 \equiv (m_q^2 + k_\perp^2) /[z (1-z)]$ is the
invariant mass of the $q\bar q$ pair ($m_q$ is the quark mass,
$k_\perp \equiv |\bm{k}_\perp |$).  For a longitudinally polarized
photon, the dominant contribution to the cross section comes from
values $z \sim 1/2$ and $k_\perp \sim Q$, for which 
$M_{\bar q q}^2 \sim Q^2$ (this will be explained in more detail in
Sec.~\ref{subsec:dipole}). For such values the time associated with
the $\gamma^\ast \rightarrow q\bar q$ transition is
\begin{equation}
\Delta t \;\; = \;\; 1/\Delta E \;\; \sim \;\;  
Q^2/P \;\; \approx \;\;  1 / (2 m_N x) .
\label{Delta_t}
\end{equation}
At small $x$, the photon converts into a $q\bar q$ pair long before
reaching the target, as illustrated in Fig.~\ref{fig:spacetime}b. Both
the quark and antiquark move essentially with the speed of light. The
distance between the point of their creation and the target, the
so-called coherence length, is given by
\begin{equation}
l_{\text{coh}} \;\; = \;\; c \Delta t .
\label{lcohlow}
\end{equation}
It is important to realize that the $q\bar q$ wave packet remains well
localized in the longitudinal direction as it travels towards the
target, and that the transverse separation between the quark and
antiquark (at a given time) is a meaningful concept. It is generally
of the order $d \sim 1/M_{q\bar q} \sim 1/Q$, and thus small if $Q^2$
is sufficiently large.

In short, high--energy $\gamma_L^\ast$--hadron scattering in the
target rest frame is essentially the scattering of a small--size
$q\bar q$ dipole from the target hadron. In this formulation, the
quantity describing the strong interaction effects is the cross
section for dipole--hadron scattering. It can be computed using
methods of QCD factorization, with the dipole size, 
$d \ll \mu$ acting as the factorization scale, see
Sec.~\ref{subsec:dipole}.  Without going into details, we can
immediately state an important property of the cross section, namely
that in the limit $d \rightarrow 0$ it vanishes as
\begin{equation}
\sigma^{\text{$q\bar q$--hadron}} 
\;\; \propto \;\; d^2 ,
\label{sigma_d2}
\end{equation}
up to logarithmic corrections in $d$. Eq.~(\ref{sigma_d2}) reflects a
fundamental property of QCD as a gauge theory --- the interaction of a
small--size color singlet object with hadronic matter is small
(``color transparency''). In this understanding of QCD factorization,
high--energy $\gamma^\ast$--hadron scattering exhibits a close
relation to the interaction of heavy quarkonia with hadronic matter
and a number of other color transparency phenomena in hadron--hadron
scattering.

To make the dipole picture quantitative, one has to take into account
the effects of QCD radiation. In particular, this is necessary in
order to determine the coefficient in Eq.~(\ref{sigma_d2}) with
logarithmic accuracy.  The importance of different types of radiation
can again be studied using the language of time--ordered perturbation
theory in the target rest frame. The characteristic time for the quark
to radiate a gluon with longitudinal momentum fraction $x_g$ and
transverse momentum $\bm{k}_{\perp, g}$, relative to the time the
$q\bar q$ pair spends between its creation and ``hitting'' the target,
(\ref{Delta_t}), is (\textit{cf.}\ Fig.~\ref{fig:spacetime}b)
\begin{equation}
\frac{\Delta t_1}{\Delta t} \;\; = \;\; 
\left[\displaystyle Q^2 + \frac{m_q^2 + k_\perp^2}{z (1 - z)}\right] 
\left/
\left[\displaystyle Q^2 + \frac{m_q^2 + k_\perp^2}{1 - z}
+\frac{m_q^2 + (\bm{k}_{\perp} - \bm{k}_{\perp g})^2}{z - x_g} 
+ \frac{k_{\perp g}^2}{x_g}\right] .
\right.
\end{equation}
If $x$ is sufficiently small, and for average values of $z$, the
emission process can be repeated several times before the evolved
system reaches the target.

There exist several kinematic domains where gluon emission during the
propagation of the $q\bar q$ wave packet is likely because of a large
phase space at small $x$.  One is the emission of partons with
transverse momenta smaller than $k_{\perp}$ of the parent parton. Each
such emission contributes a factor $\alpha_s \ln (Q^2/Q_0^2)$ in the
amplitude, where the logarithm arises from the integration over the
phase volume of the radiated gluon.  In the standard QCD description
of DIS, these are the radiative corrections summed up by the DGLAP
evolution equations for the parton distributions described above
\cite{DGLAPa,DGLAPb,DGLAPd} (see also Ref.~\cite{CCFM}).  In the
context of the dipole picture at small $x$, the summation of these
corrections in leading order (LO) corresponds to a dipole--hadron
cross section of the form \cite{BBFS93,Frankfurt:it,Frankfurt:1996ri}
\begin{equation}
\sigma^{\text{$q\bar q$--hadron}} (x, d^2) \;\; = \;\; 
\frac{\pi^2}{4} \; F^2 \; d^2 \; \alpha_s (Q^2_{\text{eff}}) \;
x G (x, Q^2_{\text{eff}}).
\label{sigma_d_DGLAP}
\end{equation}
Here $F^2 = 4/3$ is the Casimir operator of the fundamental
representation of the $SU(3)$ gauge group. Furthermore, $\alpha_s
(Q_{\text{eff}}^2 )$ is the LO running coupling constant and 
$G (x, Q_{\text{eff}}^2 )$ the LO gluon density in the target.  They
are evaluated at a scale $Q_{\text{eff}}^2 \propto d^{-2}$. The
coefficient of proportionality is not fixed within the LO
approximation, and needs to be determined from NLO calculations or
from phenomenological considerations, see Sections~\ref{subsec:dipole}
and \ref{sec:exclusive}.\footnote{There is an approach to high--energy
scattering in which the projectile particle is represented as a
superposition of eigenstates of the scattering matrix, see
\textit{e.g.} Ref.~\cite{Pumplin} and references therein.
Equation~(\ref{sigma_d_DGLAP}) implies that states with different
transverse size, $d$, should be orthogonal. However, the extension of
Eq.~\ref{sigma_d_DGLAP} to the case of elastic scattering indicates
that transitions between configurations with different $d$ are allowed
for finite $t$.  This suggests that the the eigenstate model should be
a reasonable approximation only for small values of $t$.}

Eq.~(\ref{sigma_d_DGLAP}) actually quotes a simplified expression for
the dipole--hadron cross section. The original expression involves an
integral over the gluon momentum fractions, which is concentrated in a
narrow range above $x$. Also neglected in Eq.~(\ref{sigma_d_DGLAP}) is
the contribution proportional to the quark/antiquark distribution in
the target, which at small $x$ is suppressed compared to the gluon
distribution.  This contribution would lead to a flavor dependence of
the dipole--nucleon cross section \cite{FMS2002}.

Another large contribution, specific to small $x$, comes from the
large phase space in rapidity ($\propto z_g$) for emission of gluons
without strong degrading of transverse momenta in the leading
approximation. Such emissions give rise to factors 
$\alpha_s (N_c /2\pi) \Delta y$, where $N_c = 3$ is the number of
colors in QCD, and $\Delta y = (y_i - y_{i+1})$ is the difference in
rapidities between successive partons in the ladder.  In terms of $x$,
this corresponds to corrections proportional to $\ln (x_0 / x)$, where
$x_0 \sim 0.1$ accounts for the fact that nucleon fragmentation enters
in the definition of the gluon density in the nucleon and does not
produce a logarithm in $x$.  If the rapidity interval for emissions
(\textit{i.e.}, the lifetime of the quark--gluon system) becomes very
large, one needs to sum these logarithms in addition to the $\alpha_s
\ln (Q^2/Q_0^2)$ terms \cite{Gribovcom}, see
Sec.~\ref{subsec:breakdown}.

QCD radiation generally leads to an increase of the transverse size of
the ``dressed'' dipole with decreasing $x$, and thus to an increase of
the radius of the dipole--hadron interaction with energy.  Each
individual emission shifts the transverse coordinate of the radiating
parton by $\Delta\rho \sim 1/k_{\perp}$, see
Fig.~\ref{fig:spacetime}c.  If there are $n$ successive emissions with
comparable, randomly oriented $k_{\perp}$ (this is the case in the
limit of large $\ln x$), the overall shift is \cite{Gribov:1973jg}
\begin{equation}
\Delta\rho^2 \;\; = \;\; n / k_{\perp}^2 \;\; = \;\;
y / (\Delta y \, k_{\perp}^2 ) ,
\label{diffusd}
\end{equation}
where $y$ is the rapidity of the initial parton.  A similar diffusion
mechanism for soft partons was discussed by V.~Gribov as a model for
the increase of the radius of soft hadronic interactions with energy
\cite{Gribov:ex,Gribov:1973jg}. In the case of hard processes such as
$\gamma^\ast$--hadron scattering, in the region where the DGLAP
approximation is valid, the rate of expansion with energy is much
smaller than for soft interactions, because of the larger transverse
momenta of the emitted partons and the larger rapidity intervals
between the emissions.  This manifests itself \textit{e.g.}\ in a much
weaker energy dependence of the $t$--slope of hard exclusive processes
as compared to elastic hadron--hadron scattering \cite{BFGMS94}, see
Sec.~\ref{sec:exclusive}.

For transversely polarized virtual photons the space--time picture of
the interaction is more complicated than in the longitudinal case.
Owing to the different spin structure of the 
$\gamma^\ast_T \rightarrow q\bar q$ vertex, configurations of very
different size --- from hadronic size to $1/Q$ --- contribute to the
interaction. The hadronic size configurations correspond to $z \sim 1$
or $0$, and $k_{\perp} \sim \Lambda_{\text{QCD}}$. They are dual to
two jets aligned along the virtual photon direction and are referred
to as aligned jet configurations. They are expected to interact with
the target with typical hadronic cross sections, giving the dominant
contribution to the structure function $F_2$ at $Q^2\sim \, \text{few
GeV}^2$ and $x\sim 10^{-2}$, see Sec.~\ref{subsec:dipole}.  Also, such
configurations can easily scatter elastically from the target, and
thus are an important source of diffractive scattering, see
Sec.~\ref{sec:diffraction}.
\section{Inclusive $\gamma^\ast p$ scattering at small $x$}
\label{sec:inclusive}
\subsection{DGLAP evolution and the HERA data}
\label{subsec:DGLAP}
We start our discussion of $\gamma^\ast p$ scattering at high energies
with inclusive DIS.  Inclusive DIS is the main source of information
about the parton distributions in the nucleon at small $x$.  Because
of the relatively simple structure of QCD factorization, it is also
the main testing ground for higher--order QCD calculations and
resummation approaches.

\begin{figure}[t]
\begin{center}
\includegraphics[height=16cm]{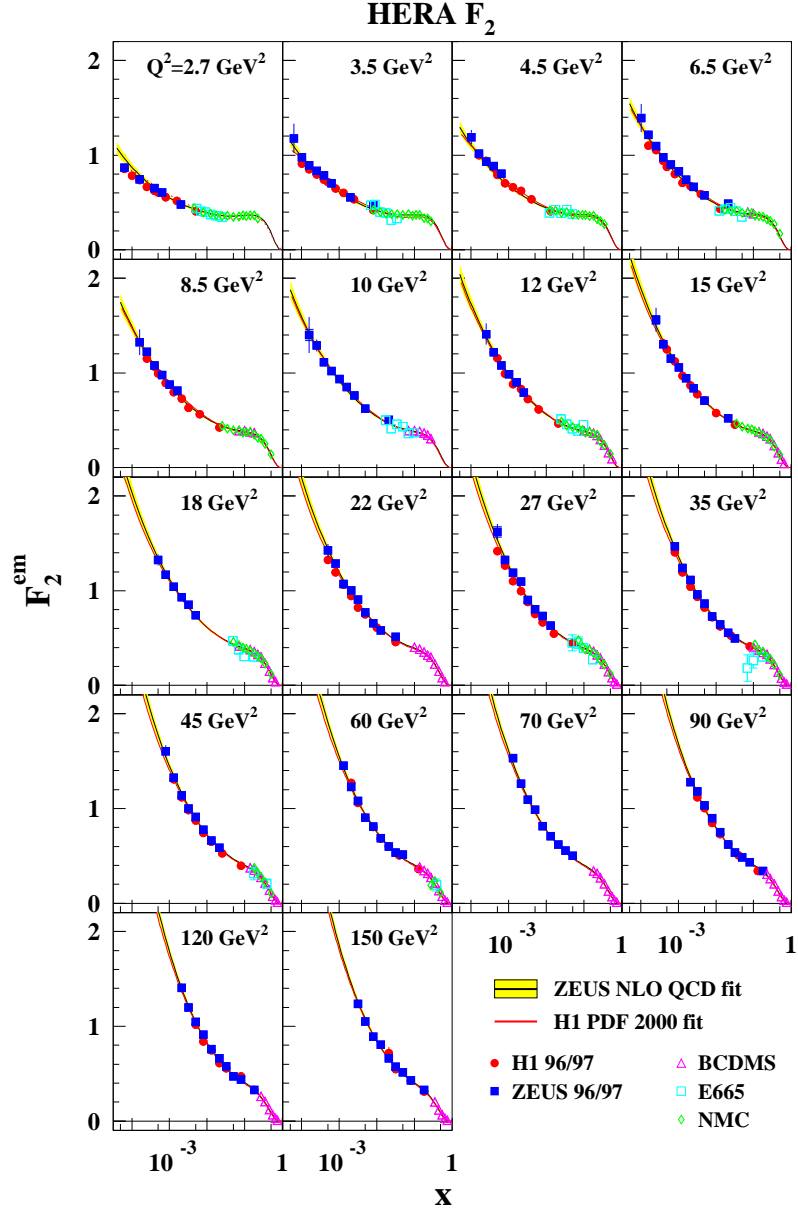}
\end{center}
\caption{The proton structure function, $F_2 (x)$, as measured by the
H1 and ZEUS experiments at HERA \cite{Moriond2004}. Also included are data
from fixed--target experiments. The lines show a QCD fit based on the
NLO DGLAP approximation.
\label{fig:HERA}}
\end{figure}

The validity of QCD factorization and DGLAP evolution for inclusive
DIS have extensively been tested in fixed--target experiments, probing
the quark/antiquark densities in the nucleon at values $x > 10^{-2}$
see \textit{e.g.}\ Ref.~\cite{Cooper-Sarkar:1997jk} for a review.
Going to smaller $x$, DGLAP evolution produces a fast increase of the
parton densities, related to the fact that the gluon has spin 1
\cite{Gross:ju,DGLAPd}, which implies a fast increase of the DIS cross
section with energy.  This prediction has spectacularly been confirmed
by the measurements with the HERA $ep$ collider. Fig.~\ref{fig:HERA}
shows a summary of the $F_2$ proton structure function data taken by
H1 and ZEUS compared to a QCD fit based on NLO DGLAP evolution
\cite{Moriond2004}.  The data clearly support the interrelation of the
$x$-- and $Q^2$--dependence as predicted by DGLAP evolution.  The
analysis of the data found that effects of next--to--next--to--leading
order (NNLO) terms of the form of $\alpha_s^2$ multiplied by a
function of $\alpha_s \ln(Q^2/Q_0^2)$ generally appear to be small. It
is remarkable that the DGLAP approximation, which does not account for
all potentially large terms containing $\ln(1/x)$, describes the
presently available high--energy data so well.

\begin{figure}[t]
\begin{center}
\includegraphics[width=10cm]{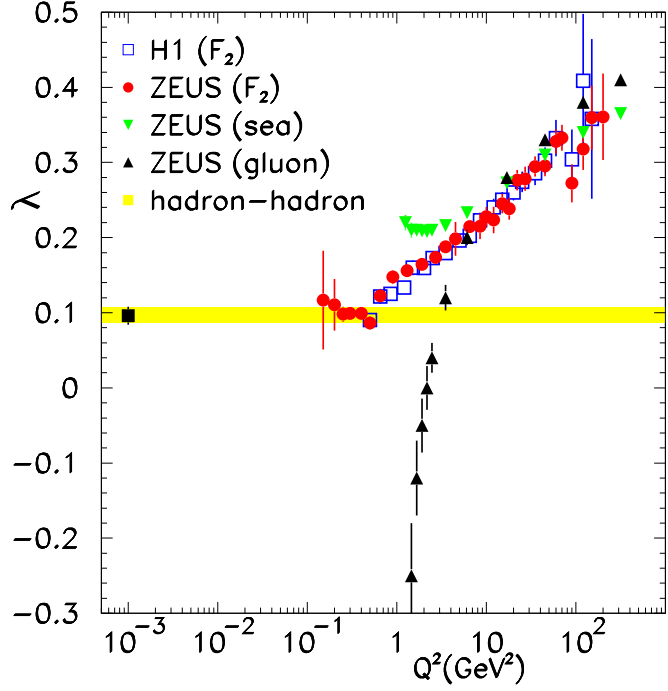}
\end{center}
\caption{The exponents characterizing the $x$--dependence of 
$F_2$, $\lambda_2$, the gluon distribution, $\lambda_g$ (black squares),
and the sea quark distributions, $\lambda_q$ (green triangles), 
\textit{cf.}\ Eq.~(\ref{power}), as extracted from the NLO DGLAP fit
to the H1 and ZEUS data (A.Levy, private communication).
\label{fig:lambda}}
\end{figure}

More detailed insights into the ``workings'' of the DGLAP approximation 
can be gained by studying the effective power behavior in $x$ of the 
structure function and the individual parton distributions in the NLO fit,
\begin{equation}
F_{2} \; \propto \; x^{-\lambda_2}, 
\hspace{2em}
x G(x) \; \propto \; x^{-\lambda_g } , 
\hspace{2em} 
\sum_f e_f^2 \, x \bar q_f (x) \; \propto \; x^{-\lambda_q}
\hspace{3em} (x < 10^{-2}),
\label{power}
\end{equation}
where the exponents depend on $Q^2$, see Fig.~\ref{fig:lambda}.  At
low $Q^2$, $\lambda_2 \approx 0.1$, reflecting the energy dependence
expected for the cross section of soft hadronic processes. Starting
from $Q^2 \approx 0.5\, \text{GeV}^2$ $\lambda_2$ grows, reaching a
value of $\sim 0.4$ at $Q^2 \sim 10\, \text{GeV}^2$ (A.Levy, private
communication).  For $Q^2 > 3\, \text{GeV}^2$, one observes that
$\lambda_g \approx \lambda_2$, indicating that in this $Q^2$--region
the $x$--dependence of the structure function is indeed driven by the
gluon distribution.  For lower $Q^2$, however, $\lambda_g$ is
significantly different from $\lambda_2$, becoming even negative at
$Q^2 \approx 2 \, \text{GeV}^2$. Thus, while the NLO DGLAP
approximation formally describes the $x$--dependence of the structure
function even at low $Q^2$, the price to be paid is the lack of a
smooth matching of the $x$--dependence of the gluon distribution to
the soft regime.  This may indicate the presence of significant
corrections to the leading--twist description of DIS at small $x$ for
$Q^2\le 3\, \text{GeV}^2$. The dynamical origin of these corrections
will be discussed in Sec.~\ref{subsec:dipole}.
%
%
%

%
%
The data show that the deviation from the soft energy dependence of
$F_2$ starts at surprisingly low scales, $Q^2 \ll 1 \, \text{GeV}^2$.
Within the DGLAP approximation this behavior can be explained by the
presence of a large non-perturbative gluon density in the nucleon at
moderate $x$ at a low scale \cite{Gluck:1994uf}. This is principally
consistent with the idea of spontaneous chiral symmetry breaking,
according to which most of the nucleon mass resides in gluon fields.
\subsection{Space--time picture of inclusive DIS}
\label{subsec:dipole}
Many of the observed features of inclusive DIS at small $x$ can be
understood within the space--time picture in the target rest frame,
\textit{cf.}\ Sec.~\ref{subsec:spacetime}.  In this formulation,
corrections to the leading--twist approximation at low $Q^2$ appear
because of the contribution from large dipole sizes.  This allows us
to quantify the region of validity of the leading--twist
approximation, and develop an ``interpolating'' approximation valid in
a wide range of $Q^2$.

Following the logic outlined in Sec.~\ref{subsec:spacetime}, one can
express the total $\gamma^\ast p$ cross section at small $x$ as a
superposition of $q\bar q$ dipole cross sections, characterized by the
longitudinal momentum fraction of the quark, $z$, and the dipole size,
$d$:
\begin{equation}
\sigma_{L, T} (x,Q^2) \;\; = \;\; \int^1_0dz \int d^2d \;
\sigma^{\text{$q\bar q$--hadron}} (z, d, x)
\left|\psi^\gamma_{L, T}(z,d,Q^2)\right|^2 ,
\label{sigma_L}
\end{equation}
where $\psi^\gamma_{L, T}(z,d)$ denotes the light--cone wave function
of the $q\bar q$ component of the virtual photon, calculable in
quantum electrodynamics.  An important question is which dipole sizes
dominate in the integral.  For a longitudinally polarized photon, the
modulus squared of the wave function is given by
\begin{equation}
\left| \psi^\gamma_L (z,d, Q^2) \right|^2
\;\; = \;\; \frac{6 \alpha_{\text{em}} Q^2}{\pi^2} \sum^{N_f}_{f=1} e_f^2 
\left[ z (1-z) K_0 (\epsilon d) \right]^2 ,
\label{psi_gamma_L}
\end{equation}
where $K_0$ is the modified Bessel function and 
$\epsilon^2 = z(1-z) Q^2 + m_f^2$ \cite{Cheng:jk}.  One can verify by
direct calculation that in this case the contributions from large
dipole sizes are suppressed at large $Q^2$, if the integral
(\ref{sigma_L}) is evaluated with the LO expression for the
dipole--nucleon cross section, Eq.~(\ref{sigma_d_DGLAP})
\cite{BFGMS94}. In fact, Eqs.~(\ref{sigma_L}, \ref{psi_gamma_L}) and
(\ref{sigma_d_DGLAP}) are formally equivalent to the LO DGLAP
approximation in QCD, \textit{cf.}\ the discussion below.  The
effective scale in the gluon distribution entering the dipole cross
section can be determined by comparing (\ref{sigma_L}) with the LO
DGLAP expression; one finds $Q^2_{\text{eff}} \approx 9/d^2$ for HERA
kinematics \cite{Frankfurt:1995jw}.  Note that the factor 
$x G(x,Q^2_{\text{eff}})$ in Eq.~(\ref{sigma_d_DGLAP}) results in a
fast increase of the cross section with energy, in contrast to the
two--gluon exchange model of Refs.~\cite{Low75,Nus75,Gunion}, where
the cross section is energy--independent.

When applying Eq.~(\ref{sigma_L}) to transversely polarized photons,
the distribution of dipole sizes is significantly wider than in the
longitudinal case. At $Q^2 \sim \text{few GeV}^2$, the transverse
cross section receives sizable contributions from dipole sizes for
which the perturbative approximation for the dipole--nucleon cross
section, \textit{cf.}\ Eq.~(\ref{sigma_d_DGLAP}), becomes invalid.
Still, at large $Q^2$ the perturbative contribution should dominate,
because of the faster increase with energy of the parton distribution
for the smaller--size quark--gluon configuration.  The contribution
from large--size $q\bar q$ configurations is strongly suppressed by
Sudakov form factors; it is actually represented by large--size 
$q\bar qg,\ldots$ configurations.

\begin{figure}[t]
\begin{center}
\includegraphics[width=7.5cm]{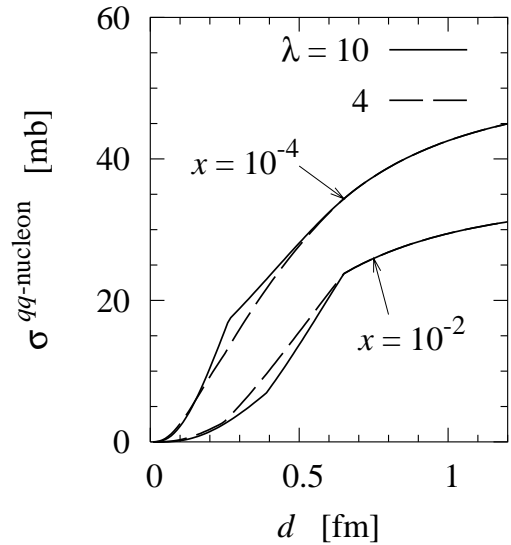}
\caption{The dipole--nucleon cross section in the ``interpolating''
model of Ref.~\cite{McDermott:1999fa}.  Shown are the results
corresponding to two different values of the parameter, $\lambda$,
determining the effective scale, $Q^2_{\text{eff}}$.
\label{fig:sighat}}
\end{center}
\end{figure}

Equation~(\ref{sigma_L}) can serve as the basis for an
``interpolating'' model that describes $\gamma^\ast p$ interactions
over a wide range of $Q^2$ for both transverse and longitudinal
polarizations \cite{McDermott:1999fa}.  There is ample evidence ---
\textit{e.g.}\ from studies of $\gamma N$ and $\pi N$ elastic
scattering --- that real photons in high--energy reactions have
transverse sizes comparable to pions. A way to ensure this within the
$q\bar q$ dipole description is to introduce a dynamical quark mass of
$\sim 300\, \text{MeV}$, which is consistent with the phenomenology of
spontaneous chiral symmetry breaking \cite{Diakonov:2002fq}.  The
cross section for the scattering of such a ``hadronic--size'' dipole
with the target can then be inferred from the $\pi N$ scattering data.
For small dipoles, $d\leq 0.4\, \text{fm}$, the cross section can be
calculated perturbatively. When evaluating the leading--twist
expression, it is important to accurately treat the kinematic limits
of the integral over the gluon momentum fractions, as this leads to an
additional dependence of the dipole cross section on $Q^2$. A dipole
cross section obtained by matching the two prescriptions is shown in
Fig.~\ref{fig:sighat}. This function is then averaged with the photon
wave function for massive quarks, \textit{cf.}
Eq.~(\ref{sigma_L}). This model reproduces well the HERA $F_{2p}$ data
for $Q^2\ge 0.1 \, \text{GeV}^2$, and correctly predicts $\sigma_L$
\cite{McDermott:1999fa}.
%
%

In order to make contact with the analysis of Sec.~\ref{subsec:DGLAP},
we need to state more precisely how the dipole picture is related to
the DGLAP approximation in QCD. In LO, it has been demonstrated
explicitly that Eqs.~(\ref{sigma_d_DGLAP}) and (\ref{sigma_L}) can be
obtained by rewriting the LO DGLAP expression for the $\gamma^\ast p$
cross section \cite{Frankfurt:1996ri}.  This simple relation appears
because in the leading logarithmic approximation the separation of the
process according to time in the target rest frame --- transition of
the virtual photon into a $q\bar q$ pair (photon wave function), and
interaction of the pair with the target (dipole cross section) ---
coincides with the separation of transverse momenta in 
$k_\perp^2 \sim Q^2$ and $k_\perp^2 \ll Q^2$ in the partonic ladder.
Beyond the leading order, one needs to explicitly include $q\bar q g$
component of the photon wave function, and the distinction between the
wave function and the dipole interaction with the target becomes more
delicate. Although in principle the leading--twist dipole picture should
be equivalent to the DGLAP approximation in any order of the
expansion, in practice the problem of formulating a consistent dipole picture 
in NLO has not been solved yet. Also, it is worth emphasizing
that the correspondence between parton distributions and amplitudes of
physical processes is not always direct. In particular, the parton
model--type contribution, which naturally leads to diffractive
processes, is hidden in the boundary condition for the QCD evolution
of the parton distributions.

In spite of the lack of an explicit NLO analysis in the dipole
picture, it still seems to be of relevance that the region of $Q^2$
where the NLO DGLAP analysis leads to a gluon exponent $\lambda_g$
dropping below the soft value (see Fig.~\ref{fig:lambda}) corresponds
in the dipole model to contributions from $d > 0.4\, \text{fm}$, where
nonperturbative effects become important. Thus, it seems that the
``anomalous'' behavior of $\lambda_g$ is a consequence of the
leading--twist DGLAP approximation trying to mock up higher--twist
corrections at low $Q^2$. The dipole picture allows us to quantify the
region of applicability of the leading--twist approximation at low
$Q^2$, and suggests a natural way to incorporate non-perturbative
effects.

Equations~(\ref{sigma_d_DGLAP}) and (\ref{sigma_L}) are valid also
within the leading $\alpha_s \ln (x_0/x)$ approximation.  Furthermore,
they can be derived from the eikonal model expression for the
propagation of a heavy quarkonium through a hadronic medium
\cite{AMueller1}.
\subsection{Breakdown of the DGLAP approximation at very small $x$}
\label{subsec:breakdown}
The observation of the fast increase of parton densities at small $x$
has stimulated theoretical discussions of the stability of the DGLAP
approximation at small $x$. In fact, in the kinematic limit of fixed
$Q^2$ and $x\rightarrow 0$ the effective parameter of the perturbative
QCD expansion is multiplied by a factor $\ln (x_0 / x)$, which arises
due to gluon emission in multi--Regge kinematics (rapidity distance
between adjacent gluons $\gg 2$), and the hierarchy of dominant terms
is changed as compared to the DGLAP approximation. The constant $x_0$
is determined by the typical momentum fraction in the initial parton
distributions; usually $x_0 \sim 0.1$.

A simple kinematic estimate shows that in typical HERA kinematics the
DGLAP approximation is still reliable. The rapidity span at HERA is
approximately $\ln (Q / x m_N) \approx 10$ for $x = 10^{-4}$ and 
$Q = 2.5\, \text{GeV}$.  To obtain a significant $\ln (x_0 / x)$ term,
the distance in rapidity between adjacent partons in the ladder should
be $\gg 2$.  Thus, the number of radiated gluons in multi--Regge
kinematics at HERA is $\ll (10 - 4)/2 -1 \approx 2$, where we took
into account that each of the fragmentation regions occupies at least
two units in rapidity. This simple estimate agrees well with a
numerical study of NLO QCD evolution, which indicates that the average
change of $x$ in the HERA region does not exceed 10, corresponding to
$\Delta y = \ln (x_0 / x) \approx 2$, provided that $Q_0^2 \geq 1\,
\text{GeV}^2$ \cite{Frankfurt:2000ty}.  Since one (two) logarithms of
$x$ are effectively taken into account by the NLO (NNLO)
approximation, there is no need for a special treatment of 
$\ln (x_0 / x)$ effects at HERA kinematics. A similar estimate shows
that at LHC kinematics the radiation of 5--6 gluons is permitted.
Thus, at LHC energies and above the resummation of $\ln (x_0/x)$ terms
becomes a practical issue.

The program of resumming leading $\alpha_{\text{em}} \ln (x_0 / x)$
terms started in quantum electrodynamics \cite{Gribovcom}. In QCD, the
reggeization of gluons slows down the energy dependence of amplitudes
of high--energy processes \cite{Sherman,Lipatov}.
\footnote{The high--energy behavior of two--body amplitudes with
color--octet quantum numbers in the crossed channel in QCD is given by
the Regge pole formula, $(1/x)^{\beta(t)}$, where $\beta(t)$ decreases
with increase of $-t$. In leading order of $\alpha_s$ QCD gives
$\beta(t) = 1$. Thus, gluons in QCD (as well as quarks) are reggeons.}
In the leading $\alpha_s \ln(x_0/x)$ approximation
(Balitsky--Fadin--Kuraev--Lipatov, or BFKL, approximation)
\cite{BFKL}, where energy--momentum conservation and the running of
the coupling constant are neglected, the reggeization of gluons is
canceled by contributions from multigluon radiation.  NLO corrections
to a large extent subtract kinematically forbidden contributions,
leading to a large negative contribution to the structure functions
\cite{Ciafaloni,Fadin:1998py}. Another feature of this approach is the
lack of an unambiguous separation between perturbative and
nonperturbative QCD effects \cite{Ciafaloni}.  Thus, this
approximation seems to be limited to the description of single--scale
hard processes where DGLAP evolution is unimportant in a wide
kinematic range, such as 
$\gamma^*(Q^2) + \gamma^*(Q^2) \rightarrow \text{hadrons}$, or
two--body processes where the hardness is controlled by proper choice
of final state like, such as 
$\gamma^* +\gamma^* \rightarrow \Upsilon\Upsilon$.

The resummation approaches of
Refs.~\cite{Ciafaloni:2003rd,Altarelli:2003hk} predict a significantly
slower increase of amplitudes with energy than the LO BFKL
approximation, and possibly even oscillations in the energy
dependence. Most of the reduction is due to the better account of
energy--momentum conservation in these approaches, and account of the
running of the coupling constant.  At extremely small $x$ (beyond the
reach of LHC) much of the LO BFKL results reappear, but with a slower
dependence on $x$.  For the parton densities in the nucleon, where
$x_0 \approx 0.1$ is a reasonable value for the constant in the
$\ln(x_0/x)$ factor, resummation effects should be small for $x \ge
10^{-4}$, that is, for the whole HERA range above 
$Q^2 \ge 2 \, \text{GeV}^2$.  At smaller $x$, the result of the
resummed evolution is close to that of NLO DGLAP evolution down to
$x\sim 10^{-6}$, but differs strongly from NNLO
\cite{Ciafaloni:2003kd}.  This suggests that NLO DGLAP evolution could
be a good guess for the parton densities down to the very small $x$
values probed at LHC, even though the underlying dynamics may change
significantly at $x \le 10^{-4}$.
\section{Exclusive processes in $\gamma^\ast p$ scattering at small $x$}
\label{sec:exclusive}
\subsection{QCD factorization for hard exclusive processes}
The concept of QCD factorization can be extended to certain exclusive
channels in $\gamma^\ast p$ scattering, namely processes of the type
\begin{equation}
\gamma_L^{*}(q) \; + \; N(p) \;\; \rightarrow  \;\;
\text{``Meson''}(q + \Delta ) \; + \; \text{``Baryon''}(p -\Delta ) ,
\label{process}
\end{equation}
at large virtuality, $Q^2 \equiv -q^2$, and center--of--mass energy,
$W^2 \equiv (p + q)^2$, with fixed $x = Q^2 / (W^2 + Q^2)$, and fixed
small invariant momentum transfer, $t \equiv \Delta^2$.  Examples
include the production of light vector mesons ($\rho, \rho'$)
\cite{BFGMS94}, heavy vector mesons ($J/\psi, \psi', \Upsilon$)
\cite{BFGMS94}, and real photons (deeply--virtual Compton scattering,
DVCS) \cite{Bartels:1981jh,Muller:1998fv,AFS,Ji:1996nm,%
Radyushkin:1997ki,Frankfurt:1997at,Collins:1998be}.  Closely related
to these processes are certain hadron--induced reactions, such as the
diffractive dissociation of pions, 
$\pi + T \rightarrow \text{2 jets} + T$, where $T$ denotes a hadronic
target (nucleon or nucleus) \cite{Frankfurt:it}.  These exclusive
processes probe the interaction of small--size color singlets with
hadronic matter in much more detail than inclusive DIS. They also
provide new information about the transverse spatial structure of the
nucleon, contained in the so-called generalized parton distributions.

%
%
\begin{figure}[b]
\begin{center}
\includegraphics[width=8cm]{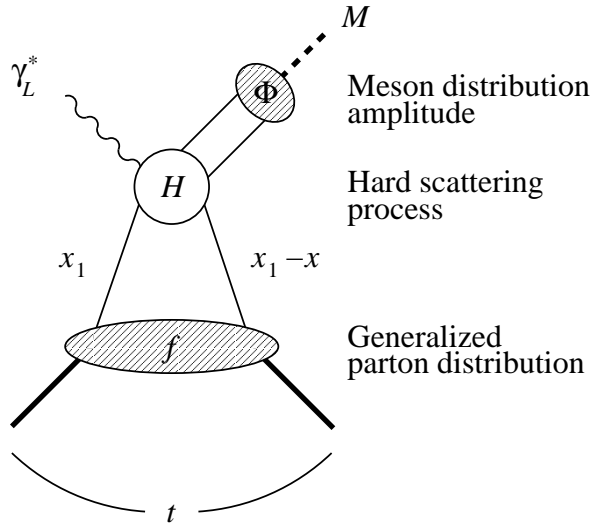}
\end{center}
\caption{Factorization of the amplitude of hard exclusive meson 
production, Eq.~(\ref{factorization}).
\label{fig:fact}}
\end{figure}
The basis for the analysis of exclusive processes (\ref{process}) is
the QCD factorization theorem \cite{Collins:1996fb}, which extends the
initial analysis of Ref.~\cite{BFGMS94} for the small--$x$ limit.  It
states that the amplitude can be represented as a convolution of three
functions, as depicted in Fig.~\ref{fig:fact}:
\begin{eqnarray}
A^{\gamma^\ast_L N \rightarrow M + B} &=&
\sum _{i,j} \int _{0}^{1}dz  \int dx_{1} \;
f_{i/p}(x_1, x - x_1, t; \mu ) \;
H_{ij}(x_1, x, z, Q^{2}; \mu ) \; \phi^{M}_{j}( z, \mu )
\nonumber \\
&+& \text{power corrections} .
\label{factorization}
\end{eqnarray}
Here, $f$ is the generalized parton distribution (GPD), which
describes the amplitude for the nucleon to ``emit'' and ``absorb'' a
parton with longitudinal momentum fractions $x_1$ and $x_2 = x_1 - x$,
respectively, accompanied by an invariant momentum transfer, $t$, and,
possibly, a transition to another baryonic state.  At zero momentum
transfer, $x_1 = x_2$ and $t = 0$, the GPD's coincide with the usual
parton densities measured in inclusive DIS.  (For a review of the
properties of GPD's and their applications, see
Refs.\cite{Diehl:2003ny,Belitsky:2005qn}.)  Furthermore, $\phi^M$ is
the distribution amplitude describing the conversion of a $q\bar q$
pair with relative longitudinal momentum fraction $z$ to the produced
meson (or photon). Finally, $H_{ij}$ denotes the amplitude of the hard
partonic scattering process, which is calculable in powers of
$\alpha_{s}(Q^2)$. The indices $i, j$ label the different parton
species. The contribution of diagrams in which the hard scattering
process involves more than the minimum number of partons is suppressed
by $1/Q^2$.  An important consequence of factorization is that the
$t$--dependence of the amplitude rests entirely in the GPD. Thus,
different processes probing the same GPD should exhibit the same
$t$--dependence.

\subsection{Space--time picture of hard exclusive processes}
\label{subsec:squeezing}
The physics of hard exclusive processes at small $x$ becomes most
transparent when following the space--time evolution in the target
rest frame. As in the case of inclusive scattering, this approach
allows one to expose the limits of the leading--twist approximation,
and to quantify power corrections due to the finite transverse size of
the produced meson.

In exclusive vector meson production, $\gamma^\ast_L N \rightarrow
VN$, one can identify three distinct stages in the time evolution in
the target rest frame. The virtual photon dissociates into a $q\bar q$
dipole of transverse size $d \sim 1/Q$ at a time 
$\tau_i = l_{\text{coh}}/c \approx 1/(m_N x)$ before interacting with
the target, \textit{cf.}\ Eq.~(\ref{lcohlow}).  The $q\bar q$ dipole
then scatters from the target, and ``lives'' for a time 
$\tau_f \gg \tau_i$ before forming the final state vector meson. The
difference in the time scales is due to the smaller transverse momenta
(virtualities) allowed by the meson wave function as compared to the
virtual photon.

In the leading logarithmic approximation in 
$\ln (Q^2 / \Lambda_{\text{QCD}}^2)$, the effects of QCD radiation can
again be absorbed in the amplitude for the scattering of the
small--size dipole off the target. It can be shown by direct
calculation of Feynman diagrams that the leading term for small dipole
sizes is proportional to the generalized gluon distribution, 
$G(x_1, x_2, t; Q^2_{\text{eff}})$, where 
$Q^2_{\text{eff}} \propto d^{-2}$ \cite{Frankfurt:1996ri}.  A simpler
approach is to infer the result for the imaginary part of the
amplitude from the expression for the cross section,
Eq.~(\ref{sigma_d_DGLAP}), via the optical theorem.  The imaginary
part is proportional to the generalized gluon distribution at $x_1 =
x$ and $x_2 = 0$. At sufficiently large $Q^2$, the generalized gluon
distribution at small $x_1$ and $x_2$ can be calculated by
perturbative evolution, starting from the ``diagonal'' generalized
gluon distribution, $x_1 = x_2 \gg x$, at a low scale
\cite{FG,Shuvaev:1999ce,Musatov:1999xp}. In applications to vector
meson production at HERA, where the effective scale is of the order
$Q^2_{\text{eff}} \sim \text{few GeV}^2$, the ``skewness'' effects
induced by the evolution are not very substantial, and one may
approximate the generalized gluon distribution by the diagonal one at
the scale $Q^2_{\text{eff}}$.  It is convenient to separate the
$t$--dependence and write the diagonal generalized gluon distribution
in the form
\begin{equation}
G (x, x, t; Q^2_{\text{eff}}) \;\; = \;\; G (x, Q^2_{\text{eff}}) \; 
F_g (x, t; Q^2_{\text{eff}}) ,
\label{twogluon}
\end{equation}
where $G (x, Q^2_{\text{eff}})$ is the usual gluon density and $F_g$
is the ``two--gluon form factor'' of the target, which satisfies 
$F_g (x, t = 0; Q^2_{\text{eff}}) = 1$.  Altogether, one obtains for
the dipole--hadron scattering amplitude in this approximation
\begin{equation}
A^{\text{$q\bar q$--$N$}} (x, d^2, t) \;\; = \;\; 
2\pi i F^2 \; W^2\, d^2 \; \alpha_s (Q^2_{\text{eff}} ) \;
x G (x, t, Q^2_{\text{eff}} ) \; F_g (x, t, Q^2_{\text{eff}} ) .
\label{amp_DGLAP}
\end{equation}
The amplitude for the hadronic process (\ref{process}) is then given
by the convolution of Eq.~(\ref{amp_DGLAP}) with the light-cone wave
function of the virtual photon, Eq.~(\ref{psi_gamma_L}), and that of
the produced vector meson, $\psi^V$. In coordinate representation,
\begin{equation}
A^{\gamma^* N \rightarrow V N} 
\;\; = \;\; \int_0^1 dz \int d^2 d \; \psi^\gamma_L (z,d) \; 
A^{\text{$q\bar q$--$N$}} (x, d^2, t) \;
\psi^V (z,d) ,
\label{conv}
\end{equation}
where the integration is over the quark longitudinal momentum
fraction, $z$, and the transverse dipole size, $d$.

In Eq.~(\ref{conv}), the wave function of the vector meson of
transverse size $1/m_V$ is convoluted with the wave function of the
virtual photon of significantly smaller transverse size, $1/Q$. One
may say that the meson in this process is ``squeezed'', \textit{i.e.},
forced to couple in a configuration much smaller than its natural
hadronic size.  In the leading--twist approximation one neglects the
spatial variation of the vector meson wave function and substitutes it
by the distribution amplitude,
\begin{equation}
\psi^V (z, d) \;\; \rightarrow \;\; \psi^V (z, 0) \; \equiv \; \phi^V (z) 
\label{psi_LT}
\end{equation}
(in momentum representation, the distribution amplitude is the
integral of the wave function over transverse momenta). The integral
over transverse sizes can then be performed explicitly, using
Eqs.~(\ref{psi_gamma_L}) and (\ref{amp_DGLAP}). After restoring the
real part of the amplitude using its analyticity properties, the
differential cross section is obtained as \cite{BFGMS94}
\begin{eqnarray}
\frac{d\sigma_L^{\gamma^* N \rightarrow V N}}{dt}
&=& \frac{3 \pi^3 \Gamma_V m_V \eta_V^2}
{N_c^2 \alpha_{\text{em}} Q^6}
\nonumber \\
&\times& \alpha_s^2(Q^2_{\text{eff}}) \;
\left| \left( 1 + \frac{i\pi}{2} \frac{d}{d\ln x} \right) 
x G(x; Q^2_{\text{eff}}) \right|^2 \; F_g^2 (x, t; Q^2_{\text{eff}}).
\label{dsigdtt}
\end{eqnarray}
Here, $\Gamma_V$ is the leptonic width of the vector meson, which
defines the normalization of the meson wave function, and
\begin{equation}
\eta_V \;\; \equiv \;\; \frac{1}{2} 
\int_0^1\, dz \frac{\phi^V (z)}{z (1 - z)} \left/
\int_0^1\, dz \, \phi^V(z) \right. ;
\label{etavdef}
\end{equation}
$\eta_V \rightarrow 1$ at asymptotically large $Q^2$.  These
expressions apply to production by a longitudinally polarized
photon. For transverse polarization, the nonperturbative contribution
is suppressed only by a Sudakov--type form factor, similar to the case
of $F_2(x,Q^2)$ in inclusive $\gamma^\ast p$ scattering. This
contribution originates from highly asymmetric $q\bar q$ pairs 
($z \sim 0$ or $1$) in the $\gamma^\ast_T$ wave function, which have
transverse size similar to that of hadrons. We note that elastic
photo/electroproduction of $J/\psi$ mesons has been evaluated also
within the LO BFKL approximation \cite{Ryskin:1992ui}. The function $x
G(x,Q^2)$ that enters there has no relation to the conventional DGLAP
gluon distribution, which is defined within the DGLAP approximation
only.

%
%
\begin{figure}
\begin{center}
\includegraphics[width=15cm]{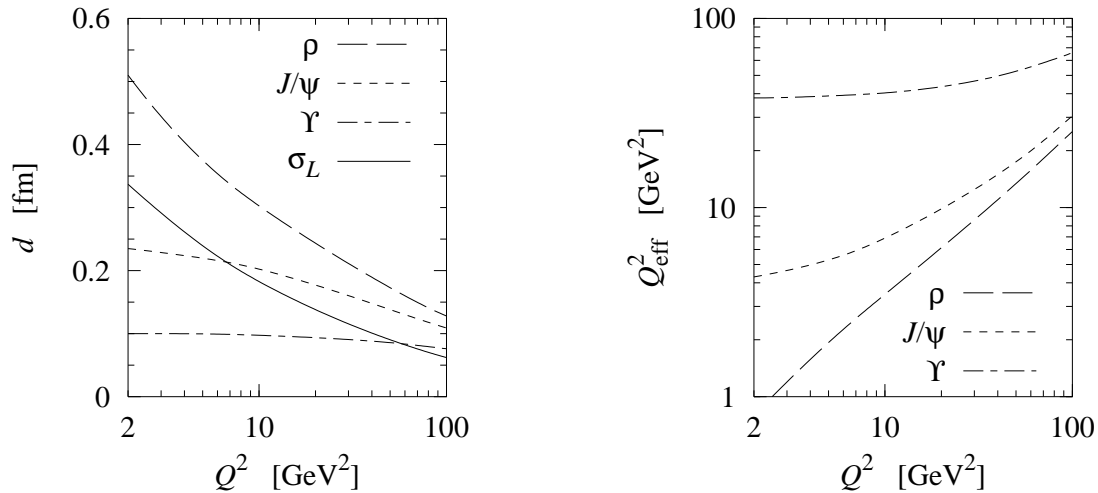}
\end{center}
\caption{The average dipole size, $d$, (left) and the effective scale,
$Q^2_{\text{eff}}$, (right) in exclusive vector meson production
($\rho, J/\psi, \Upsilon$) by longitudinally polarized photons, as a
function of $Q^2$ \cite{Frankfurt:1995jw,Frankfurt:1997fj}.  Also
shown are the shown are the average values of $d$ in the integrand 
of the expression for the inclusive cross section, $\sigma_L$.
\label{fig:bsize}}
\end{figure}
Equation~(\ref{dsigdtt}) is based on the leading logarithmic
approximation in $\ln (Q^2 / \Lambda_{\text{QCD}}^2)$, as well as on
the leading--twist approximation, Eq.~(\ref{psi_LT}). While it already
exhibits many qualitative features seen in the data (see below), two
important effects need to be taken into account before a quantitative
comparison can be attempted.  First, because the wave function of the
vector meson in Eq.~(\ref{conv}) is significantly broader than that of
the $\gamma^\ast_L$, the effective dipole sizes in the meson
production amplitude, Eq.~(\ref{amp_DGLAP}), are substantially larger
than in $\sigma_L$, Eq.~(\ref{sigma_L}), see Fig.~\ref{fig:bsize}.  As
a result, the effective scale in the gluon distribution,
$Q^2_{\text{eff}}$, is smaller in vector meson production than in
$\sigma_L$, see Fig.~\ref{fig:bsize}
\cite{Frankfurt:1995jw,Frankfurt:1997fj}.  This effect slows the $x$--
(energy) dependence of the cross section compared to the naive
estimate, $Q^2_{\text{eff}} = Q^2$.  Second, numerical studies using
model wave functions show that retaining the full $d$--dependence of
the vector meson wave function in the convolution integral
(\ref{conv}) results in a substantial decrease of the absolute cross
section at moderate $Q^2$ as compared to the leading--twist
approximation, Eq.~(\ref{psi_LT}), as well as in a slower $Q^2$
dependence \cite{Frankfurt:1995jw,Frankfurt:1997fj}.  These
higher--twist effects, related to the finite size of the vector meson,
limit the region of validity of the leading--twist approximation
(\ref{psi_LT}) and need to be taken into account in quantitative
estimates at low $Q^2$.

\subsection{Vector meson production at HERA}
\label{subsec:vector_meson}
\begin{figure}
\begin{center}
\includegraphics[width=10cm]{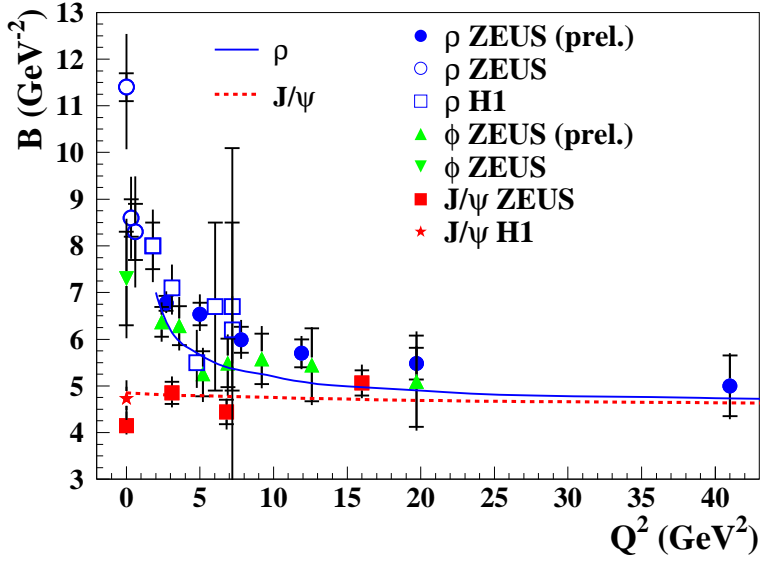}
\end{center}
\caption{The HERA H1 and ZEUS data for the $t$--slopes of the
differential cross sections for the exclusive electroproduction of
$\rho, \phi$ and $J/\psi$ mesons, as a function of $Q^2$.  The
convergence of the different slopes at large $Q^2$ indicates the
dominance of small--size configurations in the production process
(``squeezing''). The solid line shows the $Q^2$--dependence obtained
in the calculation of Ref.~\cite{Frankfurt:1995jw}. The data are from
Ref.~\cite{Levy}.
\label{fig:slopeh}}
\end{figure}

With proper choice of the effective scale, $Q^2_{\text{eff}}$, and
inclusion of higher--twist effects due to the finite transverse size
of the meson, one can quantitatively compare the results of the
leading logarithmic approximation, Eqs.~(\ref{amp_DGLAP}) and
(\ref{conv}), with the HERA data on heavy and light vector meson
production.  The data confirm in particular the following predictions
of this picture:
\begin{itemize}
\item
\textit{Increase of cross section with energy.} 
Equation~(\ref{dsigdtt}) implies that $d\sigma/dt (t = 0)$ grows with
energy as $\left[x G (x, Q^2_{\text{eff}}) \right]^2$, with
$Q^2_{\text{eff}}$ estimated to be $\sim 3\, \text{GeV}^2$. When
combined with the LO gluon density obtained from fits to DIS data,
this implies a growth $\propto W^{0.8}$. Such behavior has been
observed for $\rho$ production at $Q^2 = 10 - 20\, \text{GeV}^2$, and
for $J/\psi$ production starting from $Q^2 = 0$ \cite{Levy}.  The
later onset of the hard regime for $\rho$ electroproduction is due to
the rather slow ``squeezing'' of the $q\bar q$ configuration in the
$\rho$ meson; it reaches a size comparable to that of the $J/\psi$
only at $Q^2 \sim 20 \, \text{GeV}^2$, see
Fig.~\ref{fig:bsize}.\footnote{In the case of the $\rho-$ meson
production initiated by the transverse photon, the squeezing is
generated by the Sudakov form factor as well as by the more rapid
increase with energy of the small size contribution. The observed
behavior of $\sigma_L/\sigma_T$ can be fitted within the current
models \cite{Martin:1996bp}.} The naive choice $Q^2_{\text{eff}} =
Q^2$ would imply a too fast growth, \textit{cf.}
Fig.~\ref{fig:lambda}.  For soft interactions, on the other hand,
$d\sigma / dt (t = 0) \propto W^{0.32}$, and the growth is even
smaller for the cross section integrated over $t$.
\item 
\textit{Decrease of cross section with $Q^2$.} 
The decrease with $Q^2$ of $\sigma_L$ for $\rho$-meson production, and
of the total cross section for $J/\psi$ production, is slower than
$1/Q^6$, due to the $Q^2$--dependence of $\alpha_s G$ in
Eq.~(\ref{dsigdtt}), as well as finite--size (higher--twist) effects.
This is best observed in $J/\psi$ electroproduction, where the model
of Ref.~\cite{Ryskin:1992ui}, which neglects finite--size effects,
predicts a decrease of the cross section by a factor of $\sim 5$
faster than observed in the kinematic region covered by the H1
experiment.
\item 
\textit{Absolute cross sections.} 
The absolute cross sections for vector meson production are well
reproduced, provided that higher--twist effects due to the finite size
of the vector meson are taken into account
\cite{Frankfurt:1995jw,Frankfurt:1997fj}.
\item 
\textit{Dominance of longitudinal cross section.} 
The data on $\rho$ production indicate $\sigma_L \gg \sigma_T$ for
$Q^2\gg m_V^2$, in agreement with our picture.
\item 
\textit{Universality of $t$--dependence.} 
Comparison of $\rho$ and $J/\psi$ electroproduction data clearly show
the universality of the $t$--dependence at large $Q^2$, where the
vector mesons are ``squeezed'', and the $t$--dependence originates
solely from the two--gluon form factor, see Fig.~\ref{fig:slopeh}.
\item
\textit{Flavor symmetry.} 
Since the interaction of the $q\bar q$ dipole with the gluon
distribution is flavor blind, one expects the restoration of $SU(3)$
flavor symmetry in vector meson production for $Q^2\gg m_V^2$. For
example, $\phi:\rho=(2:9)$ in the flavor symmetry limit. The violation
of $SU(3)$ flavor symmetry due to increase of the wave function of the
vector meson at small distances with increasing quark mass leads to an
enhancement of this ratio by a factor $\sim 1.2$.
\end{itemize}
%
%

A new situation is encountered in the photoproduction of $\Upsilon$
mesons. In this case, the approximation of the generalized gluon
distribution by the usual gluon density becomes invalid (large
``skewness'' and large $Q^2_{\text{eff}}$), and the real part of the
amplitude becomes significant. Together, these effects increase the
predicted cross section by a factor of about 4 
\cite{Frankfurt:1998yf,Martinupsilon}.  For $\Upsilon$ production 
$Q^2_{\text{eff}} \approx 40\, \text{GeV}^2$, leading to an energy
dependence of the cross section as 
$d\sigma /dt (t = 0) \propto W^{1.7}$.

Closely related to vector meson production is the production of real
photons (deeply virtual Compton scattering, DVCS).  This process has
been the subject of intense theoretical study in the region of
moderate $x$, accessible in fixed--target experiments (HERMES at DESY,
COMPASS at CERN, Jefferson Lab), and is considered the main tool for
probing the generalized quark distributions in the nucleon
\cite{Brodsky:1972vv,Muller:1998fv,Ji:1996nm,Radyushkin:1997ki}.  At
small $x$, the DVCS amplitude has been computed in the leading 
$\ln (Q^2 / \Lambda_{\text{QCD}}^2)$ approximation outlined in
Sec.~\ref{subsec:squeezing}, and found to be substantially enhanced as
compared to the forward amplitude, $\gamma^*p \rightarrow \gamma^*p$
\cite{Frankfurt:1997at}.  The DVCS cross section reported by the HERA
experiments is in reasonable agreement with these predictions, as well
as with the color dipole model of Ref.~\cite{McDermott:2001pt}; see
Ref.~\cite{HERAdvcs} and references therein. The HERA data at small
$x$ are also well described by an NLO QCD analysis
\cite{Freund:2001hd,Freund:2002qf}, in which the modeling of the input
GPD's is a much more challenging problem than in LO, see
Ref.~\cite{Freund:2002qf} for details.  DVCS at small $x$ and the
closely related process of production of Z-bosons, 
$\gamma + p \rightarrow Z + p$, were also studied within the leading 
$\alpha_{s} \ln (x_0/x)$ approximation \cite{Bartels:1981jh}.

To summarize, the HERA data on exclusive electroproduction of vector
mesons clearly show the transition to the perturbative QCD regime for 
$Q^2 \geq 10-20\, \text{GeV}^2$.  This conclusion is consistent
with the observation of color transparency phenomena in several
other processes.  It establishes the study of exclusive processes 
($x, Q^2$ and $t$--dependence of the cross section) as a way to
extract detailed information about the interaction of small dipoles
with hadrons, as well as about the generalized parton distribution in
the nucleon.
\subsection{Transverse spatial distribution of gluons in the nucleon}
\label{subsec:transverse}
An important aspect of hard exclusive processes at small $x$ is that
they provide information about the transverse spatial distribution of
gluons in the nucleon. It is contained in the Fourier transform of the
two--gluon form factor, Eq.~(\ref{twogluon}),
\begin{equation}
F_g (x, \rho; Q^2_{\text{eff}}) 
\;\; \equiv \;\; \int \frac{d^2 \Delta_\perp}{(2 \pi)^2}
\; e^{i (\bm{\Delta}_\perp \bm{\rho})}
\; F_g (x, t = -\bm{\Delta}_\perp^2; Q^2_{\text{eff}}) ,
\label{rhoprof_def}
\end{equation}
where $\bm{\rho}$ is a transverse coordinate variable. The function
$F_g (x, \rho; Q^2_{\text{eff}})$ is positive definite
\cite{Pobylitsa:2002iu} and describes the spatial distribution of
gluons in the transverse plane, 
$\int d^2\rho\, F_g (x, \rho; Q^2_{\text{eff}}) = 1$.

%
%
\begin{figure}
\begin{center}
\includegraphics[width=8cm]{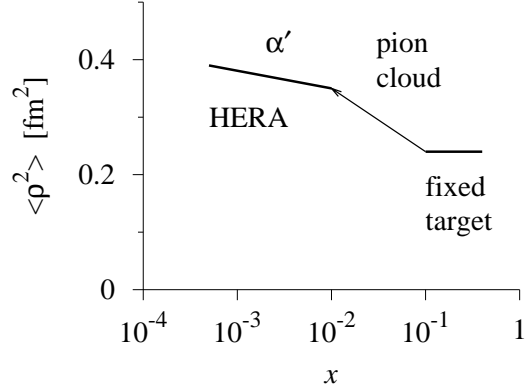}
\end{center}
\caption{The average squared transverse radius of the gluon
distribution in the nucleon, 
$\langle \rho^2 \rangle = \int d^2 \rho \, \rho^2 
F_g (x, \rho; Q^2_{\text{eff}})$, as a function of $x$, as extracted
from $J/\psi$ photoproduction data 
($Q^2_{\text{eff}} = 3\, \text{GeV}^2$) at various energies.
\label{fig:rho2}}
\end{figure}
The convergence of the $t$--slopes of $\rho$ and $J/\psi$ production
at large $Q^2$ (see Fig.~\ref{fig:slopeh}) demonstrates that the
$t$--dependence of the differential cross section is dominated by the
two--gluon form factor. The two--gluon form factor can thus be
extracted from the $J/\psi$ photoproduction data 
($Q^2_{\text{eff}} \approx 3 \, \text{GeV}^2$), with small corrections
($\sim 10\%$) due to the finite transverse size of the $J/\psi$
meson. This process has been measured over a wide range of energies;
see Refs.~\cite{Frankfurt:2002ka,DIS04} for an overview of the
data. At fixed--target energies, $x \sim 10^{-1}$, the $t$--dependence
of the data is well described by a two--gluon form factor of dipole
form,
\begin{equation}
F_g \;\; = \;\; (1 - t/m_g^2)^{-2}, 
\hspace{4em} m_g^2 \;\; = \;\; 1.1 \, \text{GeV}^2  
\hspace{4em} (x \sim 10^{-1}),
\label{twogluon_dipole}
\end{equation}
where the parameter, $m_g$, is close to that in the dipole fit
to the axial form factor of the nucleon. This corresponds to a
narrow spatial distribution of gluons in the transverse plane, with an
average transverse radius 
$\langle \rho^2 \rangle = 8/m_g^2 \approx 0.28 \, \text{fm}^2$, see
Fig.~\ref{fig:rho2}.  At HERA energies, $x \sim 10^{-2} - 10^{-3}$,
the average radius is larger, 
$\langle \rho^2 \rangle \approx 0.35 \, \text{fm}^2$.  It also
exhibits a slow growth with $\ln (1/x)$, with a slope,
$\alpha^{\prime}$, significantly smaller than the value for soft
interactions. The $J/\psi$ photoproduction data from H1 give
$\alpha^{\prime}_{\text{hard}} = 0.08 \pm 0.17\; \text{GeV}^{-2}$
\cite{Adloff:2000vm}, the ZEUS electroproduction data
$\alpha^{\prime}_{\text{hard}} =0.07 \pm 0.05 (\text{stat}) {
}^{+0.03}_{-0.04} (\text{syst}) \; \text{GeV}^{-2}$
\cite{Chekanov:2004mw}, which should be compared to $\alpha^{\prime}
\approx 0.25 \, \text{GeV}^{-2}$ for $pp$ elastic scattering. This
reflects the suppression of Gribov diffusion for partons with large
virtualities, see the discussion in Sec.~\ref{subsec:spacetime}.

The change of the nucleon's average transverse radius between 
$x\sim 10^{-1}$ and $10^{-2}$ can naturally be explained by chiral
dynamics. Pions in the nucleon wave function carry momentum fractions
of the order $m_\pi / m_N$.  For $x > m_\pi / m_N$ the pion cloud does
not contribute to the gluon distribution, and the two--gluon form
factor is similar to the nucleon axial form factor, which also does
not receive contributions from the pion cloud. For 
$x \ll m_\pi / m_N$, the pion cloud contributes to the gluon
distribution and leads to an increase of $\langle \rho^2 \rangle$ by
$20 - 30\%$, see Fig.~\ref{fig:rho2} \cite{Strikman:2003gz}.

It is worth emphasizing that for smaller $x$ the increase of the
transverse size should continue due to the Gribov diffusion. Indeed, 
a hard probe can interact with a parton of the soft ladder, responsible
for the growth of the soft radius, if the soft parton's momentum fraction 
is sufficiently small compared to $x$. At very small $x$ and fixed $Q^2$ the
rate of the growth should thus be comparable to that in the soft case
\cite{FS99}. No such effect is present in the BFKL model where
the interaction of two small dipoles is considered.

The change of the transverse spatial distribution of gluons in the
nucleon with the scale, $Q^2_{\text{eff}}$, due to DGLAP evolution
should generally be small \cite{Frankfurt:2003td}.  For
$Q^2_{\text{eff}}$ sufficiently large compared to the transverse
spatial resolution, the parton decays happen essentially locally in
transverse position. For fixed $x$, one finds that the transverse
spatial distribution shrinks with increasing scale, because the
distribution becomes sensitive to the input distribution (at the
initial scale) at higher values of $x$, where it is concentrated at
smaller transverse distances.

\subsection{Color transparency in hard processes with nuclei}
\label{subssub:hard}
QCD predicts that the spatially small quark--gluon wave packets formed
in hard $\gamma^\ast$--induced scattering processes interact weakly
with hadronic matter, because of the color neutrality of the photon.
At sufficiently small $x$, where the cross section is proportional to
the gluon density, \textit{cf.}\ Eq.~(\ref{sigma_d_DGLAP}), one
expects the ratio of the cross sections for $\gamma^\ast$ scattering
from a nucleus and a single nucleon to be equal to the ratio of the
respective gluon densities, a property known as generalized color
transparency \cite{Frankfurt:it,BFGMS94}. Because with increasing $Q^2$
gluon shadowing at fixed $x$ disappears (\textit{cf.}\ the discussion
in Sec.~\ref{subsec:shadowing_heavy_ion}), one further expects that
\begin{equation}
\sigma_{\text{tot}}^{\gamma^*A}/ (A\sigma_{\text{tot}}^{\gamma^* N}) 
\;\; \rightarrow \;\; 
1 \hspace{3em} (Q^2 \rightarrow \infty ; \; x \; \text{fixed, small}),
\label{ct}
\end{equation}
which is referred to as color transparency proper.  Conversely, at
fixed $Q^2$ and decreasing $x$, the ratio in Eq.~(\ref{ct}) should
decrease owing to the more important role of nuclear shadowing, and
color transparency phenomena should completely disappear at very small
$x$, where QCD factorization breaks down. This is in contrast to the
two-gluon exchange model of Refs.~\cite{Bertsch81,Fra81}, which
neglects the space--time evolution of the dipole. In this model
nuclear shadowing is obtained from exchanges of additional gluon
between the current and target fragmentation regions, which is a
higher--twist effect ($\propto 1/Q^2$) and disappears at large $Q^2$.

The color transparency phenomenon has been directly observed in three
experiments:
\begin{itemize}
\item 
The total cross section for $\gamma^\ast A$ scattering increases
with the atomic number as $A^{\alpha}$ with $\alpha \approx 1$, faster
than the cross section for a hadronic projectile, see
Ref.~\cite{Arneodo} for a review of the experimental data.
\item
The cross section for coherent photoproduction of $J/\psi$ mesons from
nuclei increases with $A$ much faster than that for coherent $\rho$
meson production. The Fermilab E691 experiment \cite{Sokoloff:1986bu}
observed 
$\sigma^{\gamma^\ast + A \rightarrow J/\psi + A} \propto A^{1.46}$ 
at $E_\gamma = 150\, \text{GeV}$. Color transparency predicts that the
coherent cross section integrated over $t$ is 
$\propto (A^2 / R_A^2 ) \approx A^{4/3}$. This $A$--dependence
corresponds to the coherent sum of collisions from independent
nucleons without absorption. A somewhat faster $A$--dependence emerges
because of the contribution of incoherent diffractive processes
\cite{FMS2002}.
\item
The $A$--dependence of the cross section for coherent dijet production
in pion--nucleus scattering is anomalously large, as predicted in
Ref.~\cite{Frankfurt:it}. The Fermilab E791 experiment \cite{Aitala:2000hc}
observed a dependence $\propto A^{1.54}$ at 
$E_{\pi}\approx 600\, \text{GeV}$, similar to that in coherent
$J/\psi$ production. Note that the conventional Glauber approximation
predicts $A^{1/3}$. Furthermore, the observed dependence of the cross
section on the pion momentum fraction and the jet transverse momentum
is well consistent with the perturbative QCD prediction of
Ref.~\cite{FMS2002}. Notwithstanding the fact that the absolute cross
section has not been measured, this is probably the first experimental
observation of the high--momentum tail of the pion wave function as
due to one--gluon exchange.
\end{itemize}
To summarize, there exists strong experimental evidence for color
transparency in high--energy scattering. This phenomenon could be the
basis for new ``non-destructive'' methods of investigating the
microscopic structure of hadrons and nuclei in the future.
\section{Diffraction in $\gamma^\ast p$ scattering}
\label{sec:diffraction}
\subsection{QCD factorization for hard diffractive processes}
\label{subsec:factorization_diffractive}
Measurements of DIS at HERA have established the existence of a class
of events in which the proton is observed in the final state, with a
small invariant momentum transfer, $t$, and a hadronic system of
invariant mass $M_X^2 \ll W^2$ is produced with a rapidity gap
relative to the proton. In a frame in which the nucleon is fast-moving
(\textit{i.e.}, in parton model kinematics) such processes are
characterized by the fractional energy loss of the proton,
$x_{\Pomeron} = (E_p^i - E_p^f)/E_p^i$, and the transverse momentum
transfer, $\bm{\Delta}_\perp$, with 
$t = -(\bm{\Delta}_\perp^2 + x^2_{\Pomeron} m_N^2)/(1-x_{\Pomeron})$.
In analogy with the corresponding phenomenon in hadronic collisions
one refers to such processes as diffractive, although a priori the
dynamics is not governed by soft physics.

\begin{figure}
\begin{center}
\includegraphics[width=6cm]{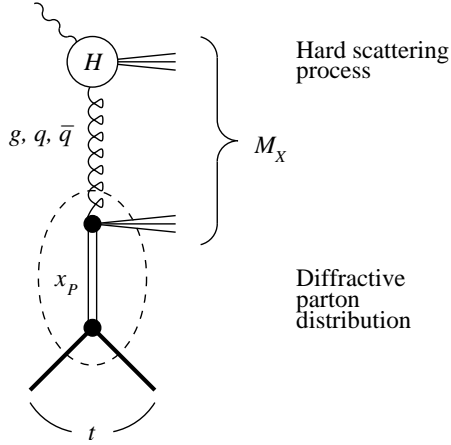}
\end{center}
\caption{Factorization in diffractive DIS. The amplitude in the dashed
blob, multiplied with its complex conjugate and summed over final
states, defines the diffractive PDF. The internal structure of the
dashed blob illustrates the assumption (\ref{factr}), which does not
follow from the QCD factorization theorem.
\label{fig:diff}}
\end{figure}

Following suggestions of earlier works, a formal QCD factorization
theorem was proved in Ref.~\cite{Trentadue:1993ka,Collins:1997sr} for
hard processes of the type
\begin{equation}
\gamma^\ast \; + \, p \;\; \rightarrow \;\; h \; + \; 
\text{(rapidity gap)} \; + \; X ,
\label{diffr}
\end{equation}
where $X$ is either an inclusive state, or a state with extra hard
activity (dijet production, heavy quark production, \textit{etc.}),
see Fig.~\ref{fig:diff}. Similar to inclusive DIS, processes
(\ref{diffr}) with a given hadron $h$ in the target fragmentation
region are characterized by so-called conditional parton distribution
functions, $f_j^h (\beta, Q^2,x_h,t)$, which are independent of the
hard process and satisfy the same DGLAP evolution equations for fixed
$x_h$ and $t$. Here $\beta \equiv x/(1 - x_h) = Q^2/(Q^2+M_X^2)$ is
the fraction of the light-cone momentum of the target available for
hard interactions in hadron $h$. Most of the current studies focus on
diffractive kinematics, where $h = p$ and 
$1 - x_h = x_{\Pomeron} \leq 0.01$.  The conditional parton
distribution functions in this case are referred to as diffractive
parton distribution functions (dPDF's), and denoted as $f^D_j$.  In
current data analysis it is usually assumed that the dependence of the
dPDF's on $x_{\Pomeron}, t$ and $\beta, Q^2$ can be factorized as
\cite{Ingelman:1984ns}
\begin{equation}
f_j^D (\beta,Q^2,x_{\Pomeron},t) \; = \; 
f_{\Pomeron/p}(x_{\Pomeron},t) \; f_j^{D} (\beta,Q^2),
\hspace{3em}
f_{\Pomeron/p}(x_{\Pomeron},t) \; = \; 
f(t) \; x_{\Pomeron}^{-2\alpha_{\Pomeron}(t)+2}.
\label{factr}
\end{equation}
This assumption is inspired by the soft Pomeron exchange model (which
does not follow from the QCD factorization theorem) and referred to as
Regge factorization, see Fig.~\ref{fig:diff}.  An additional term can
be added to Eq.~(\ref{factr}) in analogy to non-vacuum exchange in
soft physics; it gives a small contribution below 
$x_{\Pomeron} \sim 0.01$ and dominates at $x_{\Pomeron} \ge 0.05$.
%
%

Extensive studies of hard diffractive channels have been performed at
HERA. The inclusive diffractive cross section was measured both
integrated over $t$ (the so-called diffractive structure function,
$F_2^{D(3)}$), and as a function of $t$ for a limited range of
$x_{\Pomeron}$. These measurements are mostly sensitive to the quark
dPDF, while the gluon dPDF enters through scaling violations.
Diffractive dijet production for real and virtual photons, as well as
diffractive charm production, primarily probe the gluon dPDF.  The
analysis of these data on the basis of QCD evolution equations has led
to the following conclusions:
\begin{itemize}
\item 
The data at $Q^2\geq 4\, \text{GeV}^2$ are described by the universal
dPDFs, consistent with the factorization theorem.
\item 
$f_g(\beta, Q_0^2) \gg \sum_q f_q(\beta, Q_0^2)$ for the studied range
of $\beta$. This conclusion was initially based on the weak scaling
violation for $F_2^{D(3)}(\beta, Q^2)$ for large $\beta$, and was
later confirmed by the studies of diffractive dijet production and
charm production. However, the latter processes have so far been
treated only in the LO approximation, and one should await the NLO
analysis before drawing final conclusions.
\item 
The data are consistent with Regge factorization, Eq.~(\ref{factr}),
although the $x_{\Pomeron}$ dependence is faster than in soft
physics. The analysis of ZEUS and H1 diffractive data finds
$\alpha_{\Pomeron}(t=0) = 1.2\pm 0.07$ and increasing with $Q^2$,
which should be compared with the energy dependence expected in soft
hadronic collisions, $\alpha_{\Pomeron}\approx 1.1$.
\item
The absolute probability of diffraction in $\gamma^* + p$ scattering
is of the order of 10\% for moderate $Q^2$, and thus of the same
magnitude as in soft pion--nucleon collisions.  However, the rate of
increase of the diffractive cross section with energy for fixed
$M_X^2$ and $Q^2$ is significantly faster than that of the total DIS
cross section.
\end{itemize}

\begin{figure}[t]
\begin{tabular}{ll}
\includegraphics[width=7.5cm]{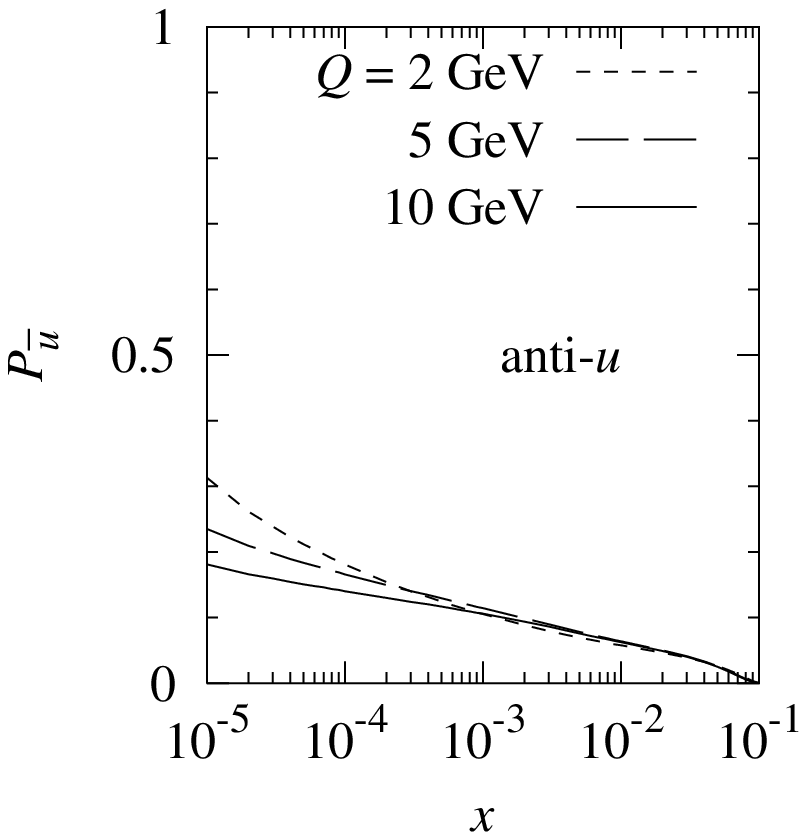}
& 
\includegraphics[width=7.5cm]{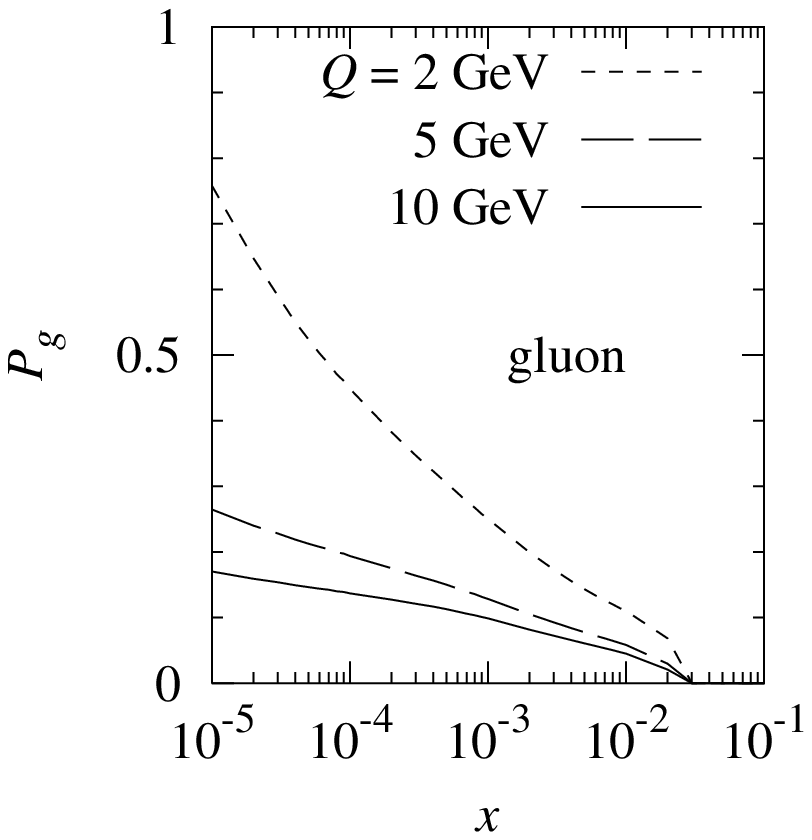}
\end{tabular}
\caption{Probability of diffractive scattering from anti--$u$ quarks
(left) and gluons (right), extracted from a fit to the H1 
data, see Ref.~\cite{Frankfurt:1998ym} (updated by V.~Guzey).
\label{fig:prob}}
\end{figure}

Another interesting characteristic of diffractive DIS is the
probability of diffractive scattering depending on the type of parton
coupling to the hard probe \cite{FS99},
\begin{equation}
P_j (x,Q^2) \;\; = \;\; \left.
\int dt \int dx_{\Pomeron} \; 
f_j^D(x / x_{\Pomeron}, Q^2, x_{\Pomeron}, t) \right/
f_j (x,Q^2) .
\end{equation}
This ratio cannot exceed the value 0.5, which corresponds to the
unitarity (black disk) limit (BDL), \textit{cf.}\ the discussion in
Sec.~\ref{sec:BDL}. Using the H1 fit to the diffractive DIS
data (see Fig.~\ref{fig:prob}), we find 
$P_g \gg P_q$, and $P_g (x \sim 10^{-3}) \approx 0.4\, (0.3)$ for 
$Q^2 = 4 \, (10)\, \text{GeV}^2$. That is, quark induced diffraction
is small, whereas gluon induced diffraction is close to the maximum value
allowed by unitarity. We shall return to this point in our discussion
of the profile function for the dipole--nucleon interaction in
Sec.~\ref{sec:BDL}. Note that the H1 fit is based on the
data at $x\geq 10^{-4}$. The fact that it leads to $P_g\geq 0.5$ at
smaller $x$ indicates that the H1 parameterization should break down
near the upper end of the HERA energy range.
%
%

\subsection{Space--time picture of hard diffractive processes}
To understand the observed pattern of hard diffraction, it is
instructive to consider the space--time evolution of such processes in
the target rest frame. Such studies reveal new information about the
interaction of small--size $q\bar q$ as well as $q\bar q g \ldots g$
configurations with hadronic matter.  In particular, the ratio of the
diffractive to the total cross section probes the interactions of such
configurations with the target without reference to the probe which
created them.

An immediate consequence of the QCD factorization theorem is that, in
the target rest frame, the number of components in the photon wave
function evolves with $Q^2$. While at low $Q^2$ an approximation by a
few components in the photon wave function may be reasonable, it is
definitely inappropriate for large $Q^2$.  This point is illustrated
by the following example: Diffractive processes induced by
longitudinally polarized photons are a leading--twist effect. If,
however, all but the $q\bar q$ component of the photon wave function
were neglected, one would erroneously conclude that diffraction is a
higher--twist effect in this channel, because the transverse size of
the longitudinal photon is $d^2\propto 1/Q^2$. The proper $Q^2$
dependence is restored by the $q\bar q g \ldots g$ configurations in
the photon wave function.

At low $Q^2$, aligned jet model--type configurations of large
transverse size dominate in the wave function of the projectile
photon, \textit{cf.}\ Sec.~\ref{subsec:spacetime}.  Such
configurations interact with a hadronic cross section, 
$\sim \sigma_{\text{tot}}(\pi N)$, and thus have a significant
probability to rescatter elastically. If $Q^2$ increases, these
configurations cannot be effectively produced without emission of
gluons. Because these gluons are predominantly emitted collinearly,
they do not change the transverse size of the diffracting system, and
hence the probability of elastic rescattering.  Assuming smooth
matching between the strength of interaction in the perturbative and
nonperturbative regimes, models reasonably describe the data on hard
diffraction in $ep$ scattering if $P_g > P_q$ at the initial scale,
$Q_0^2$ \cite{Buchmuller:1998jv,Bartels:1998ea}.  Configurations of
size $d^2 \geq \lambda/Q_0^2$, where $\lambda \approx 9$ (\textit{cf.}
the discussion in Sec.~\ref{subsec:dipole}) should be included in the
definition of the dPDF at the initial scale.  For 
$Q_0^2 = 4 \, \text{GeV}^2$, this includes rather small transverse
sizes, for which the cross section increases with energy significantly
faster than at the soft scale, which is consistent with the trend of
the HERA data mentioned in
Sec.~\ref{subsec:factorization_diffractive}.  Also, because the 
$q\bar q g$ configurations have masses considerably larger than 
$q\bar q$ configurations, they should manifest themselves in
diffraction at relatively small $\beta < 0.5$. Indications of
diffraction into non-aligned jet final states were indeed found in a
number of HERA experiments \cite{HERAdiff3jet}. To summarize, it
appears that hard diffraction at HERA with $Q^2\sim 4\, \text{GeV}^2$
represents the border between the high--$Q^2$ region where
leading--twist QCD dominates and the low--$Q^2$ region where
higher--twist effects become important.

A quantitative analysis of the HERA data within the gluon dipole
picture indicates that the interaction of gluon dipoles at HERA
energies is rather close to the BDL \cite{FS99},
\textit{cf.}\ the discussion in Sec.~\ref{sec:BDL}. However,
due to our inability to build an effective trigger on the interaction
of gluon dipoles with $d\ge 0.3 \text{fm}$ it is difficult to observe
this effect directly in the experiments.

To conclude this discussion, we briefly want to comment on the
assumption of Regge factorization, Eq.~(\ref{factr}). Because strong
deviations of the energy dependence of diffraction from the soft
regime are observed, there is a priori no reason for the validity of
this assumption.  Several effects are likely to contribute to the
breakdown of Regge factorization: \textit{(a)} different energy
dependence of the cross section for diffraction of configurations of
different transverse size, \textit{(b)} emission of gluons by 
$q\bar q$ dipoles, which at smaller $x$ occurs at large coherent
lengths (\textit{cf.}\ the discussion in Sec.~\ref{subsec:spacetime}),
and \textit{(c)} soft screening effects, which become more important with
increasing energy (these effects were observed in soft diffraction
\cite{Goulianos:wy,Erhan:1999gs}), and which should be different for
the various diffractive configurations, as they interact with
different strengths.

One way to probe the degree of validity of Regge factorization is to
check that the hadrons produced in the photon fragmentation region do
not ``talk'' with hadrons in the target fragmentation region,
\textit{i.e.}, that there are no long--range correlations in rapidity.
Such a factorization ignores the existence of color fluctuations,
which lead to processes in which the proton also dissociates.  If
Regge factorization were valid, the probability of dissociation would
not depend on the properties of the state into which the virtual
photon has diffracted. However, if screening effects are present and
non-universal (for example, due to the different strength of
interaction in quark and gluon induced diffraction), Regge
factorization should be broken.

\subsection{Diffraction and leading--twist nuclear shadowing}
\label{subsec:shadowing}
A direct relation between diffraction in high--energy hadron--hadron
collisions and the nuclear shadowing effect in hadron--nucleus
collisions was derived by V.~Gribov \cite{Gribov:1968jf,Gribov:1968gs},
in the approximation where the nucleon radius is considerably smaller
than the mean internucleon distance in nuclei. The same reasoning in
conjunction with the leading--twist approximation for hard diffractive
processes allows one to calculate nuclear shadowing of PDF's in light
nuclei \cite{FS99}.

Application of the Abramovsky--Gribov--Kancheli cutting rules
\cite{AGK}, which relate the shadowing phenomenon for the total cross
section of high energy processes to that of partial cross sections
(such as for diffraction, multiparticle production, \textit{etc.}) and
are valid in perturbative QCD (see Ref.~\cite{Treleani} for a recent
discussion), explicitly demonstrates that the interference of the
amplitudes of diffraction from a proton and a neutron leads to a
decrease of the total cross section for $\gamma^{\ast}D$ scattering.
When combined with the factorization theorem for inclusive
diffraction, one can calculate the modification of nuclear PDF's at
low values of Bjorken $x$ \cite{FS99}. In the limit of low nuclear
thickness, the nuclear shadowing corrections to the nuclear parton
densities are given by
\begin{eqnarray}
\frac{f_{j/A}(x,Q^2)}{A} &=&  f_{j/N}(x,Q^2) \; - \; \frac{1}{2}
\int\! d^2b \int_{-\infty}^{\infty}\! dz_1 \int_{z_1}^{\infty} \! dz_2 
\int_x^{x_0} \! dx_{\Pomeron} \; 
\cos\left[ x_{\Pomeron}m_N(z_1-z_2) \right]
\nonumber \\
&\times& \frac{1-\eta^2}{1+\eta^2}\; f^{D}_{j/N}
\left(\beta, Q^2,x_{\Pomeron},t\right)_{\left| \bm{k}_\perp = 0 \right.}
\; \rho_A(b,z_1) \; \rho_A(b,z_2) ,
\label{sh1}
\end{eqnarray}
where $f_{j/N}(x,Q^2)$ is the usual parton density in the
proton, $f^{D}_{j/N}(\beta,Q^2,x_{\Pomeron},t)$ the diffractive parton
density (see Sec.~\ref{subsec:factorization_diffractive}), and
$\rho_A(r)$ is the nucleon density in the nucleus with atomic number
$A$. The momentum transfer, $t$, is given by 
$-t = (\bm{k}_\perp^2+(x_{\Pomeron} m_{N})^2)/(1-x_{\Pomeron})$, where
$\bm{k}_\perp$ is the transverse component of the momentum,
transferred to the struck nucleon, and $\beta = x/x_{\Pomeron}$.  In
Eq.~(\ref{sh1}), the factor $(1-\eta^2)/(1+\eta^2)$, where 
$\eta\equiv -\pi / 2\, \partial \ln(\sqrt{f^D_{i/N}})/ \partial
\ln(1/x_{\Pomeron})=\pi / 2 \, \left[ \alpha_{\Pomeron}(t=0)-1
\right]$, accounts for the real part of the amplitude of diffractive
scattering \cite{AFS99}. One can easily modify Eq.~(\ref{sh1}) to
include the dependence of the diffractive amplitude on $t$.
Obviously, both the left-- and right--hand side of Eq.~(\ref{sh1})
satisfy the QCD evolution equations in all orders in $\alpha_s$, and
this relation does not depend on the renormalization scheme.  These
expressions represent the model--independent result for leading--twist
nuclear shadowing in the low--density limit.

In the case of heavy nuclei one may with good accuracy neglect the
fluctuations of the strength of interaction in the hadron component of
the photon wave function.  This approximation makes it possible to
extend the above formulas to the case of heavy nuclei
\cite{FS99,Guzey}. Numerical studies of shadowing using
Eq.~(\ref{sh1}) and the corresponding expression for the total DIS
cross section for heavy nuclei found large leading--twist shadowing
effects for quark and gluon distributions, with gluon shadowing being
larger up to rather high values of $Q^2$. The latter effect can be
traced to the higher probability of gluon--induced diffraction as
compared to quarks, see Fig.~\ref{fig:prob}.

The connection between diffraction and nuclear shadowing does not
depend on the twist decomposition of the cross section, and was
successfully applied also to data on nuclear shadowing of $F_{2A}$ at
intermediate $Q^2$, see \textit{e.g.}\
Refs.~\cite{Piller:1999wx,Kaidalov}. One can use this to estimate the
relative importance of leading--twist and higher--twist nuclear
shadowing at $Q^2\leq 2\, \text{GeV}^2$, using experimental
information on the leading--twist contribution to the diffractive
cross section at these values of $Q^2$. One finds that a significant
higher--twist contribution to diffractive DIS originates from $\rho$
meson production.  This implies that a significant fraction 
($\sim 40\%$) of the nuclear shadowing observed in the experiments
at CERN and Fermilab (see Ref.\cite{Arneodo} for a review)
may be due to higher--twist effects \cite{Guzey}.

Recently, leading--twist nuclear shadowing was included in the initial
conditions for the small--$x$ evolution in the McLerran--Vegnugopalan
model \cite{MV}. A distinctive feature of this model is that the
invariant masses in the nuclear vertex of the BFKL ladder should be
very large as compared to $Q^2$. That is, small $\beta$ should
dominate in the integral, in analogy to Eq.~(\ref{sh1}). A numerical
analysis of gluon shadowing using the HERA dPDF's finds
that the region $\beta \le 0.1$ becomes important only for 
$x\le 10^{-4}$. Thus, the assumption of the dominance of large
diffractive masses may give rise to important dynamical effects at the
next generation of accelerators.

\subsection{Implications of nuclear shadowing for heavy--ion collisions}
\label{subsec:shadowing_heavy_ion}
The typical $x$--values relevant for semihard production of hadrons in
heavy--ion collisions decreases with energy as 
$x_A \sim 2 p_\perp / \sqrt{s_{NN}}$ for central rapidities, and much
faster, $\propto 1/s$, for the fragmentation regions ($s_{NN}$ is the
squared center--of--mass energy of the effective nucleon--nucleon
collisions). For central rapidities and $p_\perp \geq 2\, \text{GeV}$,
gluon shadowing is still a small correction at RHIC.  However, it will
be a large effect at LHC for a wide range of $p_\perp$, because the
relevant $x_A$ are much smaller than 0.01. The expected suppression of
jet production is given by the factor 
$\left[G_A(x_A, p_\perp^2 )/A G(x_A, p_\perp^2 )\right]^2$, where
$G_A$ and $G$ are the gluon densities in the nucleus and the nucleon,
respectively. This factor can be of the order of 1/4
\cite{Frankfurt:1998ym}.

Because the current RHIC detectors have rather limited forward
coverage, they have limited sensitivity to small--$x$ phenomena. One
exception is $J/\psi$ production, which, if interpreted within
perturbative QCD, probes $x$ down to 0.003. The observed suppression
of the $J/\psi$ yield is consistent with the estimates of
Ref.\cite{Guzey}, see Ref.~\cite{Vogt:1999cu} for a review. The
$A$--dependence of forward high--$p_\perp$ hadron production was
studied by BRAHMS \cite{Jipa:2004ny}. Although at large rapidities
small $x$ contribute to the high $p_\perp$ spectra, the QCD analysis
indicates that average $x$ are $\sim 0.03$
\cite{Guzey:2004zp}. Consequently, the yields are practically not
sensitive to the shadowing effects, or, more generally, to any
initial--state modifications of the nucleus wave function.
Final--state interaction effects which could explain the data are
nonperturbative contributions to the production of leading hadrons,
due to coalescence of spectator partons and the relatively small
energy losses in the initial and final state (on the scale of 3\% ).
\section{Black--disk limit in dipole--hadron interactions}
\label{sec:BDL}
\subsection{Violation of the leading--twist approximation at small $x$}
\label{subsec:violation_LT}
A fundamentally new dynamical effect expected at high energies is the
unitarity limit, or black--disk limit (BDL), in the interaction of a
small dipole with hadronic matter.  We now describe and quantify this
effect, using the information gathered in our studies of inclusive,
exclusive and diffractive DIS in Sections~\ref{sec:inclusive},
\ref{sec:exclusive} and \ref{sec:diffraction}.

A simple argument shows that the twist expansion for the cross
sections of hard processes breaks down at sufficiently small $x$. QCD
factorization predicts that the total cross section for DIS at fixed
$Q^2$ increases with decreasing $x$ as 
$\sigma_{\text{tot}} \propto xG(x)/Q^2$.  At the same time, the cross
section of elastic dipole--hadron scattering (which corresponds to the
production of diffractive states with masses 
$M_X \propto 1/d \propto Q$) grows much faster, 
$\sigma_{\text{diff}} \propto (xG)^2 /(B Q^4)$, where $B$ is the
$t$--slope of the corresponding differential cross section
\cite{BFGMS94}. Clearly, at sufficiently small $x$ there is a
contradiction --- the total cross section should always be larger than
that for any particular channel \cite{AFS}.  The resolution of this
paradox is that the decomposition of hard amplitudes in powers of
$1/Q^2$ becomes inapplicable at sufficiently small $x$.

In order to quantify the onset of the new regime, it is instructive to
consider the effects of unitarity (conservation of probability) on a
purely theoretical scattering process, namely the scattering of a
$q\bar q$ (quark--antiquark) or $gg$ (gluon--gluon) dipole of small
transverse size, $d$, from a hadronic target. Neglecting other
constituents in the dipole is justified in a wide kinematic range by
the smallness of the coupling constant; large terms 
$\propto \alpha_s \ln (x_0 / x)$ arise only from interactions at large
rapidity intervals. The invariant amplitude for dipole--proton elastic
scattering is a function of the invariants $s \equiv W^2$, and $t$. We
write is as a Fourier integral over the dipole--proton impact
parameter, $b$,
\begin{eqnarray}
A^{dp}(s, t) &=& \frac{i \, s}{4\pi} \int d^2 b \;
e^{-i (\bm{\Delta}_\perp \bm{b})}
\; \Gamma^{dp} (s, b) 
\hspace{3em} (t = -\bm{\Delta}_\perp^2),
\label{fbs}
\end{eqnarray}
where $\Gamma^{dp} (s, b)$ is the so--called profile function.  Making
use of unitarity, one can express the total, elastic, and inelastic
(total minus elastic) cross sections in terms of the profile function
as
\begin{equation}
\left. \begin{array}{l} 
\sigma_{\text{tot}}(s) \\[1ex]
\sigma_{\text{el}}(s) \\[1ex]
\sigma_{\text{inel}}(s)
\end{array}
\right\}
\;\; = \;\; \int d^2 b \; \times 
\left\{ \begin{array}{l} 
2 \, \text{Re} \, \Gamma^{dp} (s,b) \\[1ex]
|\Gamma^{dp} (s,b)|^2 \\[1ex]
\left[ 1 - |1-\Gamma^{dp} (s,b)|^2 \right] .
\end{array}
\right.
\label{unitarity}
\end{equation}
In the situation where elastic scattering is the ``shadow'' of
inelastic scattering, the profile function at a given $b$ is
restricted to values $|\Gamma^{dp}(s, b)| \leq 1$. The value
$\Gamma^{dp} (s,b) = 1$ corresponds to complete absorption at a given
impact parameter, the so-called black disk limit (BDL).
\footnote{In non-relativistic quantum mechanics, the scattering of a
particle from a completely absorptive sphere is referred to as the
``black--body limit.'' In contrast, the high--energy limit of
scattering amplitudes in QCD is essentially two--dimensional, with the
radius of interaction increasing with energy. It is thus natural to
refer to the limit of complete absorption as the ``black disk
limit.''}

\begin{figure}[t]
\begin{center}
\includegraphics[width=16cm]{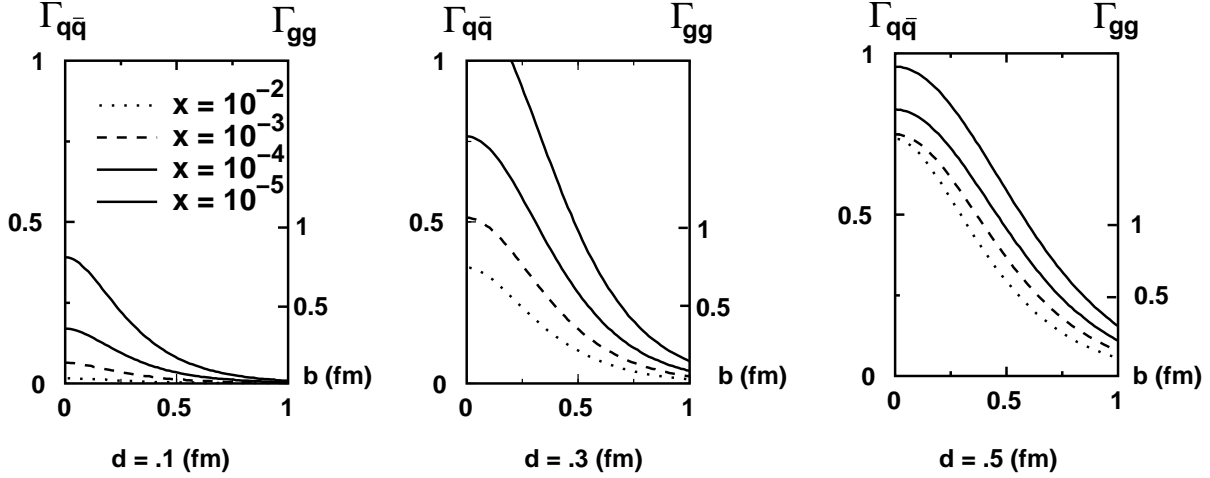}
\end{center}
\caption{The profile function of dipole--nucleon scattering,
$\Gamma^{dp}$, as a function of the impact parameter, $b$, 
for various values of the dipole size, $d$, and $x$, as 
obtained from a phenomenological estimate (see text).
\label{fig:profile}}
\end{figure}

The proximity of $\Gamma^{dp} (s,b)$ to unity is an important measure
of the strength of the interaction of the dipole with the proton.  As
outlined in Sec.~\ref{subsec:spacetime}, the analysis of 
$\gamma^\ast N$ scattering in the target rest frame allows one to
determine with reasonable accuracy (LO approximation) the total cross
section for the scattering of a $q\bar q$ dipole from the proton.
Combining this with information on the transverse spatial distribution
of gluons in the nucleon, obtained from measurements of the
$t$--dependence of exclusive $J/\psi$ photoproduction and other hard
exclusive processes (\textit{cf.}\ Sec.~\ref{subsec:transverse}), we
can calculate the profile function for dipole--nucleon scattering
\cite{Frankfurt:2003td}. A sample of the results of Ref.~\cite{Ted} is
presented in Fig.~\ref{fig:profile}. Here, $x = Q^2/s$, where 
$Q^2 = \lambda/d^2 \approx 9/d^2$ is the characteristic virtuality
corresponding to the dipole size. One sees that up to the top of the
HERA energy range, $x \sim 10^{-4}$, the profile function for a 
$q\bar q$ dipole remains small for dipole sizes 
$d \leq 0.3\, \text{fm}$, corresponding to 
$Q^2 \geq 4 \, \text{GeV}^2$, the value usually used as a starting
point for DGLAP evolution \cite{Pumplin:2002vw}. Even for larger
dipole sizes, $d\sim 0.5\, \text{fm}$, the fraction of the cross
section due to scattering with $\text{Re}\, \Gamma \geq 0.5$ remains
small.  This shows that the BDL is not reached in inclusive DIS at
HERA energies, in agreement with what we observed in
Sec.~\ref{subsec:DGLAP} and \ref{subsec:dipole}.  However, the DGLAP
evolution starting from a $q\bar q$ dipole generates $gg$ dipoles,
whose interaction at leading twist is larger by a factor of 9/4,
\textit{cf.}\ Eq.~(\ref{sigma_d_DGLAP}), and thus approaches the BDL
much earlier. The theoretical estimate shown in Fig.~\ref{fig:profile}
indicated that the strength of interaction of $gg$ dipoles is close to
maximal at the top of the HERA energy range, for a wide range of $b$,
and dipole sizes corresponding to $Q^2 \le 4\, \text{GeV}^2$.  This
implies that in gluon--induced interactions at HERA at such $Q^2$ the
probability of diffraction should be close to 1/2 --- exactly as we
found in the analysis of HERA diffractive data in
Sec.~\ref{subsec:factorization_diffractive}.
%
%

To summarize, one may expect that the leading--twist approximation for
DIS breaks down for $Q^2\le 4 \, \text{GeV}^2$, especially in the
gluon sector. Unfortunately, there are no readily available probes of
gluons at low $Q^2$, except possibly the longitudinal cross section,
$\sigma_L$.  Without such measurements, it is impossible to determine
whether the successful DGLAP fits to the HERA data down to 
$Q^2\sim 1\, \text{GeV}^2$ are an artifact of using an essentially
arbitrary gluon density at low $Q^2$ and low $x$; this function is
practically not constrained by the data at larger $Q^2$, where it is
dominated by DGLAP evolution from larger $x$.  Also, it is worth
emphasizing that although $\Gamma^{dp} (b)$ for gluon dipoles (and, at
higher energies, also for quark dipoles) reaches values close to
unity, the actual deviation of the nucleon structure function from the
DGLAP fits may still be rather small, as the contributions from
scattering at large $b$ remain dominant.
\subsection{Theoretical issues in describing the black--disk limit}
The analysis of the full DGLAP evolution equation and the resummation
approaches shows that the cross section for the scattering of a small
dipole increases effectively as $x^{-n}$, with $n\ge 0.2$ for 
$Q^2 \sim \text{few GeV}^2$, and somewhat faster at larger $Q^2$. [At
very large $Q^2$ and extremely high energies --- well above the LHC
range --- the resummation approaches predict that 
$n\propto \alpha_s(Q^2)$, and thus $n\rightarrow 0$.]  Within these
approximations, the $t$--slope of the differential cross section for
elastic dipole--nucleon scattering, $B$, increases with energy rather
slowly, \textit{cf.}\ the discussion in
Sec.~\ref{subsec:transverse}. Thus, the cross section for
dipole--nucleon elastic scattering increases with energy faster than
the total cross section, and unitarity is violated for this hard
processes within the leading--twist approximation (see the discussion
in Sec.~\ref{subsec:violation_LT}).  Probability conservation will be
violated at fixed impact parameter, $b$, in a region corresponding to
a disk of finite size. At large enough $b$, the interaction is too
small to violate the leading--twist approximation.  Note that in soft
interactions the increase of the elastic cross section does not
necessarily lead to a contradiction with unitarity, because soft
interactions generate also $\alpha'$, and therefore give rise to an
increase of the $t$--slope as $B \propto \ln s$.

The conventional assumption is that, beyond the leading--twist
approximation, taming of the growth of cross sections occurs due to
the shadowing phenomenon (this follows \textit{e.g.} from V.~Gribov's
reggeon calculus).  Specific to this phenomenon is that bare particles
may experience only one inelastic collision, but any number of elastic
interactions, without changing trajectory.  The behavior of the
amplitudes for high--energy processes in QCD differs from that given
by the eikonal approximation in non-relativistic quantum mechanics
because of the necessity to account for the non-conservation of the
number of bare particles. Application of the
Abramovsky--Gribov--Kancheli cutting rules \cite{AGK} shows that the
requirement of positive probabilities for total cross sections, single
particle densities, \textit{etc.}, impose serious restrictions on the
dynamics in the case of cross sections increasing with energy. To
satisfy these requirements in a series of multiple rescatterings in
which consecutive terms have alternating signs, the effective number
of constituents in the dipole wave function should increase with
energy. Thus, the increase of the cross section with energy leads to
resolution of constituents in the wave functions of the colliding
particles, and therefore to an evolution of final states.  Evolution
and gluon emission by dipoles are the key for generating multiple
inelastic collisions without violation of causality and
energy--momentum conservation. The evolution of a dipole in time
manifests itself in the expansion of the system, emission of gluons,
transitions between components containing different numbers of bare
particles, change of impact parameters in the intermediate states, and
the related effect of the cross section for inelastic diffraction
exceeding the elastic cross section for the scattering of small
dipoles, \textit{cf.}\ the discussion in
Sec.~\ref{sec:diffraction}. In this regime, the concept of a parton
density of the target cannot be defined in a model--independent way,
because the parton distributions in the dipole and the target are
intertwined and not restricted by probability conservation.

At energies where the dipole cross section becomes comparable or even
larger than $2\pi R_N^2$, the whole picture of rescattering becomes
inconsistent if the radius squared of the interaction is not
proportional to the dipole cross section.  For example, in this case
the Glauber approximation for hadron--deuteron scattering would lead
to negative total cross sections. Fortunately, in QCD the wave
function of a fast projectile contains many partons. This fact,
combined with the increase of the dipole cross sections with energy,
is sufficient to ensure complete absorption for central collisions,
without detailed knowledge of the hadron wave function
\cite{Frankfurt:2004fm}. To illustrate the rapid onset of complete
absorption for central collisions related to the increase of the
number of constituents, we adopt here a simple approximation, namely
that the perturbative QCD (pQCD) description of dipole--proton
scattering works up to $\text{Re} \, \Gamma^{dp} (b,x) = 1$, and that
$\Gamma^{dp} (b,x) = 1$ if the pQCD formulas lead to $\text{Re}\,
\Gamma^{dp} (b,x) > 1$.  That is,
\begin{eqnarray} 
\text{Re} \, \Gamma^{dp} (b,x) &=&
\text{Re} \, \Gamma^{dp} (b,x)_{\text{pQCD}} \;
\Theta \left[ 1-\text{Re}\, \Gamma^{dp} (b,x)_{\text{pQCD}} \right] 
\nonumber \\
&+& \Theta \left[ \text{Re} \Gamma^{dp} (b,x)_{\text{pQCD}} - 1 \right] .
\label{model}
\end{eqnarray}

Many of the models currently discussed in the literature use the
elastic eikonal approximation to describe the taming of the increase
of the dipole--hadron cross section with energy as due to the
shadowing phenomenon, see \textit{e.g.}
Refs.~\cite{BGW,Bondarenko:2003ym} for a nucleon target and
Refs.~\cite{Balitsky-Kovchegov} for a heavy nuclear target.  These
models assume that taming becomes significant for 
$\Gamma^{dp} (b) \geq 0.5$, \textit{i.e.}\ at significantly larger $x$
than the values where unitarity is explicitly violated in the pQCD
approximation. Early taming results in a slow approach to the
unitarity limit, $\Gamma^{dp}(b) = 1$.  Obviously, these conclusions
are model dependent, as such models neglect most of the QCD effects
mentioned above.

The condition of the BDL for dipole--hadron scattering,
$\Gamma^{dp}(b) = 1$, expresses the complete ``loss of memory'' of the
cross section on the structure of the projectile and the target, and
of the value of the running coupling constant, in a finite region of
transverse space. It reflects the breakdown of two--dimensional
conformal invariance (which is the basis of approximate Bjorken
scaling in DIS) because of the appearance of a dynamical scale related
to the high gluon density and the radius of the transverse
distribution of gluons. The qualitative departure from pQCD dynamics
in the BDL cannot be explained as a soft interaction effect.  This can
be understood when considering collisions of two small dipoles of same
size near the BDL, \textit{e.g.}\
$\gamma^\ast(Q^2)$--$\gamma^\ast(Q^2)$ scattering at sufficiently
large $Q^2$, in which soft interaction effects are under control and
negligible.  An interesting question is whether, from a general
perspective, the appearance of this new scale corresponds to a
spontaneous breakdown of conformal symmetry, or is related to the
conformal anomaly.
\subsection{High--energy limit of nuclear and hadronic structure functions}
The approach to the BDL in dipole--hadron scattering at high energies
has interesting implications for the theoretical behavior of hadron
and nuclear structure functions at extremely high energies, which is
subject to the Froissart bound.

We consider the scattering of a virtual photon from a heavy nucleus
(radius $R_A$) at high energies as a superposition of the scattering
of dipoles of different sizes. The interaction at impact parameters
$b\le R_A$ will be black for dipoles with sizes larger than some
critical size, $d>d(x)$, leading to a contribution to the cross
section [\textit{cf.} Eq.~(\ref{sigma_L})]
\begin{equation}
\sigma^{\gamma^* A} \;\; \approx \;\; 2\pi R_A^2 \int d^2 d 
\int_0^1 dz \; |\psi^\gamma (d,z)|^2 \;\; 
\Theta\!\left[ d - d(x) \right] .
\end{equation}
Because the profile function of dipole--nucleus scattering increases
like a power of energy in the region where it is $< 1$, one concludes
that $d(x) \propto x^{m}$, with $m>0$. When the nuclear radius
significantly exceeds the essential impact parameters in $\gamma^{*}N$
collisions, one has
\begin{equation}
F_{2A}(x, Q^2) \;\; = \;\; \frac{Q^2}{12 \pi^3} 
\left( \sum_f e_f^2 \right) \; (2\pi R_A^2 ) \; \ln \frac{x_0 (Q^2)}{x} ,
\label{F_2A_limiting}
\end{equation}
where $x_0 (Q^2)$ slowly decreases with increasing
$Q^2$.\footnote{Since in the BLD the masses of the intermediate states
are much larger than $Q^2$, the coherence length is much smaller than
the naive estimate, $l_{\text{coh}} \sim 1/(2 m_N x)$.  If the gluon
density in the approach to the BDL grows as $x^{-\lambda(Q^2)}$, one
expects that $l_{\text{coh}} \propto 1/(m_N x^{1 - \lambda})$.}  For
illustrative purposes, we neglect here the contributions from
peripheral collisions, which grow with the atomic number as
$A^{1/3}$. Although formally these contributions increase with energy
faster than those from central collisions (which are nearly
energy--independent), they are still comparatively small at all
achievable energies.  The gross violation of Bjorken scaling in
Eq.~(\ref{F_2A_limiting}), $F_{2A} \propto Q^2 \ln (x_0 / x)$, and the
numerical coefficient follow from the normalization of the photon wave
function to the $Q^2$--derivative of the photon polarization operator,
as opposed to unity as for hadron wave functions.
Equation~(\ref{F_2A_limiting}) represents a QCD modification of the
Gribov BDL \cite{Gribov:1968gs}, which assumed that all configurations
in the virtual photon with masses $M^2 \le 2 m_N x/ R_A$ interact with
the heavy nucleus with maximum strength.

In DIS from a proton target scattering at large impact parameters is
always important in the regime where the pQCD interaction becomes
strong. Indeed, based on the studies of the transverse spatial
distribution of gluons in hard exclusive processes (see
Sec.~\ref{subsec:transverse}) one expects that 
$\Gamma^{dp} (s,b) \propto \exp(-\mu b)$ at large $b$, with
$\mu\approx m_{g}$ for moderately small $x$ [\textit{cf.}\
Eq.~(\ref{twogluon_dipole})], and $\mu\rightarrow 2 m_{\pi}$ in the
limit of infinitely large energies.  It follows from the unitarity
bound, Eq.~(\ref{unitarity}), that that the essential impact
parameters increase with energy as 
$b^2 \propto \mu^{-2} \ln^2 \left[ \sigma(s,d) / (8\pi B) \right]$,
where $B$ is the $t$--slope of the differential cross section of
dipole--nucleon scattering, which is almost energy--independent within
the leading--twist approximation. The cross section of dipole--nucleon
scattering therefore increases with energy as \cite{FGMS}
\begin{equation}
\sigma \;\; \propto \;\; \mu^{-2} \; \ln^2 \frac{\sigma(s,d)}{8\pi B} .
\end{equation}
In general, this behavior differs from the Froissart limit for soft
hadronic interactions, because of the more complicated dependence of
the dipole cross section, $\sigma(s,d)$, on the energy, as described
by the resummation approaches. To simplify the formulas, below we
shall use the observation that effectively 
$\sigma(s,d) \propto s^{n(d)}$, with $n\geq 0.2$ for small $d$.  This
approximation leads to the limiting behavior familiar from soft
hadronic interactions \cite{Froissart}.  The leading asymptotic term
in $x$ for fixed $Q^2$ for the nucleon structure function is
\begin{equation}
F_2 (x, Q^2) \;\; = \;\; 
\frac{Q^2}{12 \pi^3} \left(\sum_f e_f^2 \right) 
\sigma \ln \frac{s}{s_0}
\;\; \propto  \;\;
\ln^3\frac{s}{s_0},
\label{log3}
\end{equation}
where two logarithms originate from the dipole--nucleon cross section,
and one from the integral over the photon wave function, similar to
the case of scattering from nuclei.

In hard exclusive processes in $\gamma^\ast p$ scattering, the
approach to the BDL implies that the $t$--slope increases with energy
$\propto \ln^2(s/s_0)$ (\textit{cf.}\ also
Sec.~\ref{subsec:BDL_elastic}). A promising strategy in searching for
BDL effects would be to extract partial waves for small impact
parameters from the cross sections of processes such as DVCS from the
nucleon, $\rho$-meson production \cite{Munier:2001nr}, coherent
photoproduction of high $p_\perp$ dijets from the nucleon, and
coherent photoproduction of $J/\psi,\psi',\psi''$ mesons.  This would
allow one to probe small $b$, where the gluon density is maximal and
unitarity effects should manifest themselves early.  Another possible
strategy is to study the structure functions of heavy nuclei, where
due to higher gluon densities the BDL effects are enhanced by a factor
$A^{1/3}$\cite{Mueller:wy}. Note that this enhancement is partly
compensated by nuclear shadowing; \textit{cf.}\ the discussion in
Sec.~\ref{subsec:shadowing}.

It is interesting also to explore the behavior of nuclear structure
functions at extremely high energies, where the radius of the
$\gamma^* N$ interaction becomes comparable to or even exceeds the
nuclear radius. In this case, first the edge of the nucleus
contributes terms $\propto A^{1/3}\ln (x_0/x)^3$ to the cross
section. Ultimately, for $s\to \infty$ and fixed $Q^2$, one would
reach the universality regime where
$F_{2A}(x,Q^2)/F_{2p}(x,Q^2)\rightarrow 1$ \cite{Frankfurt:2004fm}.
However, the relevant scale is comparable to the gravitational scale.
\subsection{Black--disk limit in hard diffractive scattering from 
heavy nuclei}
\label{subsec:black_heavy}
An important consequence of the BDL is that, in diffractive scattering
at high energies, non-diagonal transitions between diffractive
eigenstates are forbidden \cite{Gribov:1968gs}. This follows from the
orthogonality of the eigenstates --- if every configuration in the
projectile interacts with the same strength, the relative proportion
between different configurations remains the same. This implies that
half of the nuclear DIS cross section should be due to diffraction,
with the nucleus remaining intact, and a ``jetty'' diffractive final
state resembling that of $e^+e^- \rightarrow \text{hadrons}$. In
contrast, in the leading--twist approximation this cross section
should be negligible.

At the onset of the BDL regime, where the contributions from
configurations in the virtual photon interacting with the BDL strength
and those for which pQCD is applicable are comparable, one can
calculate the differential cross section for diffraction to final
states of small mass, $M_X$, for which the interaction is already
black,
\begin{equation}
\frac{d F_T^{\gamma^*_T\rightarrow X}(x,Q^2, M_X^2)}{d M_X^2 d\Omega_X}
\;\; = \;\;
\frac{\pi R_A^2}{12 \pi^3} 
\frac{Q^2 M_X^2}{(M_X^2+Q^2)^2}
\frac{d\sigma(e^+e^-\to X)/ d\Omega_X}{\sigma(e^+e^-\to \mu^+\mu^-)} .
\end{equation}
This shows a much slower decrease with $Q^2$ than in the
leading--twist approximation, and corresponds to ``jetty'' final
states (mostly diffraction to $q\bar q$ and $q\bar q g$ jet states).
The earliest signal for the change of $Q^2$ dependence should be in
$\rho$ meson production, where the $Q^2$ dependence of the dominant
longitudinal cross section should change from $1/Q^6$ 
(see Sec.~\ref {subsec:vector_meson}) to $1/Q^2$.

Theoretical studies show that at HERA kinematics the fraction of the
cross section due to diffraction should be much closer to 1/2 for
nuclear targets than for the proton \cite{Frankfurt:2003gx}.  The use
of nuclear beams would greatly facilitate the exploration of the BDL
regime. Possibilities for such measurements in ultraperipheral
collisions at LHC will be discussed in Sec.~\ref{sec:outlook}.
\section{Small--$x$ dynamics in hadron--hadron collisions}
\label{sec:hadron}
\subsection{Transverse radius of hard and soft interactions}
\label{subsec:twoscale}
Our studies of $\gamma^\ast p$ scattering at HERA have taught us
several important lessons about the gluon density at small $x$, the
transverse spatial distribution of gluons, and about the interaction
of small--size color singlets with hadrons (see Sec.~\ref{sec:intro}
for a summary).  We now explore the implications for the physics of
$pp /p\bar p$ and $pA$ collisions at high energies.

In hadron--hadron collisions, hard processes arise from binary
collisions of partons in the colliding hadrons, in which a system of
large invariant mass, $M^2 \gg \Lambda_{\text{QCD}}^2$, is
produced. Examples are the production of dijets, dilepton pairs
(Drell--Yan process), and the production of heavy particles such as
Higgs bosons or SUSY particles. Such hard partonic processes are
generally accompanied by a rich spectrum of soft interactions, which
dominate the total hadronic cross section and determine the overall
characteristics of hadron production in the final state.
Understanding the interplay of hard and soft interactions is the main
challenge in describing hadron--hadron collisions with hard processes.

%
%
\begin{figure}[t]
\begin{center}
\begin{tabular}{l}
(a) \hspace{1em} \includegraphics[width=7cm]{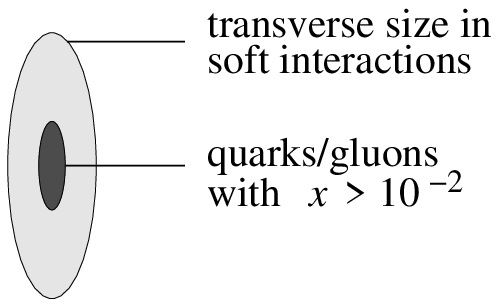} 
\\[1cm]
(b) \hspace{1em} \includegraphics[width=12cm]{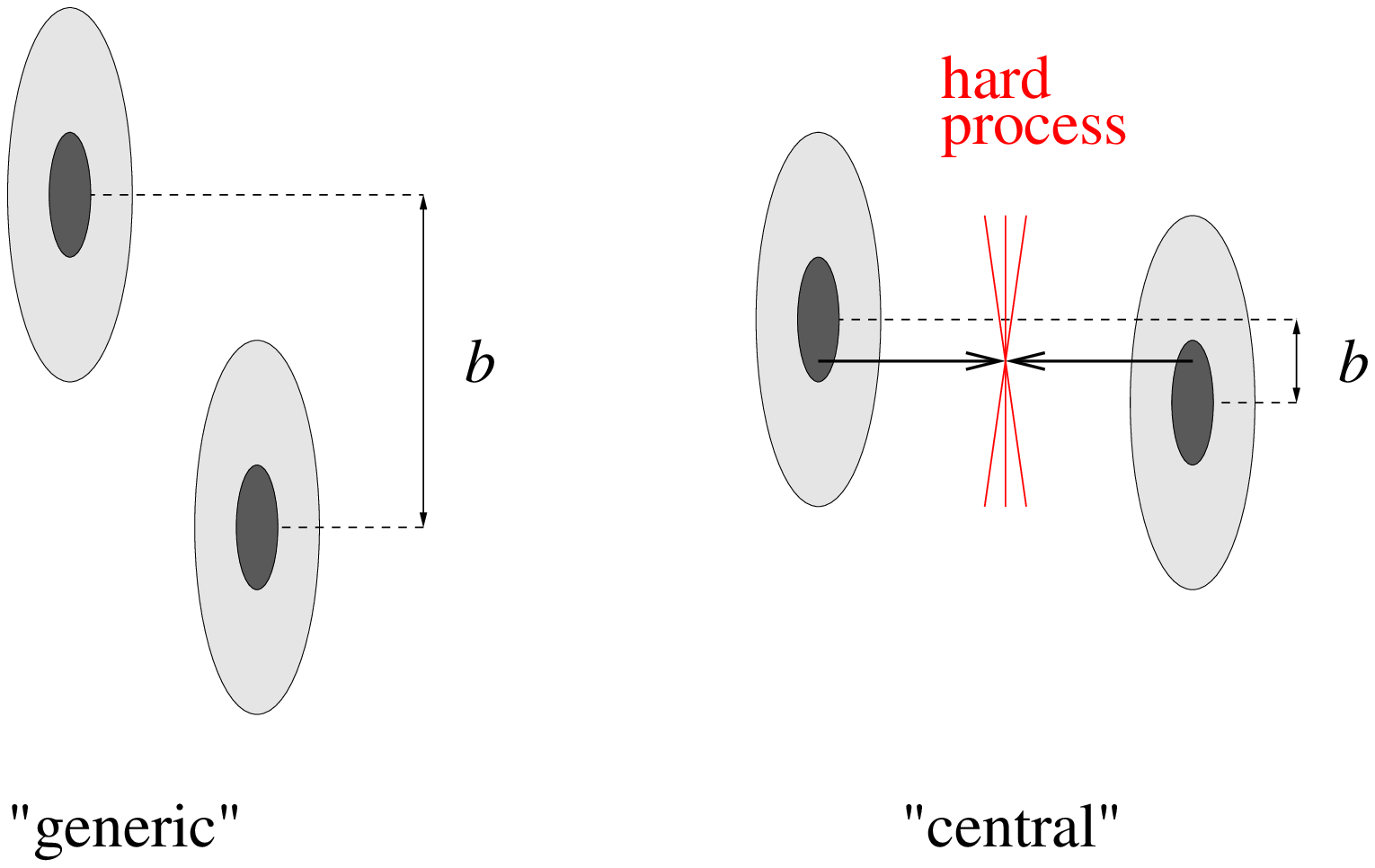}
\end{tabular}
\end{center}
\caption[]{(a) The two--scale picture of the transverse structure
of the nucleon in high--energy collisions. (b) The resulting 
classification of $pp/\bar p p$ events in ``generic'' and ``central''
collisions.
\label{fig:coll}}
\end{figure}
A crucial observation in studies of hard exclusive processes in
$\gamma^\ast p$ scattering is that gluons with significant momentum
fraction ($x > 10^{-3}$) are concentrated in a transverse area much
smaller than that associated with the nucleon in $pp$ elastic
scattering at high energies, which is dominated by soft
interactions. The difference between the two areas becomes more
pronounced with increasing energy.  When considering the production of
a system of fixed mass, $M^2$, in collision of partons with 
$x_1 x_2 = M^2 / s$, the transverse area of the hard partons grows
with energy as $\langle \rho^2 \rangle = \alpha'_{\text{hard}} \ln s$,
whereas the transverse area for soft interactions grows at a much
faster rate, $\alpha'_{\text{soft}} \gg \alpha'_{\text{hard}}$.  The
cause of this difference is the suppression of Gribov diffusion for
partons of large virtuality, as described in
Sec.~\ref{subsec:spacetime}. Thus, in high--energy $pp$ collisions one
is dealing with an ``onion--like'' transverse structure of the nucleon
(two--scale picture), see Fig.~\ref{fig:coll}a.

The two--scale picture of the transverse structure of the nucleon
implies a classification of $pp / \bar p p$ events in ``central'' and
``generic'' collisions, depending on whether the transverse areas
occupied by the large--$x$ partons in the two protons overlap or not,
see Fig.~\ref{fig:coll}b \cite{Frankfurt:2003td}.  Generic collisions
give the dominant contribution to the overall inelastic cross section.
Hard processes, such as heavy particle production at central
rapidities, will practically happen in central collisions
only. (Obviously, in these collisions multiparton interactions due to
the small--$x$ gluon fields are strongly enhanced, giving rise to the
dynamical effects described in the following subsections.)

\begin{figure}[t]
\begin{tabular}{cc}
\includegraphics[width=7.5cm]{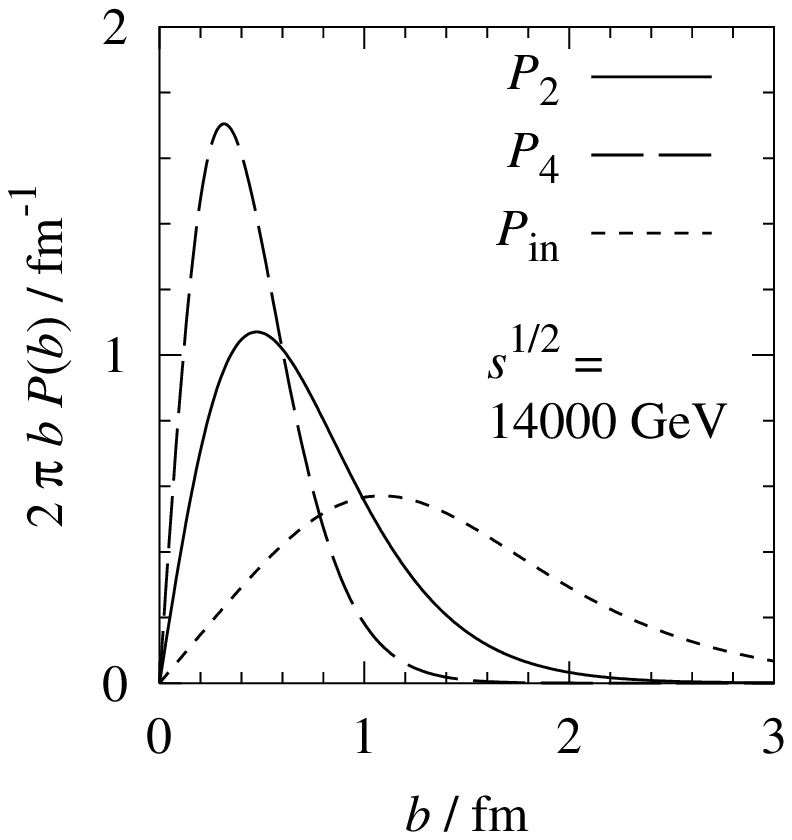}
&
\includegraphics[width=7.5cm]{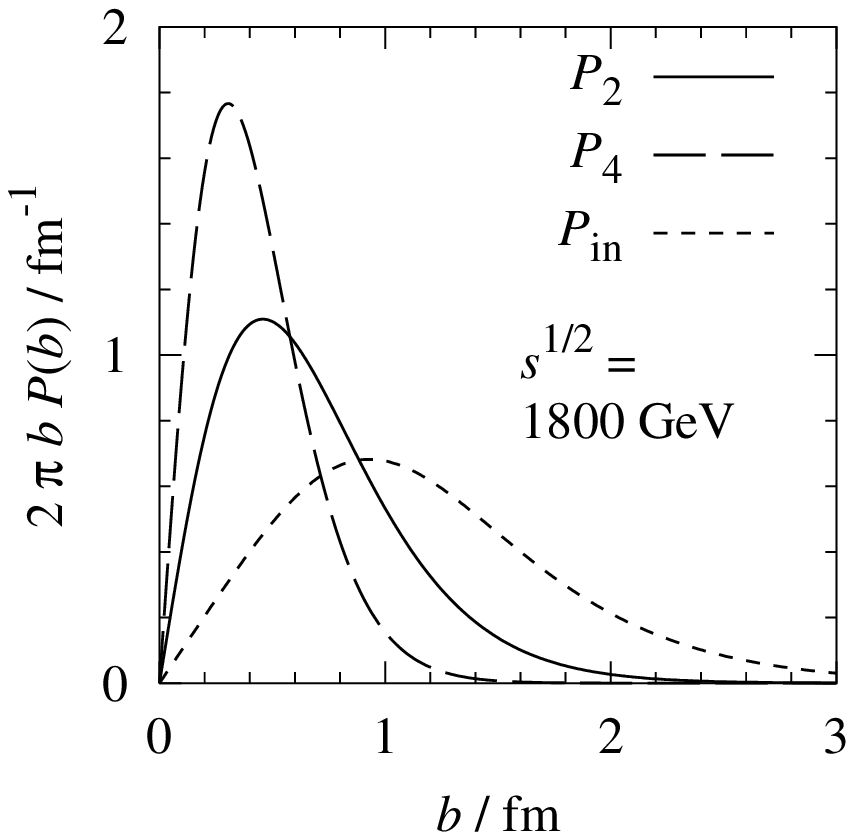}
\end{tabular}
\caption[]{\textit{Solid lines:} Impact parameter distributions in
events with hard dijet production, $P_2 (b)$, in $pp$ collisions at
LHC (left panel) and $\bar p p$ collisions at the Tevatron (right
panel), for a back--to--back dijet at zero rapidity with transverse
momentum $q_\perp = 25 \, \text{GeV}$.
\textit{Long--dashed lines:} Same for double dijet events,
$P_4 (b)$. \textit{Short--dashed lines}: Impact parameter distributions in
generic inelastic collisions, obtained from the parameterization of the
elastic $pp$ amplitude of Islam \textit{et al.} \cite{Islam:2002au}.
The plots show the ``radial'' distributions in the impact parameter
plane, $2 \pi b \, P (b)$.
\label{fig:pb}} 
\end{figure}
To quantify the distinction between generic and central collisions, we
estimate the distribution of the probability for both types of events
over the impact parameter of the $pp$ collision, $b$.  For generic
collisions, the distribution is determined by the $b$--dependent
probability of inelastic interaction, obtained via unitarity from the
elastic $pp$ amplitude in the impact parameter representation,
Eq.~(\ref{unitarity}). We define a normalized $b$--distribution as
\begin{equation}
P_{\text{in}} (s, b) \;\; = \;\; 
\left[2 \text{Re}\; \Gamma^{pp} (s, b) - |\Gamma^{pp} (s, b)|^2\right]/
\sigma_{\text{in}} (s) ,
\label{P_in_def}
\end{equation}
where $\sigma_{\text{in}}(s)$ is the inelastic cross section, which is
given by the integral $\int d^2 b$ of the expression in the numerator.
For collisions with a hard process, on the other hand, the
$b$--distribution is determined by the overlap integral of the
distribution of hard partons in the two colliding protons (see
Fig.~\ref{fig:coll}b),
\begin{equation}
P_2 (b) \;\; \equiv \;\; \int d^2\rho_1 \int d^2\rho_2 \; 
\delta^{(2)} 
(\bm{b} - \bm{\rho}_1 + \bm{\rho}_2 )
\; F_g (x_1, \rho_1 ) \; F_g (x_2, \rho_2 ) .
\label{P_2}
\end{equation}
Numerical estimates can be performed with our parametrization of the
transverse spatial distribution of hard gluons, see
Sec.~\ref{subsec:transverse}, which takes into account the change
of the distribution with $x$ and the scale of the hard process. The
two $b$--distributions are compared in Fig.~\ref{fig:pb} for Tevatron
and LHC energies. For the hard process we have taken the production of
a dijet with transverse momentum $q_\perp = 25\,\text{GeV}$
at rapidity $y = 0$ in the center--of--mass frame; in the case of Higgs
production at LHC the distribution $P_2 (b)$ would be even
narrower. The results clearly show that events with hard processes
have a much narrower impact parameter distribution than generic
inelastic events.

One expects that at LHC energies the rate of production of two pairs
of jets will be very high. It is interesting to consider the
$b$--distribution also for the production of two dijets in two binary
parton--parton collisions. Neglecting possible correlations between
the partons in the transverse plane it is given by
\begin{equation}
P_4 (b) \;\; \equiv \;\; \left[ P_2 (b)\right]^2 \left/
\int d^2 b \; \left[ P_2 (b)\right]^2 . \right.
\label{P_4}
\end{equation}
Fig.~\ref{fig:pb} shows that this distribution is significantly
narrower than $P_2$, \textit{i.e.}, the requirement of two hard
processes results in a further reduction of effective impact
parameters.  

Correlations in the transverse positions of partons can be probed by
studying $pp/\bar p p$ events with two hard processes, involving two
binary collisions of partons. At the Tevatron such a process ---
production of three jets and a photon --- was studied by the CDF
collaboration \cite{Abe:1997bp}.  The observed cross section is by a
factor of 4 larger than the naive estimate based on the assumption
that the partons are distributed uniformly in the transverse plane,
over an area whose size was inferred from the proton electromagnetic
form factor. The effect of correlations in the transverse position of
partons (\textit{i.e.}, their localization in ``spots'' of
significantly smaller size than the radius of their distribution
within the nucleon) reduces this discrepancy by a factor of 2. This
hints at the presence of significant correlations in the parton
transverse positions for $x\ge 0.05$. A possible explanation of such
correlations is the localization of the non-perturbative gluon fields
in ``constituent'' quarks (and antiquarks), as suggested by the
instanton vacuum model of chiral symmetry breaking in QCD and
supported by a large body of information on low--energy hadron
structure, see Ref.~\cite{Diakonov:2002fq} for a review. We find that
the parton correlations implied by this model indeed give rise to a
further enhancement of the cross section for two hard processes by a
factor $\sim 2$, see Refs.~\cite{Frankfurt:2003td,Frankfurt:2004kn}
for details.
\subsection{Black--disk limit in high--energy $pA$ and $pp$ collisions}
\label{subsec:BDL}
A new effect encountered in high--energy $pA$ and $pp$ collisions is
that the interaction of leading partons in the proton with the gluon
field in the nucleus (or other proton) approaches the maximum strength
allowed by $s$--channel unitarity, the BDL. This leads to certain
qualitative modifications of the hadronic final state, which will be
observable in central $pp$ and $pA$ collisions at LHC. In particular,
this effect dramatically changes the strong interaction environment
for new heavy particle production in central $pp$ collisions at LHC.

%
%
\begin{figure}[b]
\begin{center}
\includegraphics[width=6cm]{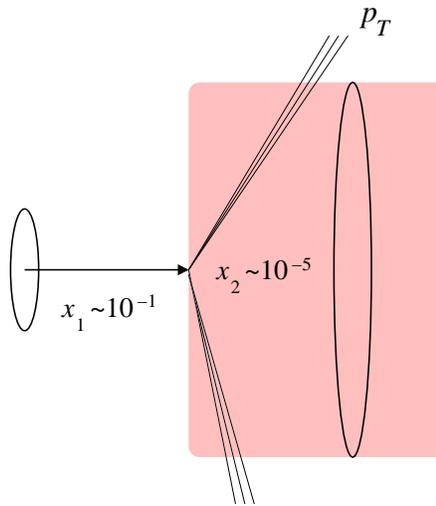}
\end{center}
\caption[]{The black--disk limit (BDL) in central $pA $ collisions: Leading
partons in the proton, $x_1 \sim 10^{-1}$, interact with a dense
medium of small--$x_2$ gluons in the nucleus (shaded area), acquiring
a large transverse momentum, $p_{\perp}$.
\label{fig:bbl}}
\end{figure}
In a high--energy $pA$ collision, consider a leading parton in the
proton, with longitudinal momentum fraction $x_1 \sim 10^{-1}$ and
typical transverse momentum of the order of the inverse hadron size;
see Fig.~\ref{fig:bbl}. Assuming a leading--twist two-body scattering
process with transverse momentum $p_\perp$ in the final state, this
leading parton can interact with partons in the nucleus of momentum
fraction
\begin{equation}
x_2 \;\; = \;\; \frac{4 \, p_\perp^2}{x_1 \, s} 
\label{x_resolved}
\end{equation}
($x_2$ and $s$ here refer to the effective $pN$ collision).  If $s$
becomes sufficiently large, $x_2$ can reach very small values even for
sizable transverse momenta, $p_\perp^2 \gg \Lambda_{\text{QCD}}^2$.
For example, at LHC $x_2 \sim 10^{-6}$ is reached for 
$p_{\perp} \approx 2\, \text{GeV}$. At such values of $x_2$, the gluon
density in the nucleus becomes large. The leading parton can be
thought of as propagating through a dense ``medium'' of gluons.  In
this situation, the probability for the leading parton to split into
several partons and scatter inelastically approaches unity,
corresponding to the scattering from a ``black'' object.  As a result
the leading parton effectively (via splittings) undergoes inelastic
collisions, losing energy and acquiring transverse momentum, until its
transverse momentum is so large that the interaction probability
becomes small, and the nucleus no longer appears ``black''. To
summarize, we can say that in $pA$ collisions the leading partons
acquire transverse momenta of the order of the maximum transverse
momentum for which their interaction with the nucleus at their
respective $x_1$ is close to the BDL, $p_{\perp,\text{BDL}}$.  This
transverse momentum represents a new hard scale in high--energy
hadron--hadron collisions, which appears because of the combined
effect of the rise of the gluon density at small $x$ and the unitarity
condition.\footnote{The kinematics of the final state produced in the
interaction of the large--$x_1$ parton with the small--$x_2$ gluon
field resembles the backscattering of a laser beam off a high--energy
electron beam. The large--$x_1$ parton gets a significant transverse
momentum and loses a certain fraction of its longitudinal momentum,
accelerating at the same time a small--$x_2$ parton.}

%
%
\begin{figure}[t]
\begin{tabular}{ll}
\includegraphics[width=7cm]{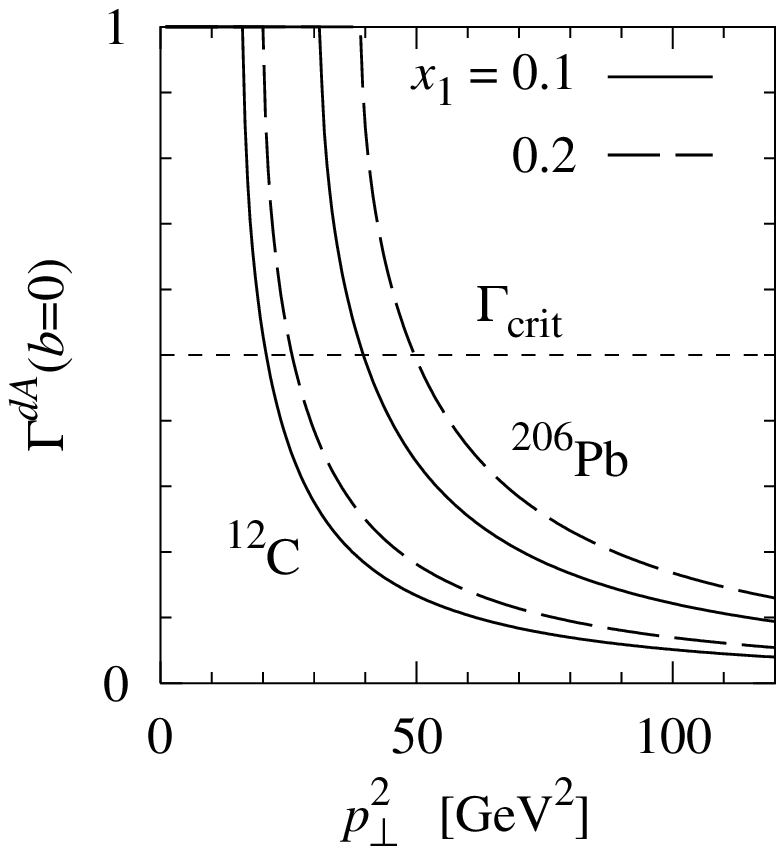}
&
\includegraphics[width=7cm]{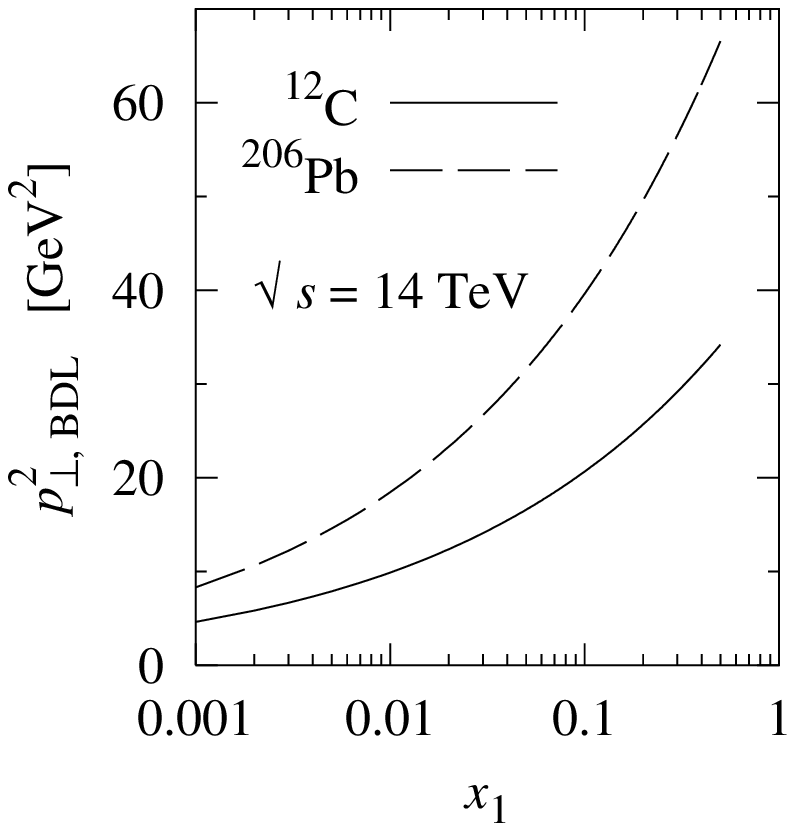}
\\
(a) & (b)
\end{tabular}
\caption[]{Black--disk limit in central $pA$ collisions at LHC: (a)~The
profile function for the scattering of a leading gluon in the proton
(regarded as a constituent of a $gg$ dipole) from the nucleus at zero
impact parameter, $\Gamma^{dA}(b = 0)$, as a function of the
transverse momentum squared, $p_\perp^2$.  (b)~The maximum transverse
momentum squared, $p_{\perp , \text{BDL}}^2$, for which the
interaction of the leading gluon is ``black'', 
$\Gamma^{dA} > \Gamma_{\text{crit}}$, as a function of the gluon's
momentum fraction, $x_1$. Here we assume $\sqrt{s} = 14 \, \text{TeV}$ 
for the effective $NN$ collisions, in order to facilitate comparison 
with the case of central $pp$ collisions in Fig.~\ref{fig:pt}.
\label{fig:pt_pA}}
\end{figure}
To estimate the maximum transverse momentum for interactions close to
the BDL, we can treat the leading parton as one of the constituents of
a small dipole scattering from the target.  This ``trick'' allows us
to apply the results of Sec.~\ref{sec:BDL} to hadron--hadron
scattering.  In this analogy, the effective scale in the gluon
distribution is $Q^2_{\text{eff}} = 4 p_\perp^2$, corresponding to an
effective dipole size of $d \approx 3 /(2 p_\perp )$. For simplicity,
we first consider the case of central collisions of a proton with a
large nucleus, which allows us to neglect the spatial variation of the
gluon density in the target in the transverse direction.  This amounts
to approximating the transverse spatial distribution of gluons in the
nucleus by
\begin{equation}
G_A (x, \rho ; Q^2_{\text{eff}}) \;\; \approx \;\; 
\frac{G_A (x; Q^2_{\text{eff}})}{\pi R_A^2}
\hspace{4em} (\rho < R_A) .
\end{equation}
As a criterion for the proximity to the BDL, we require that the
profile function of the dipole--nucleus amplitude at zero impact
parameter satisfy $\Gamma^{dA}(b = 0) > \Gamma_{\text{crit}}$, see
Fig.~\ref{fig:pt_pA}a.  For an estimate, we choose
$\Gamma_{\text{crit}} = 0.5$, corresponding to a probability for
inelastic interaction of 0.75, reasonably close to unity. We then
determine the maximum $p_\perp$ for which the criterion is
satisfied. Fig.~\ref{fig:pt_pA} shows the result for
$p_{\perp,\text{BDL}}^2$ for a leading gluon, as a function of the
gluon momentum fraction, $x_1$; for leading quarks, the result for
$p_{\perp,\text{BDL}}^2$ is approximately 0.5 times the value for
gluons. The numerical estimates show that leading partons indeed
receive substantial transverse momenta when traversing the
small--$x_2$ gluon medium of the nucleus. We emphasize that our
estimate of $p_{\perp,\text{BDL}}$ applies equally well to the
interaction of leading partons in the central region of $AA$
collisions.

\begin{figure}[t]
\begin{tabular}{cc}
\includegraphics[width=7cm]{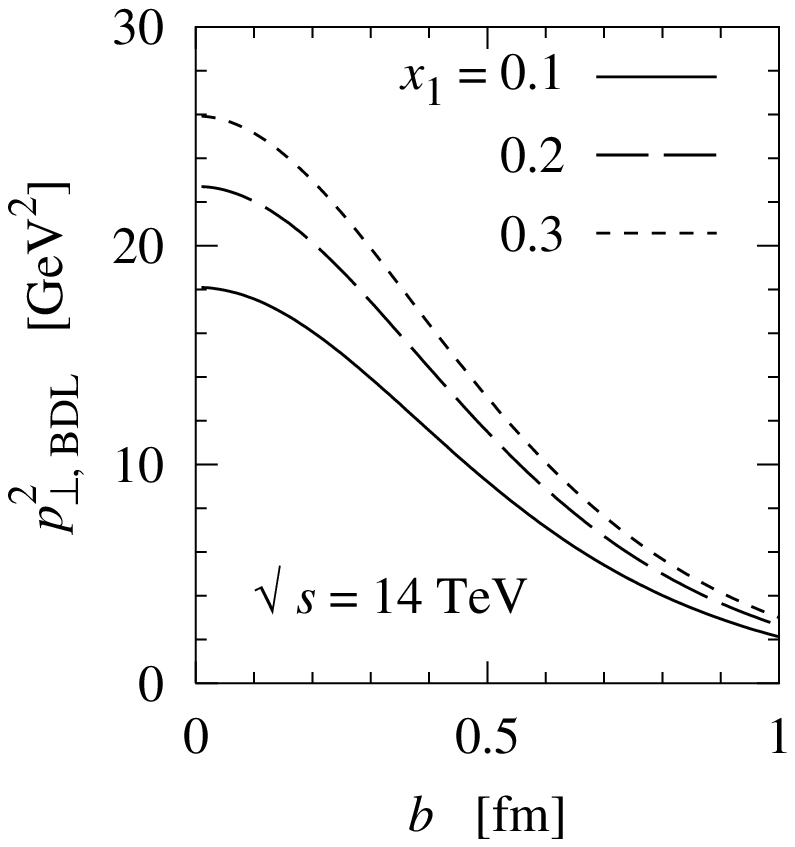}
&
\includegraphics[width=7cm]{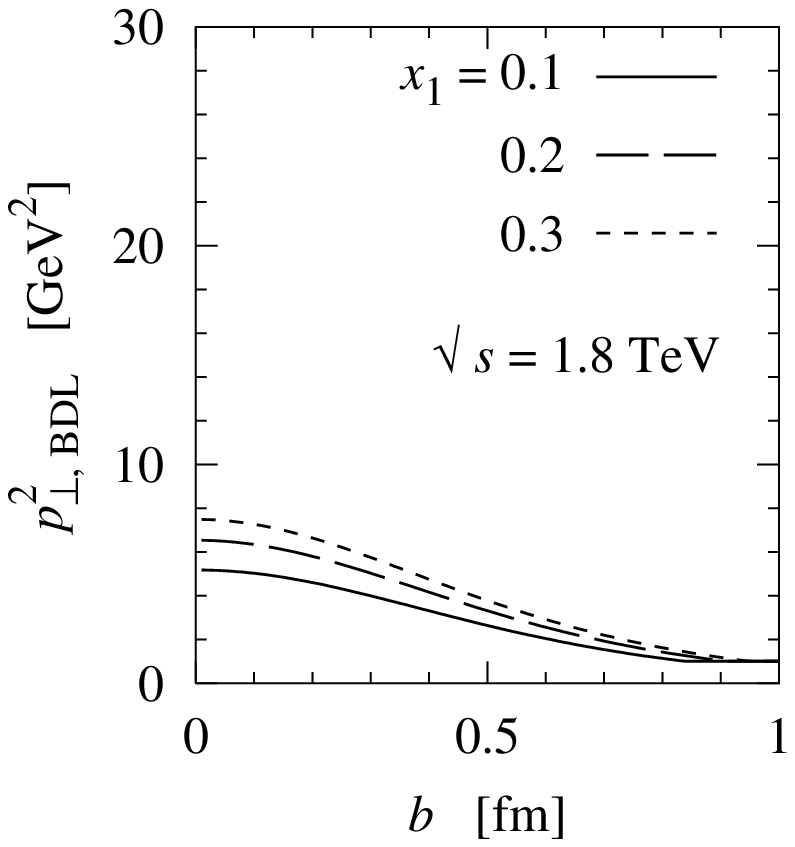}
\\
\includegraphics[width=7cm]{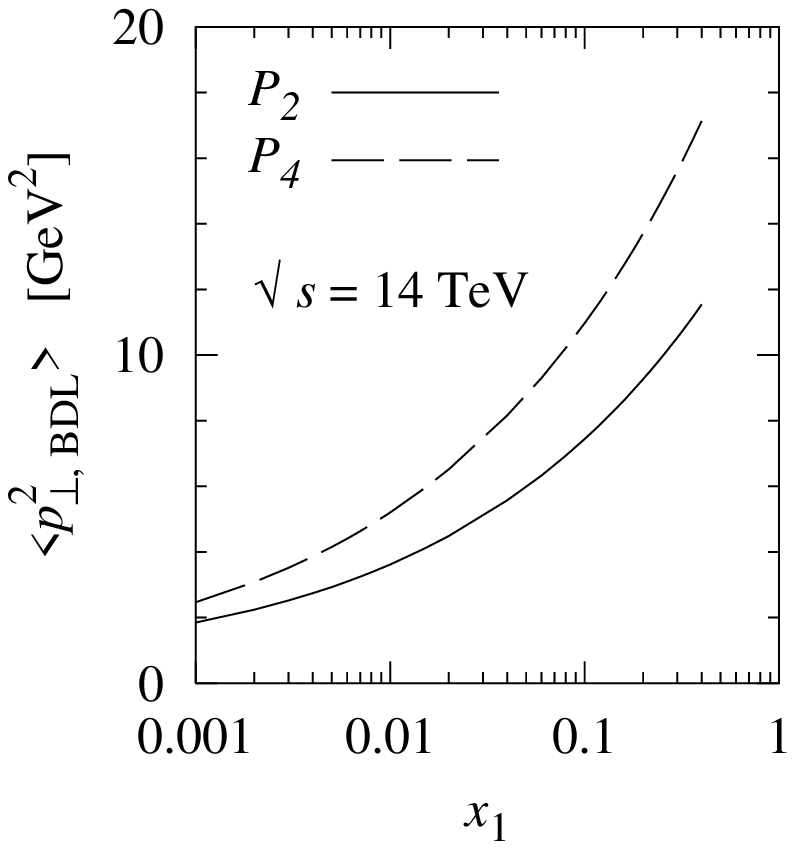}
&
\includegraphics[width=7cm]{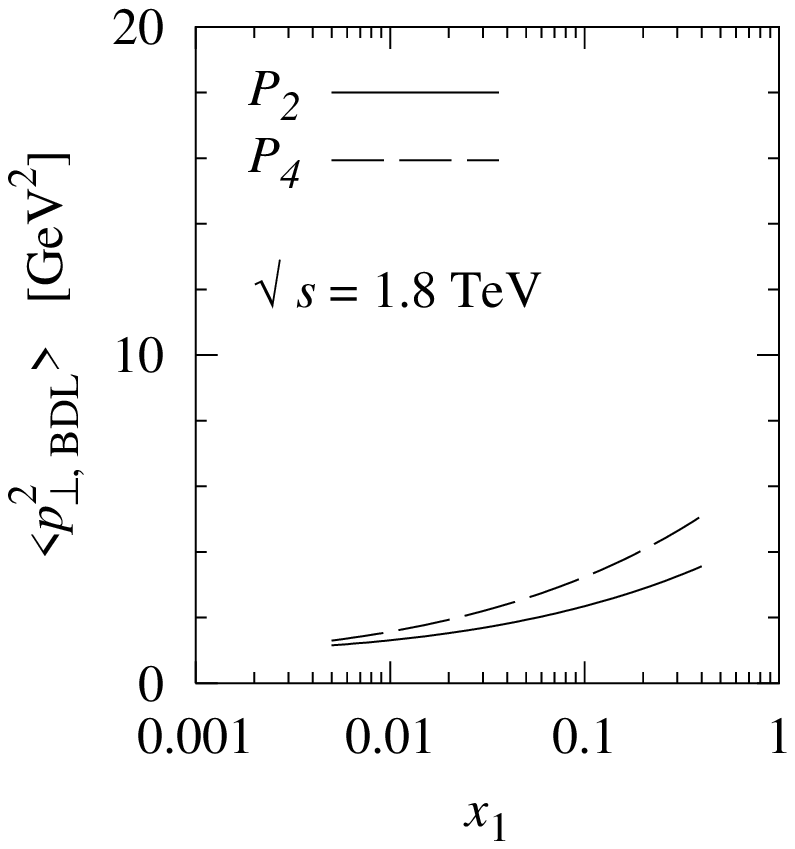}
\end{tabular}
\caption[]{\textit{Upper row:} The transverse momentum squared,
$p_{\perp , \text{BDL}}^2$, acquired by a leading gluon (momentum
fraction $x_1$) through interactions with the small--$x_2$ gluon field
in the other proton near the BDL, as a function of the impact
parameter of the $pp$ collision, $b$.  Shown are the estimates for LHC
(left panel) and Tevatron energies (right panel). \textit{Lower row:}
Average values of $p_{\perp , \text{BDL}}^2$ in $pp$ collisions with a
single hard process (impact parameter distribution $P_2$) and two hard
processes (distribution $P_4$), \textit{cf.}\ Fig.~\ref{fig:pb}. For
leading quarks, the values of $\langle p_{\perp , \text{BDL}}^2
\rangle$ are about half of those for gluons shown here.
\label{fig:ptav}}
\label{fig:pt}
\end{figure}
Turning now to $pp$ collisions, we have to take into account the
transverse spatial structure of the colliding hadrons.  A crucial
point is that high--energy interactions do not significantly change
the transverse position of the leading partons, so that their
interaction with the small--$x_2$ gluons is primarily determined by
the gluon density at this transverse position. Because the leading
partons in the ``projectile'' proton are concentrated in a small
transverse area, and the small--$x_2$ gluon density in the ``target''
proton decreases with transverse distance from the center, it is clear
that the maximum transverse momentum for interactions close to the
BDL, $p_{\perp , \text{BDL}}^2$, decreases with the impact parameter
of the $pp$ collision, $b$.  Fig.~\ref{fig:pt} (upper row) shows the
dependence of $p_{\perp , \text{BDL}}^2$ on $b$, as obtained with the
parametrization of the transverse spatial distribution of gluons based
on analysis of the HERA exclusive data, see Section
\ref{subsec:transverse} \cite{Frankfurt:2003td}.  One sees that
$p_{\perp,\text{BDL}} \sim \text{several GeV}$ in central collisions
at LHC. Substantially smaller values are obtained at the Tevatron
energy.

To determine the typical transverse momenta of leading partons in
events with new particle (or hard dijet) production, we need to
average the results for $p_{\perp,\text{BDL}}^2$ over $pp$ impact
parameters, with the distribution implied by the hard production
process, $P_2(b)$, Eq.~(\ref{P_2}), or, in the case of four jet
production, with $P_4(b)$, Eq.~(\ref{P_4}). The resulting average
values of $p_{\perp,\text{BDL}}^2$ are shown in Fig.~\ref{fig:pt}
(lower row). We find that the suppression of large impact parameters
implied by the hard process, described in Sec.~\ref{subsec:twoscale},
is sufficient to keep $p_{\perp,\text{BDL}}$ above $1\, \text{GeV}$ in
more than 99\% of events at LHC.

To summarize, our estimates show that in generic central $pA$ and
central (triggered) $pp$ collisions at LHC the leading partons acquire
substantial transverse momenta due to interactions near the BDL. A
much weaker effect is found at the Tevatron energy. The origin of this
difference is the increase in the gluon density due to the decrease of
$x_2$ between Tevatron and LHC energies, \textit{cf.}\
Eq.~(\ref{x_resolved}).
\subsection{Final state properties in central $pp$ collisions at LHC}
\label{subsec:final}
The approach to the BDL in the interaction of leading partons implies
certain qualitative chan\-ges in the hadronic final state in central
$pp$ and $pA$ collisions. In particular, these effects will profoundly
influence the strong interaction environment for the production of new
heavy particles (Higgs boson, \textit{etc.}) at LHC.

The main effect of the BDL is that the leading partons in the
projectile acquire substantial transverse momenta, of the order
$p_{\perp,\text{BDL}}^2$, when propagating through the dense medium of
small--$x_2$ gluons in the target.  As a result, the projectile
becomes ``shattered'': The leading partons lose coherence and fragment
independently over a wide range of rapidities close to the maximal
rapidity, corresponding to hadron momentum fractions 
$z \sim x_1 \mu / p_{\perp,\text{BDL}}$ ($\mu$ is a typical hadronic
mass scale).  The differential multiplicity of leading hadrons,
integrated over $p_{\perp}$, is approximately given by the convolution
of the nucleon parton density, $f_a$, with the corresponding parton
fragmentation function, $D_{h/a}$, at the scale 
$Q^2_{\text{eff}} = 4 p_{\perp,\text{BDL}}^2$,
\begin{equation}
\frac{1}{N} \left( \frac{d N}{dz} \right)^{pp \rightarrow h + X} 
\;\; = \;\;
\sum_{a = q,g} \int dx_1 \, x_1 \; f_a (x_1, Q^2_{\text{eff}}) \;
D_{h/a}(z/x_1, Q^2_{\text{eff}}) ,
\label{eq:forward}
\end{equation}
where $N$ is total number of inelastic events
\cite{Berera,FGMS,Dumitru:2002wd}.  This corresponds to a very strong
suppression of forward hadron production as compared to generic
inelastic $pp$ events.  The suppression is particularly pronounced for
nucleons; one expects that for $z\ge 0.1$ the differential
multiplicity of pions should exceed that of nucleons. At the same time
the transverse spectrum of the leading hadrons will be much broader,
extending up to $p_{\perp,\text{BDL}} \gg 1\, \text{GeV}$.  Finally,
the independent fragmentation mechanism implies that there will be no
correlations between the transverse momenta of leading hadrons (some
correlations will remain, however, because two partons produced in
collisions of large--$x_1$ and small--$x_2$ partons may end up at
similar rapidities).

In central $pp$ collisions at LHC, where leading particles are
suppressed in both the forward and backward direction, one expects a
large fraction of events with no particles with $z\ge 0.02 - 0.05$
in both fragmentation regions. This amounts to the appearance of
long--range rapidity correlations. Such events should show a large
energy release at rapidities $y = 4-6$. However, similar to the case 
of diffractive processes in $ep$ scattering (\textit{cf.}\ the discussion 
in Section \ref{sec:harddiff}), one should expect that there is a 
$\sim 10 \%$ probability for dijets to be produced in $pp$ collisions 
at large impact parameters without additional interactions between 
the constituents of the nucleons.

Another important effect of the BDL is a significantly increased
energy loss of the leading partons, due to the larger probability of
inelastic collisions, and the wider distribution of the propagating
parton (dipole) over transverse momenta.  In particular, a 2\% energy
loss would explain the suppression of forward pion production in
deuteron--nucleus collisions at RHIC at $p_{\perp} \sim 4 \, \text{GeV}$
\cite{Guzey:2004zp}. Studies of this effect would be possible both at
RHIC and LHC, in particular if the forward capabilities of the current
detectors were upgraded, as discussed in several proposals presently
under consideration.  Note that energy loss is neglected in
Eq.~(\ref{eq:forward}).  This corresponds to the usual assumption of
models in which parton propagation is treated as multiple elastic
rescattering of the parton's accompanying gluon field from the medium.
A consistent treatment of energy loss and transverse momentum
broadening near the BDL remains a challenge for theory.
We note that the pattern of energy loss in our approach is 
qualitatively different from models in which the leading partons 
scatter from a classical gluon field 
(in that case energy loss is negligible) \cite{Venugopalan:2004dj}.

The approach to the BDL has consequences also for hadron production in
the central rapidity region. The multiple scattering of large--$x_1$
projectile partons from the small--$x_2$ gluons in the target shifts a
large number of these gluons to larger rapidities, leaving numerous
``holes'' in the target wave function.  Furthermore, multiple
interactions of partons with moderately small $x_1 \sim x_2$ also
occur with large probability.  (Unitarity effects should be important
for these interactions as well, but have not been studied so far.)
Both effects lead to the creation of a substantial amount of color
charge, which should result in an increase of soft particle
multiplicities over a broad range of rapidities as compared to the
situation far from the BDL.  This increase should in fact be
observable already at Tevatron energies, in central events selected by
a trigger on two--jet or $Z^0$ production.  An increase of the
multiplicity at rapidities $|y|\le 1.0$ in such events compared to
minimum bias events was indeed reported in
Ref.~\cite{Field:2002vt}. It would be extremely interesting to extend
such studies to higher rapidities.

Our findings imply that new heavy particles at LHC will be produced in
a much more ``violent'' strong--interaction environment than one would
expect from the extrapolation of the properties of minimum bias events
at the Tevatron. Even the extrapolation of properties of hard dijet
events should not be smooth, as the transverse momenta acquired by
leading partons are estimated to be substantially larger at LHC than
at Tevatron, see Fig.~\ref{fig:pt}.
\subsection{Black--disk limit in elastic $pp$ scattering}
\label{subsec:BDL_elastic}
The assumption of the BDL in the interaction of leading partons,
combined with the complex structure of the proton wave function in
QCD, allows us to estimate the profile function of the $pp$ elastic
amplitude at small impact parameters. This simple estimate nicely
explains the observed ``blackness'' of the phenomenological $pp$
profile function at $b = 0$ at the Tevatron energy, and allows us to
extrapolate the profile function at small $b$ to higher energies. It
also raises the question whether the observed blackness could be
explained on the basis of soft interactions.

In $pp$ collisions at small impact parameters, leading quarks on
average receive substantial transverse momenta when passing through
the small--$x$ gluon field of the other proton; see Fig.~\ref{fig:pt}.
When a single leading quark gets a transverse momentum, $p_{\perp}$,
the probability for the nucleon to remain intact is approximately
given by the square of the nucleon form factor, $F_N^2(p_{\perp}^2)$,
which is $\le 0.1 $ for $p_{\perp}> 1 \, \text{GeV}$, \textit{i.e.},
very small. One may thus conclude that the probability of the survival
averaged over $p_{\perp}$ should be less than 1/2 (on average, half of
the time the quark should receive a transverse momentum larger than
the average transverse momentum, 
$p_{\perp}> 1 \, \text{GeV}$). Because there are six leading quarks
(plus a number of leading gluons), the probability for the two protons
to stay intact, $|1 - \Gamma^{pp}(s, b)|^2$, \textit{cf.}\
Eq.~(\ref{unitarity}), should go as a high power of the survival
probability for the case of single parton removal, and thus be very
small. This crude estimate shows that already at Tevatron energies 
$|1 - \Gamma^{pp}(s, b =0)|^2$ should be close to zero owing to hard
interactions. The conclusion that the small impact parameter
hadron-hadron interactions should become black at high energies
follows principally from the composite structure of the hadrons and
does not depend on any details. In particular, if taming effects
stopped the growth of the dipole--nucleon interaction at a fraction of
the BDL, our result would not change. Our conclusion that
$\Gamma^{pp}(\sqrt{s}\ge 2\, \text{TeV}, b = 0) \approx 1$ agrees well
with the current analysis of the Tevatron data, see \textit{e.g.}\
Ref.~\cite{Islam:2002au}.

\begin{figure}[t]
\begin{center}
\includegraphics[width=10cm]{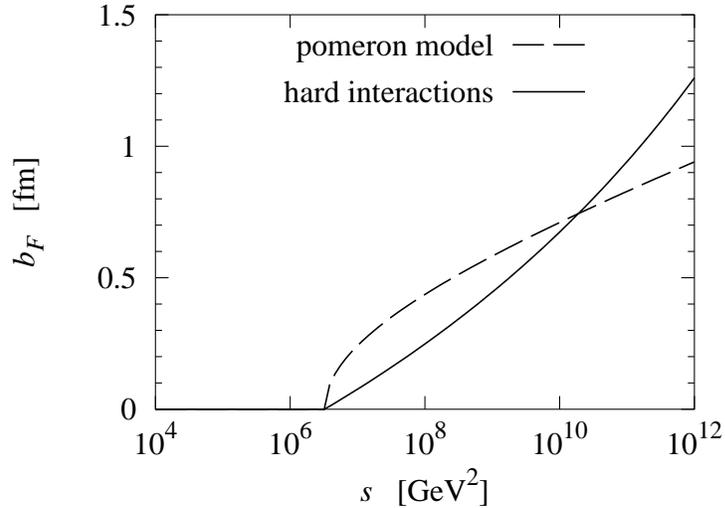}
\end{center}
\caption[]{Energy dependence of the maximum impact parameter for
``black'' $pp$ interactions, $b_F$. Solid line: Estimate based on the
pomeron model fit of $pp$ elastic scattering of \cite{Donnachie:1992ny}.
Dashed line: Estimate based on hard interactions, Eq.~(\ref{pwave4}).
\label{fig:bf}}
\end{figure}

One can estimate the maximum impact parameter, $b_F$, up to which hard
interactions cause the $pp$ interaction to be ``black''.  The
probability for a leading parton with $x_1 \sim 10^{-1}$ to experience
a hard inelastic interaction increases with the collision energy at
least as fast as dictated by the increase of the gluon density in
the other proton at $x_2 = 4 p_\perp^2 / (s x_1)$, \textit{cf.}\
Eq.~(\ref{x_resolved}) below. Because
$x_2 G(x_2,Q^2) \propto x_2^{-n_h}$ with $n_h\ge 0.2$,
the probability should grow as $s^{n_h}$.\footnote{The HERA data
on dipole--nucleon scattering suggest that the taming of the gluon
density starts only when the probability of inelastic interactions
becomes large, $\ge 1/2$. However, for such probabilities of single
parton interactions, multiparton interactions ensure that the overall
interaction is practically black.}  The dipole parametrization of the
transverse spatial distribution of gluons,
Eq.~(\ref{twogluon_dipole}), suggests that the gluon density decreases
with the distance from the center of the nucleon approximately as
$\sim \exp \left[ -m_g(x_2 )\rho \right]$. If one neglects the
transverse spread of the large--$x_1$ partons as compared to that of
the small--$x_2$ gluons one arrives at an estimate of the energy
dependence of $b_F$ as due to hard interactions
\cite{Frankfurt:2004fm},
\begin{equation}
b_F \;\; \approx \;\; \frac{n_h \ln (s/s_T)}{m_g (x_2 )} ,
\label{pwave4}
\end{equation}
where $\sqrt{s_T} = 2\, \text{TeV}$ is the Tevatron energy.  In
principle, $n_h$ may decrease at very large virtualities, which would
become important at extremely high energies. However, this effect is
likely to be compensated by the increased number of constituents
in the nucleon wave function affected by the BDL.  The above estimate is
consistent with the popular Pomeron model parameterization of the
$pp$ elastic amplitude \cite{Donnachie:1992ny}.  In this model
$\Gamma^{pp}(0) \approx 1 $ for $s = s_{T}$ and s--channel unitarity
is violated at $b < b_F$ for $s> s_T$.  The dependence of $b_F$ on the
energy in the Pomeron model is similar to our estimate
(\ref{pwave4}), see Fig.~\ref{fig:bf}. This shows that the two--scale
picture of the transverse structure of the nucleon --- and the ensuing
picture of hard and soft interactions --- are self--consistent.

Overall, the analysis of Ref.~\cite{Frankfurt:2004fm} summarized in
this subsection suggests that at high energies hadron--hadron
interactions should be ``black'' in a range of impact parameters
growing approximately like $\ln s$, with a coefficient growing only
very slowly with energy due to decrease of $m_g(x_2)$ with
energy. This corresponds to the Froissart regime, with interactions in
the black region dominated by semi--hard interactions, and
interactions at large impact parameters dominated by the single
Pomeron exchange. At the same time, this analysis indicates that the
concept of summing multi--Pomeron exchanges, which should give the
dominant contribution at small $b$, breaks down as the soft physics is
gradually squeezed out to large $b$.

It is worth emphasizing here that, in principle, the BDL can emerge in
hadron-hadron interactions already at the level of the soft
interactions \cite{Migdal:1973gz,Amati:1976ck}.  However, it is hardly
possible to reconcile it with the pre--QCD Feynman parton model
description of high energy processes, if one would require it to be
valid both in the rest frame of the target and in the center--of--mass
frame.  Really, within the parton model one cannot generate complete
absorption of the projectile in central collisions in the target rest
frame, where the target consists of few partons \cite{Kancheli}.  This
puzzle is resolved in QCD, where radiation leads to the ``blackening''
of the hard interactions at central impact parameters.  At extremely
high energies, as a result of this effect, all memory of the colliding
hadrons is lost. Hence the universal behavior of total cross sections,
$\sigma_{\text{tot}} \propto \ln^2 (s/s_0)$, with universal
coefficient for all hadrons (nuclei) \cite{Frankfurt:2004fm}.
\footnote{The increase of the interaction with energy, and the related
increase of essential impact parameters, show that the theoretical
description of high--energy hadronic collisions should be closer to
classical mechanics than to quantum mechanics (V.~Gribov, private
communication to Yu.~Dokshitzer). An example is the cross section for
the scattering of a high--energy particle from a potential rapidly
decreasing with impact parameter. The essential impact parameters ---
and therefore the cross section --- are infinite within classical
mechanics, but finite in quantum mechanics, while they increase with
energy in QCD.}
\subsection{Ion--induced quark--gluon implosion}
The small--$x$ phenomena outlined above --- the approach to the BDL,
and large leading--twist gluon shadowing --- play an important role
also in the heavy ion collisions at LHC energies. Here we consider
just one example, the so-called ion--induced quark--gluon implosion in
the nucleus fragmentation region.  For a review of other effects in
the framework of the color glass condensate model, see
Ref.~\cite{Venugopalan:2004dj}.

In generic central $AA$ collisions at collider energies, in analogy to
the central $pA$ collisions discussed in Section \ref{subsec:BDL}, all
the leading partons of the individual nucleons are stripped off
``soft'' partons and form a collection of quarks and
gluons with large $p_\perp$.  In the rest frame of the fragmenting
nucleus, the incoming nucleus has a ``pancake'' shape with
longitudinal length $\sim 1\, \text{fm}$ for soft partons, and 
$R_N (m_N/p_N)(x_V/x) \ll R_N$ for hard partons, where $x_V \sim 0.2$
is the average $x$ value for the valence quarks. That is, the nucleons
in the nucleus at rest at different locations along the collision axis
are hit by the hard partons in the incoming nucleus one after another.
In the BDL, no spectators are left. The hit partons are produced with
practically the same $x$ that they had in the nucleus (because the
fractional energy loss is small), transverse momenta 
$\sim p_{\perp,\text{BDL}}$, and virtualities 
$\leq p_{\perp,\text{BDL}}^2$. The partons move in the direction of
the projectile nucleus. Because they are emitted at finite angles,
their longitudinal velocity is smaller than the speed of light, and
they are left behind the projectile wave. However, because the
emission angles are small, a shock wave is formed, compressing the
produced system in the nucleus rest frame. In the frame co-moving with
the shock wave, valence quarks and gluons are streaming in the
opposite directions.  The resulting pattern of fragmentation of the
colliding nuclei leads to an ``implosion'' of the quark and gluon
constituents of the nuclei. The non-equilibrium state produced at the
initial stage in the nucleus fragmentation region is estimated to have
densities $\propto p_{\perp,\text{BDL}}^2$, which is 
$\geq 50\, \text{GeV}/\text{fm}^3$ at LHC, and probably 
$\geq 10\, \text{GeV}/\text{fm}^3$ at RHIC. It seems likely that the
partons would rescatter strongly at the second stage, although much
more detailed modeling is required to find out whether the system
would reach thermal equilibrium. Such large--angle rescattering of
partons would lead to production of partons at higher rapidities, and
re-population of the cool region. In particular, two gluons from the
pancake could have the right energies to produce near--threshold
$c\bar c$ pairs and $\chi_c$-mesons with small transverse momenta and
$x_F(c\bar c) \sim 2 x_g \sim 0.1$.
\subsection{Cosmic ray physics near the GZK cutoff}
An extensive program of studies of cosmic rays at energies close to
the Greisen--Zatsepin--Kuzmin (GZK) cutoff \cite{gzk}, $E_{\text{GZK}}
\simeq 6\times 10^{10} \, \text{GeV}$, is under way, using several
cosmic ray detectors. These experiments detect cosmic rays indirectly,
via the air showers induced when they enter the atmosphere. The
properties of the primary particle need to be inferred from those of
the observed shower. For this, a good understanding of the physics of
high--energy interactions in the atmosphere is mandatory.  The
observed characteristics of the shower are predominantly sensitive to
leading hadron production ($x_F \ge 10^{-2}$), which, according to our
discussion above, at these energies probes small--$x$ dynamics down to
$x \sim 10^{-10}$, deep inside the regime affected by the approach to
the BDL. First studies of these effects were performed in
Ref.~\cite{Drescher:2004sd}. It was found that the steeper
$x_F$--distribution of leading hadrons as compared to low--energy
collisions, caused by the strong increase of the gluon densities at
small $x$ (see Section \ref{subsec:BDL_elastic}), leads to a reduction
of the position of the shower maximum, $X_{\text{max}}$.  Account of
this effect in the models currently used for the interpretation of the
data may shift fits of the composition of the cosmic ray spectrum near
the GZK cutoff towards lighter elements.  Also, it appears that the
present data on $X_{\text{max}}(E)$ exclude the possibility that the
prediction of a rapid growth of the critical $x$--value where the BDL
becomes effective ($\sim 1/x^{0.3}$), which is compatible with RHIC
and HERA data, would persist up to the GZK cutoff energy.
\section{Hard diffraction at hadron colliders}
\label{sec:harddiff}
\subsection{Diffractive proton dissociation into three jets}
\label{3jetsec}
LHC will offer an opportunity to study a variety of hard diffractive
processes in $pp$ and $pA$ scattering. One interesting aspect of such
processes is that they allow to probe rare small--size configurations
in the nucleon wave function.A proton in such a configuration can
scatter elastically off the target and fragment into three jets,
corresponding to the process
\begin{equation}
p + p (A) \;\; \rightarrow \;\; 
\text{jet1} + \text{jet2} + \text{jet3} + p(A).
\label{pp3jets}
\end{equation}

The cross section for the diffractive process (\ref{pp3jets}) can be
evaluated based on the kind of QCD factorization theorem derived in
Ref.~\cite{FMS2002}. The cross section is proportional to the square of the 
gluon density in the nucleon at $x\approx M^2 (\text{3 jets})/ s$, and
virtuality $Q^2 \sim (1 - 2) \, p_\perp^2$ \cite{Frankfurt:1998eu}.
The distribution over the fractions of the proton longitudinal
momentum carried by the jets is proportional to the square of the
light--cone wave function of the $|qqq \rangle$ configuration. The
numerical estimates suggest \cite{FELIX} that the process could be
observed at the LHC energies provided one would be able to measure
jets with $p_{\perp}\sim 10\, \text{GeV}$ at very
high rapidities, $y_{\text{jet}}(p_\perp = 10\, \text{GeV}) \sim 6$,
and with a large background from leading--twist hard diffraction.  The
latter will be suppressed in $pA$ collisions, because the coherent
3--jet process has a much stronger $A$--dependence than the background
due to soft and hard diffraction induced by strong interactions. 
The main background will be due to hard electromagnetic 
interactions of the proton with the Coulomb field of the nucleus.
We note that the discussed mechanism of hard diffraction requires 
that the interaction of the spatially small three--quark color 
singlet configuration with the proton be far from the BDL
at small impact parameters. Otherwise production at small impact 
parameters would be suppressed, leading to a dip in the $t$--dependence 
of the differential cross section for the production of three jets 
with moderate $p_{\perp}$.

The detection of the three--jet final state produced by diffractive
scattering of a $qqq$ configuration from a proton should be easier
than that resulting from $e^+e^-$ annihilation into $q\bar q g$, as in
the former case all color charges are in the triplet representation,
leading to less radiation between the jets. Finally, it would also be
possible to study the process $pp \rightarrow pn + \text{two jets}$,
which is similar to pion dissociation into two jets. Experimentally,
this would require the measurement of jets at rapidities $y\sim 4$,
together with the detection of a leading neutron by a zero--degree
calorimeter, as is present in several of the LHC detectors (ALICE,
ATLAS, CMS).
\subsection{Exclusive diffractive Higgs production}
\label{sec:higgs}
\begin{figure}[t]
\begin{tabular}{l}
(a) \hspace{1em} \includegraphics[width=10cm]{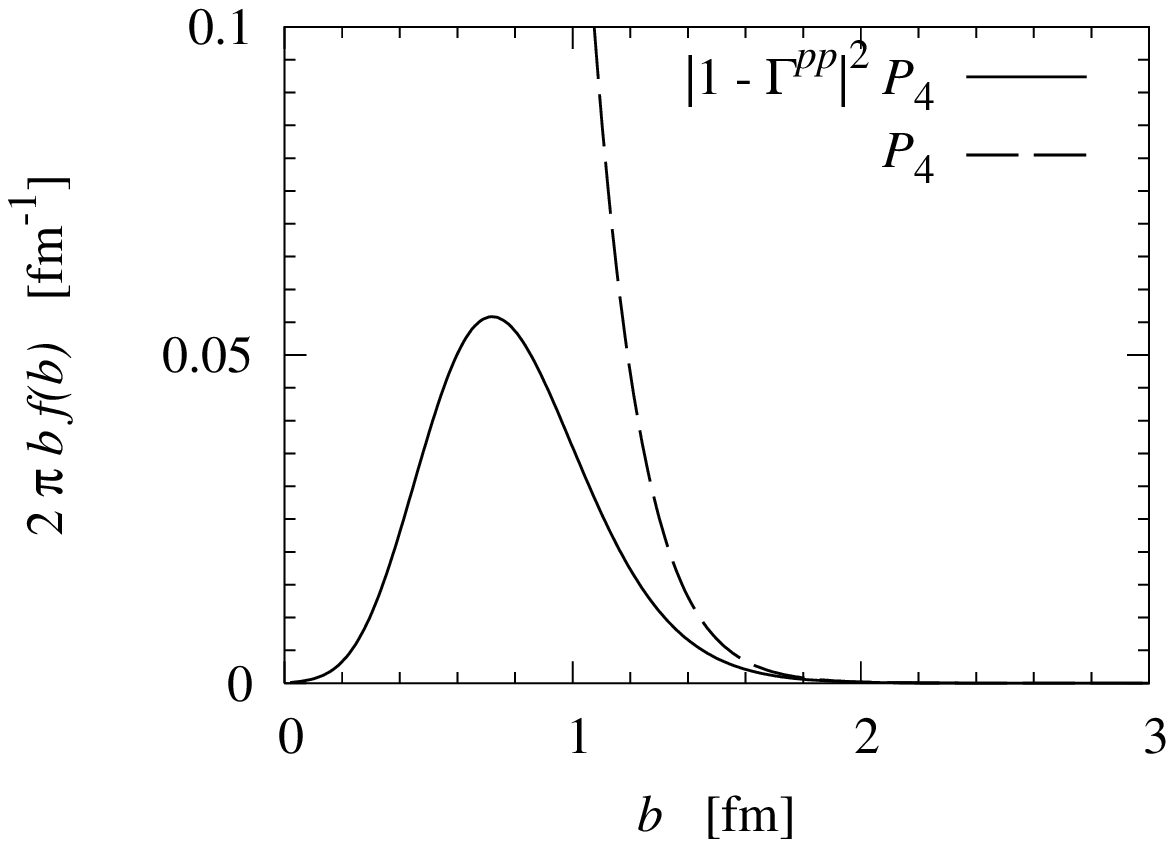}
\\
(b) \hspace{1em} \includegraphics[width=10cm]{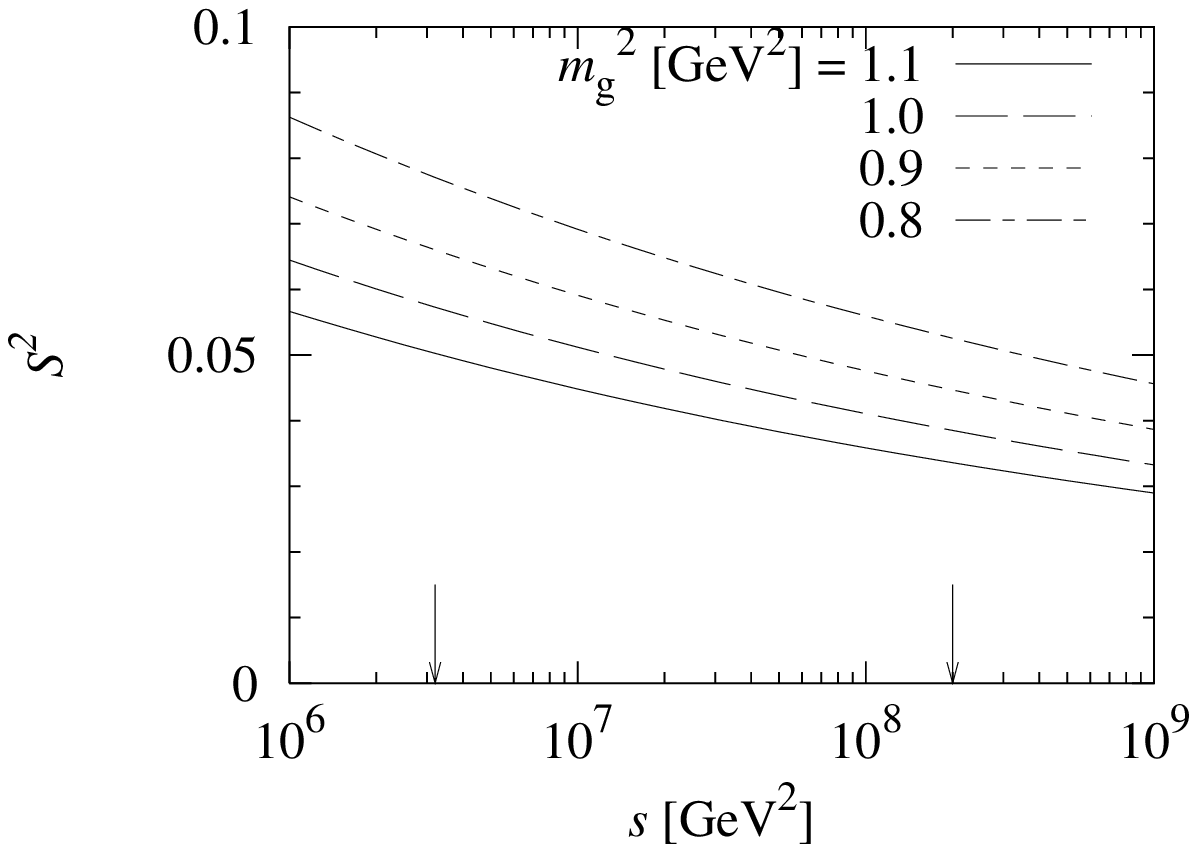}
\end{tabular}
\caption[]{(a) The impact parameter ($b$--) distribution of the cross 
section for diffractive Higgs production at LHC 
($\sqrt{s} = 14\, \text{TeV}$). Dashed line: $b$--distribution of the hard
process, $P_4 (b)$, Eq.~(\ref{P_4}), \textit{cf.}\ Fig.~\ref{fig:pb}.
Solid line: $b$--distribution of the total process, $|1 - \Gamma^{pp}
(s, b)|^2 P_4 (b)$.  Here, $m_g^2 = 1 \, \text{GeV}^2$.  (b) The
rapidity gap survival probability, $S^2$, Eq.~(\ref{surv})
\cite{Frankfurt:2004kn}.  Shown is the result as a function of $s$,
for various values of the mass parameter in the two--gluon form
factor, $m_g^2$.  The Tevatron and LHC energies are marked by arrows.
\label{fig:surv}}
\end{figure}

Hard diffractive processes are also being considered in connection
with the production of new heavy particles in $pp$ collisions at LHC.
In particular, the exclusive diffractive production of Higgs bosons,
\begin{equation}
p + p \;\; \rightarrow \;\; p \; + \; \text{(rapidity gap)} \; + \; H \; 
+ \; \text{(rapidity gap)} \; + \; p ,
\end{equation}
is regarded as a promising candidate for the Higgs search; see
Ref.~\cite{Kaidalov:2003fw} and references therein.  From the point of
view of strong interactions, this process involves a delicate
interplay between ``hard'' and ``soft'' interactions, which can be
described within our two--scale picture of the transverse structure of
the nucleon \cite{Frankfurt:2004kn}.  The Higgs boson is produced in a
hard partonic process, involving the exchange of two hard gluons
between the nucleons.  The impact parameter distribution of the cross
section for this process is described by the square of the convolution
of the transverse spatial distributions of gluons in the in and out
states, $P_4 (b)$, defined in Eq.~(\ref{P_4}), where the scale is of
the order of the gluon transverse momentum squared, 
$\sim M_H^2 / 4$. In addition, the soft interactions between the the
spectator systems have to conspire in such a way as not to fill the
rapidity gaps left open by the hard process.  The probability for this
to happen is approximately given by one minus the probability of an
inelastic $pp$ interaction at a given impact parameter, or 
$|1 - \Gamma^{pp} (s, b)|^2$.  The product of the two probabilities,
which determines the $b$--distribution for the total process, is shown
in Fig.~\ref{fig:surv}a.  At small $b$ the probability for no
inelastic interaction is very small $|1 - \Gamma^{pp}|^2 \approx 0$,
leading to a strong suppression of small $b$ in the overall
distribution.

The so-called rapidity gap survival probability, which measures the
``price'' to be paid for leaving the protons intact, is given by the
integral \cite{Frankfurt:2004kn}
\begin{equation}
S^2 \;\; \equiv \;\; \int d^2 b \; |1 - \Gamma^{pp} (s, b)|^2 \; P_4 (b) .
\label{surv}
\end{equation}
Fig.~\ref{fig:surv}b shows our result for this quantity, with $s$
ranging between Tevatron and LHC energies, for various values of the
dipole mass in the two--gluon form factor of the nucleon, $m_g^2$,
Eq.~(\ref{twogluon_dipole}).  The survival probability decreases with
$s$ because the size of the ``black'' region at small impact
parameters (in which inelastic interactions happen with high
probability) grows with the collision energy. Note that the effective
$x$ values in the gluon distribution decrease with the energy (for
fixed mass of the produced Higgs boson), resulting in smaller
effective values of $m_g^2$. This makes the actual drop of the
survival probability with energy slower than appears from the
fixed--$m_g^2$ curves of Fig.~\ref{fig:surv}b.  Our estimates of $S^2$
are in reasonable agreement with those obtained by Khoze et al.\
\cite{Khoze:2000wk} in a multi--Pomeron model, as well as with those
reported by Maor et al.\ \cite{Maor}.  In view of the different
theoretical input to these approaches this is very encouraging.

Our results for the rapidity gap survival probability apply equally
well to the production of two hard dijets instead of a Higgs boson.
For this process, one expects much larger cross section, and it would
be possible to investigate experimentally the interplay of hard
physics and absorptive effects, which leads to a rich, distinctive
structure of the cross section as a function of the transverse momenta
of the two protons, $\bm{\Delta}_{1\perp}$ and $\bm{\Delta}_{2\perp}$
\cite{Khoze:2002nf}. This structure should also rather strongly depend
on the rapidities of the jets, due to the $x$--dependence of the
transverse spatial distribution of gluons, see
Sec.~\ref{subsec:transverse} (L.~Frankfurt \textit{et al.}, in
preparation).
\subsection{Inclusive hard diffractive processes}
Inclusive hard diffractive processes, such as 
\begin{equation}
\begin{array}{lcl}
p + p &\rightarrow& p \; + \; \text{(rapidity gap)} \; + \; 
\text{2 jets} \; + \; X,
\\[1ex]
p + p &\rightarrow& p \; + \; \text{(rapidity gap)} \; + \; 
\text{2 jets} \; + \; X \; + \; \text{(rapidity gap)} \; + \; p, 
\end{array}
\end{equation}
offer a possibility to probe the ``periphery'' of the proton with hard
scattering processes. The cross section for these processes is again
suppressed compared to the naive estimate based on the diffractive
parton densities of the proton measured in $ep$ scattering at HERA.
As in the case of exclusive diffractive Higgs production, the cause of
this is the very small probability for the nucleons not to interact
inelastically at small impact parameters.  The suppression factors can
be estimated by generalizing the approach to the description of hard
and soft interactions outlined in Sec.~\ref{sec:higgs}. Simple
estimates along the lines of Eq.~(\ref{surv}) naturally reproduce the
suppression factors of the order $0.1 - 0.2$ observed at
Tevatron. However, the results in this case are more sensitive to the
details of the impact parameter dependence of the hard scattering
process and the soft spectator interactions.
 
Measurements of inclusive hard diffractive processes at LHC would
allow one to perform many interesting tests of the diffractive
reaction mechanism.  In particular, one could \textit{(a)} investigate
how the overall increase of the nucleon size with energy leads to a
suppression of hard diffraction, \textit{(b)} check how the rate of
suppression depends on the $x$--value of the parton involved in the
hard process, \textit{(c)} look for the breakdown of Regge
factorization, that is, the change of the diffractive parton
distributions with $x_{\Pomeron}$.
\section{Summary and Outlook}
\label{sec:outlook}
\subsection{From HERA to LHC}
The HERA experiments and the theoretical investigations they
stimulated have greatly advanced our knowledge of small--$x$ dynamics.
The key result of these studies are: \textit{(a)} The rapid increase
with energy of the cross section for the scattering of small $q\bar q$
wave packets from the nucleon. HERA energies are not sufficient to
reach the BDL in the dipole--nucleon interaction in average
configurations. The interaction of gluon dipoles in diffractive
scattering appears to be close to the unitarity limit for 
$Q^2 \sim 4\, \text{GeV}^2$, but this can hardly be verified directly
because of the lack of a trigger for such configurations.
\textit{(b)} The establishment of a three--dimensional picture of the
partonic structure of the nucleon. The leading partons are
concentrated in a much smaller transverse area than the area
associated with the nucleon in soft hadronic processes at high
energies.

We have demonstrated that these elements of small--$x$ dynamics are of
utmost importance for building a realistic description of $pp/pA$
collisions at LHC. The BDL will be commonplace in central $pp/pA$
collisions at LHC, affecting average configurations in the colliding
protons (nuclei), with numerous consequences for the hadronic final
state.  In particular, these phenomena qualitatively change the strong
interaction environment for new particle production.

We have identified several directions for future theoretical research,
necessary for describing the expected new phenomena at LHC.  These
include the resummation approaches to QCD radiation (combining
logarithms of $Q^2$ and $1/x$), the account for energy loss in the
interaction of leading partons with the small--$x$ gluon medium, and
the development of realistic models of hadron production in central
$pp/pA$ collisions with interactions close to the BDL.

An overarching goal of future theoretical research on the structure of
the nucleon and small--$x$ dynamics should be to bring together the
approaches starting from ``soft'' (hadronic) and ``hard'' (partonic)
physics, as envisioned in Gribov's space--time picture of high--energy
interactions. We have pointed out several instances in which ``soft''
and ``hard'' dynamics match smoothly or exhibit a delicate interplay,
\textit{e.g.}\ $pp$ elastic scattering at small impact parameters, or
diffractive processes.

A natural question is what are the most promising directions for
future experimental studies of small--$x$ dynamics.  This question
needs to be discussed with regard to the general long--term
perspectives in high--energy physics.  We assume here that the
decision by the DESY management to stop HERA operations in 2007 will
be enacted. This would clearly be a great loss, as many insights could
be obtained from further measurements at HERA in both $ep$ and $eA$
mode, see \textit{e.g.}\ the proposals put forward for the HERA III
run. We shall thus focus on the possibilities offered by LHC, with
comments on the possible future program at Tevatron, as well as on the
electron--ion collider envisaged in the U.S.\ government's long--range
plan \cite{EIC}. The small--$x$ investigations at LHC described in the
following subsections are meant to complement the studies of central
inelastic $pp/pA$ collisions (Sec.~\ref{subsec:BDL_elastic}) and
diffractive phenomena in $pp/pA$ scattering, which are the main topic
of this review.
\subsection{Measurement of parton densities in $pp$ and $pA$ collisions 
at LHC}
\label{subsec:parton}
Measurements at LHC could greatly expand the $x$--range in which the
parton densities are known. This would require measurements of hard
processes such as
\begin{equation}
\begin{array}{lcll}
pp & \rightarrow  & 
   \text{jet1} + \text{jet2} + X & \text{dijet production}          \\
&& \text{jet} + \gamma + X, \; \gamma + \gamma  +X   
                                 & \text{photon production}          \\
&& Q + {\bar Q} + X              & \text{heavy quark production}     \\ 
&& l^+ + l^- +X                  & \text{Drell--Yan pair production} \\
&& W^\pm (Z) + X                 & \text{weak boson production},     \\
\end{array}
\label{processes}
\end{equation}
in the region where one of the colliding partons carries small
momentum fraction. The cross sections of all these processes remain
large down to the very edge of the LHC kinematics, corresponding to 
$x \approx 3 \times 10^{-7}$ for Drell--Yan pair production with
$M_{\mu^+\mu^-} = 5 \, \text{GeV}$ \cite{Alvero:1998cb,FELIX}.  The
main limitations come from the need to identify relatively
low--$p_{\perp}$ jets, and from the detector acceptance.  The smaller
the $x$ one wants to probe, the more forward one must look, as the
momentum fractions of colliding partons are related to the rapidities
of the produced jets as
\begin{equation}
x_{1, 2} \;\; = \;\; 
\frac{p_{\perp}}{\sqrt{s}} \left( e^{\pm y_1} + e^{y_2} \right) .
\end{equation}
The presently planned configuration of the CMS detector would allow
for the measurement of dijet production down to 
$x \approx 3 \times 10^{-6}$ at $p_{\perp} = 10 \, \text{GeV}$. This
would push parton distribution measurements deep into the region where
unitarity effects play an important role in the dynamics of hard
processes, and where evolution effect in both $\ln (1/x)$ and 
$\ln Q^2$ need to be taken into account. When the BDL is reached, the
$M^2$--dependence of the cross section, $d\sigma/dx_1dx_2$, is
predicted to be much slower than $\propto 1/M^2$ as in the leading
twist approximation, similarly to the case of the inclusive deep
inelastic scattering \cite{implosion}.

If several of the reactions (\ref{processes}) were measured, it would
allow for independent tests of the QCD factorization, which may be
violated at intermediate virtualities owing to the strong interaction of
the propagating system with the small--$x$ gluon medium, see
Sec.~\ref{sec:hadron}. The latter will be strongly enhanced in the
region of $x_1$ close to 1. These effects, which are of great interest
in themselves, can be probed by comparing the production cross
sections for fixed, large $x_1$ and various values of $x_2$, including
relatively large ones where the parton densities are known.
 
\subsection{Small--$x$ phenomena in ultraperipheral collisions at LHC}
\label{subsec:ultraperipheral}
It has been long known that nuclei in high--energy collisions generate
a large flux of equivalent photons, which are spread in the transverse
plane over distances substantially larger than twice the nuclear
radius --- the maximal distance at which strong interactions are
possible. Scattering processes induced by these photons are referred
to as ultraperipheral collisions. They have a distinctive signature,
which allows them to be separated from the more frequent events caused
by strong interactions. Experimentally, one selects events in which
one of the nuclei remains intact, or emits one or a few neutrons by
way of dipole excitation. Such events are extremely rare in scattering
at impact parameters smaller than twice the nuclear radius.

The experiments at HERA have shown that photon--induced processes
provide a well--understood probe of the gluon density in the proton.
At LHC, such processes could be studied up to invariant $\gamma p$
energies (\textit{i.e.}, $\gamma A$ energies per nucleon) exceeding
the maximal HERA energy by a factor of 10. This would allow one to use
dijet (charm, \textit{etc.}) production to measure the gluon density
in the proton/nucleus down to $x \approx 3\times 10^{-5}$ for 
$p_\perp \sim \text{few GeV}$ \cite{Klein:2002wm,White}, as well as
the diffractive gluon density. Among other things, measurements of
diffractive channels would allow one to perform critical tests of the
HERA observation of a large probability of gluon--induced diffraction
(see Sec.~\ref{sec:diffraction}), and the prediction of its further
enhancement for nuclear targets, see Ref.~\cite{Frankfurt:2003gx} and
references therein.  Another important measurement would be the
$t$--dependence of gluon--induced diffraction and its change with
energy, using CMS--TOTEM in $pA$ mode. We remind the reader that the
lack of direct information on the $t$--dependence of diffraction leads
to a large uncertainty in the predictions for leading--twist nuclear
shadowing (see Sec.~\ref{subsec:shadowing}).

Ultraperipheral collisions would also allow one to study the coherent
production of heavy quarkonia, $\gamma A \to J/\psi \, (\Upsilon) + A$
at $x \leq 10^{-2}$, and to investigate the propagation of small
dipoles through the nuclear medium at high energies. The QCD
factorization theorem predicts that the $A$--dependence of the
amplitude for this process should change between the color
transparency regime (observed at FNAL \cite{Sokoloff:1986bu}), where
it is $\propto A$, and the perturbative color opacity regime, where it
is proportional to the leading--twist shadowed gluon density.  It
would be possible also to use coherent diffraction from nuclei to
study the approach to the BDL in $\gamma A\rightarrow X + A$, by
comparing the measured cross section to the BDL prediction,
\textit{cf.}\ Sec.~\ref{subsec:black_heavy}. The most promising
channels are $J/\psi$ and dijet photoproduction; see
Ref.~\cite{Frankfurt:2003wv} for a review and discussion.  In $AA$
collisions, it is difficult to separate processes induced by
the photons generated by the left-- and right--moving nucleus.  Away
from zero rapidity, a low--energy contribution tends to dominate,
limiting the range of $x$ which could be explored for production of a
state with mass $M$ to $x \geq M A/(2 p_A)$, where $p_A$ is the
momentum of the colliding nuclei. However, it seems that selection of
events in which the heavy nucleus undergoes a dipole excitation
enhances the contribution of hard photons \cite{Baltz:2002pp},
allowing one to extend the $x$--range of the measurements (by a factor
of up to 10 in the case of $J/\psi$ production). The challenge is to
trigger both on events with and without break--up. In $pA$ mode, the
dominant process will be the production of heavy quarkonia. Such
measurements would extend the $W$--range of the HERA measurements by a
factor of three, and make it possible to measure directly the
$t$--dependence of the cross section in a very broad range of
rapidities, using the proposed 420m proton tagger \cite{FP420},
which is critical for a more accurate determination of the
$x$--dependence of the nucleon's transverse structure, see
Sec.~\ref{subsec:transverse}. Note also that in $pp$ scattering it
is possible to detect protons at very small momentum transfers, where
Coulomb exchange dominates \cite{Krzystof}.  This would allow one to
measure exclusive photoproduction of heavy quarkonia in $pp$
scattering with good statistics \cite{Khoze:2002dc,Klein:2003vd}.
\subsection{Small--$x$ physics at RHIC and an electron--ion collider}
The LHC measurements described in Sections~\ref{subsec:parton} and
\ref{subsec:ultraperipheral} will probe small--$x$ dynamics at least
down to $x \sim 10^{-5}$. However, most of these measurements are
restricted to scales $Q \geq 5\, \text{GeV}$, and it would be
difficult to connect them with the physics at smaller scales
(virtualities), $Q \sim 2-3 \, \text{GeV}$, relevant for the overall
structure of central $pp/pA$ collisions at LHC (see
Sec.~\ref{sec:hadron}).  The gap could be filled, to some extent, by
experiments at RHIC and the proposed electron--ion collider
\cite{EIC,Deshpande:2005wd}.

Extension of the forward acceptance of the current RHIC detectors
would make it possible to measure Drell--Yan pair production at 
$x\sim 10^{-3}$ in $pp$ and $pA/dA$ scattering. This would allow one
to test the predictions for leading--twist nuclear shadowing and look
for deviations from the leading--twist prediction in the $p_{\perp}$
distributions of the dileptons. Qualitatively, one expects a
suppression of the low transverse momentum part of the distribution up
to $p_{\perp} \sim p_{\perp,\text{BDL}}$. As mentioned above, such
measurements would also allow one to probe the role of final state
interactions by varying the $x$--values of the leading partons in the
proton. If absorption effects were significant, one would have to
introduce a cut on $x_p\le 0.3$ to suppress these effects, which would
reduce somewhat the $x_A$--range where the parton densities can be
probed.

The eRHIC design for a future electron--ion collider envisages an
$ep/eA$ collider with $\sqrt{s}\leq 100\, \text{GeV}$, with
significantly higher luminosity than HERA, and the ability to
continuously vary the beam energies over a wide range
\cite{EIC,Deshpande:2005wd}.  With such a facility one could
systematically study a variety of color transparency phenomena and use
them to disentangle the quark--gluon structure of hadrons and nuclei;
one could also measure longitudinal cross sections, which provide
stringent tests of the range of the validity of the leading--twist
approximation at small $x$ (see Sections~\ref{sec:inclusive},
\ref{sec:exclusive} and \ref{sec:diffraction}).  In $eA$ collisions,
one could study the transition of the nonperturbative shadowing at low
$Q^2$ to the regime of leading--twist shadowing at high $Q^2$, and
explore whether there exists an ``intermediate'' regime characterized
by weak coupling but large parton densities. The ability to perform
such measurements with a range of nuclear beams would allow one to
study these effects as a function of the nuclear thickness, reaching
values 1.5 times larger than the average thickness of the heavy
nuclei.  No other planned facilities would be able to cover this
important kinematic region. Finally, eRHIC would make it possible to
measure the $t$--dependence of a variety of hard exclusive processes
in a wide range of $x$, $0.1> x \ge 0.003$.  This would probe the
transverse structure of the proton directly in the $x$--range relevant
for understanding nucleon fragmentation in central $pp/pA$ collisions
at LHC.
\\[1cm]
{\bf Acknowledgments.}
We would like to thank our colleagues, many of them collaborators of
many years, for their contributions to the studies discussed in this
review and many enjoyable discussions, in particular H.~Abramowicz,
J.~Bjorken, S.~Brodsky, J.~Collins, J.~Dainton, Yu.~Dokshitzer,
A.~DeRoeck, H.~Drescher, A.~Dumitru, K.~Eggert, A.~Freund, V.~Gribov,
V.~Guzey, A.~Levy, L.~Lipatov, M.~McDermott, G.~Miller, A.~Mueller,
A.~Radyushkin, T.~Rogers, W.~Vogelsang, R.~Vogt, H.~Weigert, S.~White,
and M.~Zhalov. This work is supported by U.S. Department of Energy
Contract DE-AC05-84ER40150, under which the Southeastern Universities
Research Association (SURA) operates the Thomas Jefferson National
Accelerator Facility. L.~F.\ and M.~S.\ acknowledge support by the
Binational Scientific Foundation. The research of M.~S.\ was supported
by DOE. M.S.~thanks the Frankfurt Institute for Advanced Studies at
Frankfurt University for the hospitality during the time when this
work was completed.
\vfill\eject
\end{document}